\documentclass[12pt]{article}
\newlength{\abstractwidth}
\renewcommand{\thefootnote}{\fnsymbol{footnote}}
\renewcommand{\thanks}[1]{\footnote{#1}}
\newcommand{\starttext}{
\setcounter{footnote}{0}
\renewcommand{\thefootnote}{\arabic{footnote}}}

\newcommand{\bea}{\begin{eqnarray}}
\newcommand{\eea}{\end{eqnarray}}
\newcommand{\<}{\langle}
\renewcommand{\>}{\rangle}

\def\A{{\cal A}}
\def\B{{\cal B}}
\def\C{{\cal C}}

\def\E{{\cal E}}

\def\L{{\cal L}}
\def\M{{\cal M}}

\def\O{{\cal O}}
\def\P{{\cal P}}
\def\Q{{\cal Q}}
\def\R{{\cal R}}

\def\V{{\cal V}}
\def\W{{\cal W}}

\def\Y{{\cal Y}}
\def\Z{{\cal Z}}
\def\B{{\cal B}}

\def\Im{{\rm Im}}
\def\tr{{\rm tr}}
\def\det{{\rm det}}

\def\half{ {1\over 2}}

\def\p{\partial}

\def\tet{\vartheta}
\def\chiz{{\chi _{\bar z}{} ^+}}

\def\muhat{{\hat \mu _{\bar z} {}^z}}
\def\e{\epsilon}
\def\ep{\varepsilon}
\def\no{\nonumber}
\def\zet{\zeta}
\def\I{{\cal I}}
\def\J{{\cal J}}

\begin{document}
\starttext
\baselineskip=16pt
\setcounter{footnote}{0}

\begin{flushright}
UCLA/05/TEP/3 \\
Columbia/Math/05 \\
2005 January 24 \\
\end{flushright}

\bigskip

\begin{center}
{\Large\bf TWO-LOOP SUPERSTRINGS VI}\\
\bigskip
{\large \bf Non-Renormalization Theorems and the 4-Point Function}
\footnote{Research supported in part by National Science
Foundation grants PHY-01-40151 and DMS-02-45371.}

\bigskip\bigskip
{\large Eric D'Hoker$^a$ and D.H. Phong$^b$}

\bigskip
$^a$ \sl Department of Physics and Astronomy \\
\sl University of California, Los Angeles, CA 90095, USA\\
$^b$ \sl Department of Mathematics\\
\sl Columbia University, New York, NY 10027, USA

\end{center}

\bigskip\bigskip

\begin{abstract}

The N-point amplitudes for the Type II  and Heterotic superstrings 
at two-loop order and for $N  \leq 4$ massless NS bosons are evaluated 
explicitly from first principles, using the method of projection onto super period 
matrices introduced and developed in the first five papers of this series.
The gauge-dependent corrections to the vertex operators, identified in 
paper V, are carefully taken into account, and the crucial counterterms 
which are Dolbeault exact  in one insertion point and de Rham closed in 
the remaining points are constructed explicitly. This procedure maintains 
gauge slice independence at every stage of the evaluation.
Analysis of the resulting amplitudes demonstrates,
from first principles, that for $N\leq 3$,
no two-loop corrections occur, while for $N=4$, no two-loop 
corrections  to the low energy effective action occur for 
$R^4$ terms in the Type II superstrings, and for $F^4$, $F^2F^2$, $F^2R^2$,
and $R^4$ terms in the Heterotic strings.

\end{abstract}

\vfill\eject

\baselineskip=15pt
\setcounter{equation}{0}
\setcounter{footnote}{0}

\section{Introduction}
\setcounter{equation}{0}

The gauge fixing procedure of \cite{I,II,III,IV}, based on the projection of supergeometries to their super period matrices, has produced a gauge slice 
independent chiral superstring measure\footnote{A survey is
given in \cite{dp02}.}. For the $N$-point function, the same procedure applies, 
after insertion of the corresponding vertex operators. 
As shown in \cite{V}, the procedure produces readily gauge slice independent 
$N$-point functions, provided some important subtleties in the vertex operators 
are treated with proper care.
In the present paper, we work it out explicitly for the $N$-point function for 
$N\leq 4$ and massless NS bosons, for both the Type II and Heterotic superstrings.
The resulting formulas exhibit clearly the mechanism by which all dependence on gauge slice choices cancels. Central to this mechanism is a new
scalar function $\Lambda(z)$, resulting from the subtleties in the vertex operators.

\medskip

The $N$-point functions for $N\leq 4$ and massless bosons to higher loop 
orders have been considered by many authors over the years. In the earlier 
literature dating from the 1980's  on multi-loops \cite{friedan,old0,vv1,bis2}, (more 
extensive bibliographies were given in \cite{II,dp88})
and specifically 2-loop \cite{bis1,bis3,lech,iz} amplitudes,
the gauge-fixing procedure was known to suffer from ambiguities, that is, the
gauge-fixed amplitudes ended up being gauge slice dependent
\cite{vv1,ambiguities}. In retrospect, we know now that an ill-defined projection from
supergeometries to their standard period matrices was implicitly used. 
More recently, Zheng, Wu, and Zhu \cite{zwzI,zwzII,zwzIII} have taken up anew the
evaluation of the $N$-point function for $N\leq 4$, using this time the gauge 
slice independent superstring measure  of \cite{I,II,III,IV}. However, the 
ensuing subtleties in the vertex operators  were not taken into account,
and the gauge slice independence of the amplitudes was again  broken 
along the way. In particular, a gauge-fixed formula was found which remained dependent on the gauge slice chosen.
A final gauge invariant formula could only be obtained
after certain ad hoc remedies.
Although for the 4-point function, the correct answer (see (\ref{TypeII}) below) 
can be recovered in this way, this may not be the case for other amplitudes, 
and we need a reliable evaluation procedure insuring gauge slice independence.

\medskip

As explained in \cite{V}, the subtleties with vertex operators come from the
following two interrelated sources.
First, to insure covariance under local worldsheet supersymmetry, the 
naive (un-integrated) vertex operator 
${\cal V}^{(0)}(z,\epsilon,k)$ for emission of the supergraviton multiplet
\bea
{\cal V}^{(0)}(z,\epsilon,k)=
\epsilon^\mu dz(\p_zx_+^\mu-ik^\nu\psi_+^\mu\psi_+^\nu)(z)
\,e^{ik\cdot x(z)}
\eea
has to be completed into the following vertex operator ${\cal V}(z,\epsilon,k)$
\bea
\label{V}
{\cal V}(z,\epsilon,k)=
{\cal V}^{(0)}(z,\epsilon,k)
+
{\cal V}^{(1)}(z,\epsilon,k)
+
{\cal V}^{(2)}(z,\epsilon,k)
\eea
with the corrections ${\cal V}^{(1)}$ and ${\cal V}^{(2)}$
incorporating the chiral volume element on the (super) worldsheet
and the Beltrami differential $\muhat $
for the deformation of complex structure from period matrix $\Omega_{IJ}$ to
super period matrix
$\hat\Omega_{IJ}$
\bea
\label{V1V2}
{\cal V}^{(1)}(z,\epsilon,k)&=& -\half\epsilon^\mu d\bar
z\chiz\psi_+^\mu(z)\,e^{ik\cdot x(z)}\nonumber\\
{\cal V}^{(2)}(z,\epsilon,k)&=&
-
\epsilon^\mu\hat\mu_{\bar z}{}^z d\bar z(\p_zx_+^\mu
-ik^\nu\psi_+^\mu\psi_+^\nu)(z)\,
e^{ik\cdot x_+(z)}.
\eea
Second, even though the $(x_+^\mu,\psi_+^\mu)$-correlation functions
$\<\prod_{i=1}^N{\cal V}(z_i,\epsilon_i,k_i)\>$
are {\it chiral} (in the sense of
depending only on $z_i$, $\hat\Omega_{IJ}$ and $\chiz$
and not on the conjugates $\bar z_i$, $\hat\Omega_{IJ}^*$
and $\chi_z{}^-$), they are {\it not holomorphic}
with respect to the complex structure defined by the super period matrix
$\hat\Omega_{IJ}$. This is most readily seen from the
fact that the corrections ${\cal V}^{(1)}$ and ${\cal V}^{(2)}$
are $(0,1)$-forms.
On $(0,1)$-forms, the derivative $\nabla_{\bar z}$ requires a
connection, and thus the condition $\nabla_{\bar z}\phi=0$ does not provide an
appropriate notion of holomorphicity.
The appearance of these non-trivial $(0,1)$-components is a major difficulty
in descending from {\it superholomorphic} correlation functions
on the super worldsheet to {\it holomorphic} correlation
functions on the worldsheet itself.

\subsection{Chiral and Holomorphic Superstring Amplitudes}

The key to the solution is to show that all terms from
$\<\prod_{i=1}^N{\cal V}(z_i,\epsilon_i,k_i)\>$ which are not $(1,0)$-forms in
each insertion point $z_i$, can be grouped into
terms which are Dolbeault $\bar\p$-exact differentials in one insertion point,
and de~Rham $d$-closed forms in all the other insertion points (see
\cite{V}, eq.(1.12)). This can be done for each even spin structure on the
worldsheet \cite{VII}. However, the process is quite involved
in practical calculations. If we are concerned solely with the final 
GSO-projected superstring amplitude, only a sum over spin structures is needed.
In the unitary gauge, where the supports
$q_{\alpha}$ of the gravitino slice 
$\chi(z)=\sum_{\alpha=1}^2\zeta^\alpha\delta(z,q_\alpha)$
are chosen to
be the zeroes of a certain holomorphic $(1,0)$-form $\varpi$, 
\bea
\varpi(q_{\alpha})=0,
\qquad \alpha=1,2,
\eea
the sum over spin structures simplifies considerably if $N\leq 4$, and the
process of identifying the Dolbeault $\bar\p$-exact forms becomes 
much more transparent.
In the present paper, we shall restrict ourselves to this case.
The gauge slice independence of the amplitudes corresponds then
to the independence from the choice of $\muhat $ and points $q_\alpha$, or
equivalently, from the choice of $\muhat $ and holomorphic form
$\varpi(z)$.

\medskip

To describe our results more precisely, we recall
that the chiral superstring measure determined in \cite{I,II,III,IV} is given by
\bea
\label{chiralmeasure}
d\mu_0[\delta](\Omega)+\zeta^1\zeta^2\,d\mu_2[\delta](\Omega),
\eea
with (to simplify the notation, in accord with previous practice,
we denote $\hat\Omega_{IJ}$ simply by $\Omega_{IJ}$ after the
deformation of complex structures has already been carried out)
\bea
\label{dmu0dmu2}
d\mu_0[\delta](\Omega)
&=& \Z[\delta]\,\prod_{I\leq J}d\Omega_{IJ},
\nonumber\\
d\mu_2[\delta](\Omega)
&=&
{\tet[\delta](0,\Omega)^4\Xi_6[\delta](\Omega)
\over 16\pi^6\Psi_{10}(\Omega)}\,\prod_{I\leq J}d\Omega_{IJ}
\eea
and the chiral partition function $\Z [\delta]$ is given by,
\bea
\Z[\delta]={\<\prod_{a=1}^3b(p_a)
\prod_{\alpha=1}^2\delta(\beta(q_\alpha))\>\over
\det\,\omega_I\omega_J(p_a)}
\eea
Here $\zeta^\alpha$ are the odd supermoduli parameters,
$p_a$, $1\leq a\leq 3$, are arbitrary points of which the measure is manifestly
independent, and $q_\alpha$, $1\leq\alpha\leq 2$,
are arbitrary points at which the gravitino gauge slice
$\chiz=\sum_{\alpha=1}^2\zeta^\alpha\delta(z,q_\alpha)$ is supported.
As indicated above, in this paper, we choose the $q_\alpha$'s
to be the zeroes of a holomorphic 1-form $\varpi(z)$.
The spin structure is denoted by $\delta$, and $\omega_I(z)$
are a canonical basis of holomorphic $(1,0)$-forms.
In \cite{V}, it was shown that the chiral amplitude for the $N$-point function
is given by
\bea
\label{BcBd}
{\cal B}[\delta] ~~
&=&{\cal B}[\delta]^{(d)}+{\cal B}[\delta]^{(c)}
\nonumber\\
{\cal B}[\delta]^{(d)}
&=&
d\mu_2[\delta]\,\bigg\<Q(p_I)\prod_{i=1}^N\V^{(0)}(z_i,\epsilon_i,k_i)\bigg\>
\nonumber\\
{\cal B}[\delta]^{(c)}
&=&
d\mu_0[\delta]\int \! d^2 \! \zeta \,
\big(\Y_1+\Y_2+\Y_3+\Y_4+\Y_5)
\eea
The $p_I$ are the internal loop momenta, and $Q(p_I)$ is the insertion
required by chiral splitting
\bea
\label{Q}
Q(p_I)= \exp \,\bigg \{ ip_I^\mu\oint_{B_I}dz\,\p_zx_+^\mu(z) \bigg \}
\eea
The fields $x_+^\mu$ are effective chiral bosons, with propagator 
$\<x_+^\mu(z)x_+^\nu(z)\>=-\delta^{\mu\nu}\ln E(z,w)$, where $E(z,w)$ 
is the prime form. The precise formulas for $\Y_1,\cdots,\Y_5$ are given in 
section \S 2. In essence, $\Y_1$ is the familiar term
corresponding to the naive vertex operator $\V^{(0)}$,
$\Y_2$ is due to the deformation of complex structures,
and $\Y_3,\Y_4,\Y_5$ are due to the corrections $\V^{(1)}$ and
$\V^{(2)}$ in the vertex operators. We note that both $d\mu_2[\delta]$ as well
as all the $\Y_k$'s depend on the $q_\alpha$'s. Henceforth, we shall suppress
the measure
$\prod_{I\leq J}d\Omega_{IJ}$ from our notation since it appears as a common
factor in all our formulas.

\medskip

The GSO projection introduces a unique set of phases
for the summation over spin structures, which were shown in \cite{IV}
to all be equal to 1. Thus we consider the sums $\sum_\delta {\cal B}[\delta]^{(c)}$
and $\sum_\delta{\cal B}[\delta]^{(d)}$. 
For the ${\cal B}[\delta]^{(d)}$ component, the gauge slice independence 
as well as the holomorphicity is manifest. In \cite{V}, we have seen that,
although this gauge slice independence and holomorphicity does not hold for
${\cal B}[\delta]^{(c)}$, it will hold for a certain holomorphic component
of ${\cal B}[\delta]^{(c)}$ which is all that matters.
Here we shall exhibit this phenomenon explicitly for the sum over spin
structures $\sum_\delta{\cal B}[\delta]^{(c)}$.
Let $\Sigma$ be the worldsheet, and let
\bea
T^{z_i}(\Sigma)=T_{1,0}^{z_i}(\Sigma)\oplus T_{0,1}^{z_i}(\Sigma)
\eea
denote the space of $1$-forms in $z_i$ on $\Sigma$, decomposed 
into $(1,0)$ and $(0,1)$ forms. For gravitino slices supported at 
$q_1$ and $q_2$, the Beltrami differential $\hat\mu$ which deforms the
period matrix $\Omega_{IJ}$ to the super period matrix $\hat\Omega_{IJ}$ 
is of the form $\muhat =S_\delta(q_1,q_2)\,\mu(z)$. A first important 
observation is that $\mu(z)\,\varpi(z)$ is a $(0,1)$ form which pairs to 
$0$ against any holomorphic differential (see section \S 4). Thus the function 
$\Lambda(z)$ defined (up to an additive constant) by
\bea
\label{Lambda}
\p_{\bar z}\Lambda(z)=\mu(z)\,\varpi(z),
\eea
is single-valued and smooth if $\mu$ is chosen to be smooth.
Now both ${\cal B}[\delta]^{(c)}$ and
${\cal B}[\delta]^{(d)}$ are closed 1-forms in each variable $z_i$. However,
${\cal B}[\delta]^{(d)}$
is in $\otimes_{i=1}^NT_{1,0}^{z_i}(\Sigma)$,
while we only have
${\cal B}[\delta]^{(c)}\in \otimes_{i=1}^NT^{z_i}(\Sigma)$.
The crucial Dolbeault cohomology statement which we need is the following: for
$N\leq 3$,
$\sum_\delta {\cal B}[\delta]^{(c)}$
is $0$,
while for $N=4$, we have
\bea
\label{Dolbeault}
\sum_\delta {\cal B}[\delta]^{(c)} (z_i; \epsilon_i,k_i, p_I)  
- \sum_{j=1}^4d\bar z_j\p_{\bar z_j} {\cal S}_j (z_i; \epsilon_i,k_i, p_I)
\in\otimes_{i=1}^4T_{1,0}^{z_i}(\Sigma)
\eea
where the forms ${\cal S}_j$ are given by,
\bea
{\cal S}_j (z_i; \epsilon_i,k_i, p_I) 
= 
\half\, K \,\Z_0 
\bigg( \int \! d^2 \zeta \, \Lambda(z_j)
\,\<Q(p_I)\prod_{m=1}^4e^{ik_m\cdot x_+(z_m)}\>\prod_{m\not=j}\varpi
(z_m)\bigg),
\eea
Here, ${\cal Z}_0$ is a $\delta$ and $z_i$ independent factor,
and $K = K(\epsilon , k) = t_8 f_1 f_2 f_3 f_4$ is the well-known 
kinematic factor common  to tree level and one-loop (and $f_i ^{\mu \nu}
= \epsilon _i ^\mu k_i ^\nu - \epsilon _i ^\nu k_i ^\mu$). Thus, if we 
complete $\bar\p_j{\cal S}_j$ into a de~Rham exact form by 
$\bar\p_j{\cal S}_j=d_j{\cal S}_j-\p_j{\cal S}_j$ and write $\sum_\delta{\cal B}[\delta]$ as
\bea
\label{H}
\sum_\delta{\cal B}[\delta] (z_i; \epsilon_i,k_i, p_I)
=
{\cal H}(z_i; \epsilon_i,k_i, p_I) + 
\sum_{j=1}^4d_j {\cal S}_j (z_i; \epsilon_i,k_i, p_I)
\eea
the form ${\cal H}(z_i; \epsilon_i,k_i, p_I)$ will be in
$\otimes_{i=1}^4T_{1,0}^{z_i}(\Sigma)$. It will be automatically
holomorphic since ${\cal B}[\delta]$ is a closed $1$-form in each variable
$z_i$. The form ${\cal H}$ can then be evaluated explicitly. All dependences on
gauge choices cancel in ${\cal H}$, and we find
\bea
\label{explicitH}
{\cal H}
=
{1\over 64\pi^2} K
\,{\cal Y}_S \,
\exp \bigg \{ i\pi p_I^\mu\Omega_{IJ}p_J^\mu
+
2\pi i\sum_jp_I^\mu k_j^\mu\int_{z_0}^{z_j}\omega_I \bigg \}
\prod_{i<j}E(z_i,z_j)^{k_i\cdot k_j},
\eea
where $\omega_I$ is a canonical basis of holomorphic differentials, $E(z_i,z_j)$
is the prime form, and the factor ${\cal Y}_S$ is given by
\bea
\label{Ystar}
3\,{\cal Y}_S
&=&
+ (k_1-k_2)\cdot(k_3-k_4) \, \Delta (z_1,z_2) \Delta (z_3,z_4) 
\nonumber\\
&&
+
(k_1-k_3)\cdot(k_2-k_4) \, \Delta (z_1,z_3) \Delta (z_2,z_4)
\nonumber\\
&&
+
(k_1-k_4)\cdot(k_2-k_3) \, \Delta (z_1,z_4) \Delta (z_2,z_3)
\eea
where the basic antisymmetric biholomorphic 1-form $\Delta$ is defined by,
\bea
\label{Delta}
\Delta (z,w) \equiv 
\omega _1(z) \omega _2 (w) - \omega _1(w) \omega _2 (z). 
\eea
The formulas (\ref{H}) and (\ref{explicitH}) show clearly
the essential step in the evaluation of the $N$-point function: the {\it chiral}
amplitude $\sum_\delta{\cal B}[\delta]$ is not gauge slice
independent. Only the {\it holomorphic} amplitude ${\cal H}$ is, which is
obtained by combining the $\otimes_{i=1}^4T_{1,0}^{z_i}(\Sigma)$
component of $\sum_\delta{\cal B}[\delta]$
with the vertex operator corrections
\bea
-\sum_{j=1}^4dz_j\,\p_{z_j}
\bigg( \int \! d^2 \zeta \, \Lambda(z_j)
\,\<Q(p_I)\prod_{m=1}^4e^{ik_m\cdot x_+(z_m)}\>\prod_{m\not=j}\varpi
(z_m)\bigg).
\eea
Thus, perhaps paradoxically, the seemingly undesirable
$(0,1)$-forms in the chiral amplitudes have supplied
the crucial counterterms for both the gauge slice
independence and the holomorphicity of ${\cal H}$.

\subsection{Type II Amplitudes}

The physical scattering amplitudes can be obtained by pairing the
chiral amplitudes of left with those of right movers, at common internal loop
momenta $p_I^\mu$. As shown in \cite{V}, \S 5, the holomorphicity of ${\cal
H}$, together with the fact that ${\cal H}$ as well as the $d$-exact terms have
the same monodromy, implies that the $d$-exact terms in (\ref{H}) drop out. Combining all these ingredients, one finds
the following expression for the Type II amplitude,
\bea
\label{dolbeaultcancellation}
\int dp_I^\mu
\int_{\Sigma^4}
\bigg|\sum_\delta {\cal B}[\delta](z_i,\epsilon_i,k_i;p_I^\mu)\bigg|^2
=
\int dp_I^\mu
\int_{\Sigma^4}
\bigg|{\cal H}(z_i,\epsilon_i,k_i;p_I^\mu)\bigg|^2
\eea
We obtain then for the $4$-point function of the Type II superstring
\bea
\label{TypeII}
{\bf A}_{II} (\epsilon_i, k_i)
&=&
{K \bar K \over 2^{12} \pi ^4}  \int |\prod_{I\leq J}d\Omega_{IJ}|^2
\int_{\Sigma^4}
\,
|{\cal Y}_S|^2\,
\prod_{i<j}|E(z_i,z_j)|^{2k_i\cdot k_j}
\\
&&
\qquad\times
\int dp_I
\,
\bigg|{\rm exp}(i\pi p_I^\mu\Omega_{IJ}p_J^\mu
+
2\pi i\sum_jp_I^\mu k_j^\mu\int_{z_0}^{z_j}\omega_I)\bigg|^2
\no \\
{\bf A}_{II} (\epsilon_i, k_i)
&=&
{ K \bar K  \over 2^{12} \pi ^4} \int {|\prod_{I\leq J}d\Omega_{IJ}|^2
\over
({\rm det}\,{\rm Im}\,\Omega)^5}
\int_{\Sigma^4}
|{\cal Y}_S|^2
{\rm exp}\bigg(-\sum_{i<j}k_i\cdot k_j\,G(z_i,z_j)\bigg),
\nonumber
\eea
where $G(z,w)$ is the conformally invariant Green's function
\bea
\label{scalarprop}
G(z,w)=-\ln |E(z,w)|^2+2\pi ({\rm Im}\,\Omega)_{IJ}^{-1} \, 
\biggl ( {\rm Im}\int_z^w\omega_I \biggr )
\biggl ( {\rm Im}\int_z^w\omega_J \biggr )
\eea
Holomorphicity is well known to be essential for the construction of the
Heterotic string. It may be worth stressing that
it is here also necessary for
the gauge slice independence of the gauge-fixed amplitudes.

\medskip
The formula (\ref{TypeII}) shows that the
amplitude is finite. For purely imaginary values of the
Mandelstam variables $s_{ij}=-2k_i\cdot k_j$,
the integrals converge, since the volume of the moduli space of Riemann surfaces of genus 2 with respect to the $SL(4,{\bf Z})$
invariant measure $|\prod_{I\leq J}d\Omega_{IJ}|^2({\rm det}\,
{\rm Im}\,\Omega)^{-3}$
is finite.
The finiteness for general $s_{ij}$ has to be understood in the sense of analytic continuation, as shown in \cite{dp92}
for one-loop amplitudes.

\medskip

Translated into the hyperelliptic formalism,
the formula (\ref{TypeII}) agrees with the
formula (109) of Zheng, Wu, and Zhu \cite{zwzIII}.
Not surprisingly, point by point over the moduli space of
Riemann surfaces,
it differs from the ones of the
earlier literature for the 4-point function
\cite{old0, bis3, iz}. In \cite{wz},
it is argued that, although pointwise different, the
formula (\ref{TypeII}) and the one from
\cite{iz} have the same integral over the whole of
moduli space. However, the gauge-fixed superstring integrand
is a well-defined notion pointwise on moduli space, and
should be uniquely prescribed by the holomorphic component of
(\ref{H}).

\subsection{Heterotic Amplitudes}

To obtain Heterotic string amplitudes, we pair the form (\ref{explicitH}) 
with the holomorphic blocks, at common loop momenta $p_I^\mu$, of the 
$10$-dimensional bosonic string, coupled with 32 worldsheet chiral fermions $\lambda^I(z)$, $I=1,\cdots,32$ \cite{Gross:1985fr}. By the cancelled 
propagator argument, we can ignore the poles at coincident insertion points
of these blocks \footnote{Mathematically, this corresponds to an
analytic continuation in the Mandelstam variables. See for example \cite{dp92}
for the analytic continuation process in
the case of one-loop.}. The monodromies of both left and right movers are still
the same. Thus the $d$-exact terms
again drop out by holomorphicity. Let $HO$ and $HE$ be respectively the
$Spin(32)/{\bf Z}_2$ and the $E_8\times E_8$
Heterotic strings. We obtain in this manner the following amplitudes: for the
scattering of 4 gauge bosons,
we have
\bea
\label{bfAF4}
{\bf A}_{F^4}&=&
{\bar K\over 64\pi^{14}}
\int_{{\cal M}_2}{|\prod_{I\leq J}d\Omega_{IJ}|^2
\over ({\rm det}\,{\rm Im}\,\Omega)^5\Psi_{10}(\Omega)}
\int_{\Sigma^4}
\W_{(F^4)}\,\bar\Y_S
\,
{\rm exp}\bigg(-\sum_{i<j}k_i\cdot k_j\,G(z_i,z_j)\bigg),
\nonumber\\
\eea
where, in the notation of (\ref{WF4HO}-\ref{WF4HE2}),
the term $\W_{(F^4)}$ is given by 
$\W_{(F^4)}=\W_{(F^4)}^{HO}$
for the $Spin(32)/{\bf Z}_2$ string, and by either
$\W_{(F^4)}=\W_{(F^4)}^{HE}$ or
$\W_{(F^2 \, F^2)}=$
$\W_{(F^2\, F^2)}^{HE}$ in the case of the $E_8\times E_8$ string,
depending on whether
all four gauge bosons are in the same $E_8$, or whether the first
two gauge bosons are in one $E_8$ and the remaining two in the other.

\medskip

For the scattering of two gauge bosons and two gravitons,
we find
\bea
\label{bfAR2F2}
{\bf A}_{R^2F^2} &=&
{\bar K\over 64\pi^{14}}
\int_{{\cal M}_2}{|\prod_{I\leq J}d\Omega_{IJ}|^2
\over ({\rm det}\,{\rm Im}\,\Omega)^5\Psi_{10}(\Omega)}
\int_{\Sigma^4}
\W_{(R^2)}(z_1,z_2)\W_{(F^2)}(z_3,z_4)\,\overline{\Y_S(z_1,z_2,z_3,z_4)}
\,
\no\\
&&
\hskip 1.6in
\times\
{\rm exp}\bigg(-\sum_{i<j}k_i\cdot k_j\,G(z_i,z_j)\bigg),
\eea
where the Heterotic graviton part $\W_{(R^2)}$ is given by
(\ref{WR2}), and the gauge boson part $\W_{(F^2)}$ is again given
by two distinct formulas, (\ref{WF2HO}) or (\ref{WF2HE}),
depending on the Heterotic theory under consideration.

\medskip

For the scattering of four gravitons, the simplest formulation of the amplitude is
as
\bea
\label{AR4}
{\bf A}_{R^4}&=&
{\bar K\over 64\pi^{14}}
\int_{{\cal M}_2}{|\prod_{I\leq J}d\Omega_{IJ}|^2
\over ({\rm det}\,{\rm Im}\,\Omega)^5\Psi_{10}(\Omega)}
\int_{\Sigma^4}
\W_{(R^4)}\,\bar\Y_S
\,
{\rm exp}\bigg(-\sum_{i<j}k_i\cdot k_j\,G(z_i,z_j)\bigg),
\nonumber\\
\eea
where $\W_{(R^4)}=\<\prod_{j=1}^4\epsilon_j^\mu\p
x^\mu(z_j)e^{ik_j\cdot x(z_j)}\>/\<\prod_{j=1}^4e^{ik_j\cdot x(z_j)}\>$, where
$x(z,\bar z)$ is a non-chiral scalar field,
whose propagator is the single-valued Green's function
$G(z,w)$.

\subsection{Non-Renormalization Theorems}

An immediate consequence of the vanishing of the $N\leq 3$ 
chiral superstring amplitudes, established in \S 5 from  first 
principles, is that the  massless NS  boson  scattering amplitudes 
in Type II and Heterotic superstrings receive no corrections at 
two-loop order. This non-renormalization theorem holds pointwise
in moduli and  in the vertex operator points. It had been 
conjectured to hold long ago, in part thanks to space-time
supersymmetry \cite{old1}.
Prior to the present derivation from first principles,
many arguments in favor of this conjecture had been proposed in the literature \cite{bis1,lech, ambiguities,zwzI,zwzII,
berkovits}.

\subsubsection{Effective action $R^4$ terms in the Type II superstring}

It is of  interest to consider contributions generated
to the low energy effective action. It was shown in
\cite{Gross:1986iv} that novel corrections to the Einstein-Hilbert
action (and its supergravity extension) arise from tree-level
and one-loop orders in Type II superstring theory. Using the 
$SL(2,{\bf Z})$ or $S$-duality of Type IIB and the $T$-duality
of toroidal compactifications, it was argued in \cite{Green:1997as}
that the order in the derivative expansion at which a given term
first enters in the effective action is related to the order in
string perturbation theory (see also \cite{Russo:1997fi}). 
In particular, these considerations lead to the conjecture that no 
corrections to this term arise at two loop order.
As in the case of the $N$-point function with $N\leq 3$,
there are many arguments in support of this conjecture
\cite{bis3, iz, zwzIII}.  
It may now be 
proven from our first principles formulas.

\medskip

From the form of the Type II amplitude for the scattering of 4 NS-NS
supergravitons in (\ref{TypeII}) and the form of $\Y_S$ in (\ref{Ystar}),
as well as the expression for $K$ in terms of the chiral field
strengths $f_i ^{\mu \nu}$ via $K = t_8 f_1f_2f_3 f_4$, it is manifest
that the contribution to the term $R^4$ in the Type II superstring
vanishes. This non-renormalization theorem holds pointwise
in moduli and in the vertex operator insertion points.

\subsubsection{Effective action $F^4$ and $F^2F^2$ terms in the 
Heterotic string}

In \cite{Cai:1986sa} analogous corrections to tree level and one-loop 
were calculated in the Heterotic strings. With less supersymmetry,
the number of corrections proliferates. Nonetheless, further 
non-renormalization conjectures have emerged. A first driving force
is the conjectured duality \cite{Witten:1995ex} between 
the Heterotic  $Spin(32)/Z_2$  theory and the Type I superstring,
and its lower dimensional extensions \cite{Bachas:1997xn}
(for a review see \cite{Bachas:1998rg}). A second is from the duality 
\cite{Horava:1995qa} between the Heterotic $E_8\times E_8$ theory 
and M-theory \cite{Lukas:1998ew}, \cite{Kiritsis:2000zi}.
A third is from the interplay between space-time supersymmetry and 
space-time anomalies   \cite{Tseytlin:1995fy}.
A discussion of such non-renormalization effects 
order by order in the string coupling was given in
\cite{STI,STII}. The forms $F^4$ and $F^6$
are expected to receive no 2-loop corrections. This conjecture was
partially verified for $F^4$ in \cite{STII} using the measure of \cite{I,IV} 
for the  disconnected parts, and arguments using gauge slice dependent
measures were given for both the cases $F^4$ and $F^6$ in \cite{STI,STII}.
The validity of the $F^4$ conjecture is proven here from first principles.
It would be interesting to extend the proof to the case of $F^6$.

\medskip

From the form of the Heterotic amplitude for the scattering of 4 NS
gauge particles in (\ref{bfAF4}), the form of $\Y_S$ in (\ref{Ystar}),
the fact that $\W_{F^4}$ and $\W_{F^2F^2}$ do not depend on the 
momenta $k_i$, and the expression $K = t_8 f_1f_2f_3 f_4$, 
it is manifest that the contribution to the terms $F^4$ and $F^2 F^2$ 
in the Heterotic string vanish. This non-renormalization theorem also 
holds pointwise in moduli and in the vertex operator insertion points.

\subsubsection{Effective action $R^2F^2$ terms in the Heterotic string}

The interplay between space-time supersymmetry and space-time 
anomalies leads to the conjecture that the $R^2F^2$
terms are also not renormalized \cite{Tseytlin:1995fy}.
This is proven below.

\medskip

The low-energy effective action corrections for both $R^2 F^2$ and $R^4$ 
terms involve the space-time polarization tensors not only from the 
superstring $\bar K=t_8 \bar f_1 \bar f_2 \bar f_3 \bar f_4$, but now also 
from the bosonic  chiral half. This presents a new  complication, as the 
corresponding chiral polarization vectors $\epsilon _i ^\mu$ from the 
bosonic half must first be converted to gauge invariant field strengths 
$f_i^{\mu\nu}=\epsilon_i^\mu k_i^\nu-\epsilon_i^\nu k_i^\mu$,
before the limit $k_i^\mu\to 0$ can be safely taken. This conversion 
leads to apparent first order poles in the Mandelstam variables $s$, $t$, or $u$.
Actually, it will be shown in \S 12.4 that, in the full Heterotic amplitudes, 
{\sl all such poles} are precisely compensated by the linear dependence in 
$s$, $t$ and $u$ of  the superstring chiral half factor $\bar \Y_S$.
For example, it will be shown that the following rearrangement
formula holds\footnote{We sometimes abbreviate the insertion points $z_1,z_2,z_3,z_4$ by 1,2,3,4.}
\bea
s\W_{(R^2)}(3,4)
=
2(f_3f_4)\p_3\p_4\,G(3,4)
-
2
\sum_{ij}k_i^\mu f_3^{\mu\nu}f_4^{\nu\rho}k_j^\rho
\p_3G(3,i)\p_4G(4,j),
\eea
with analogous expressions for $t\W_{(R^2)}$, $u\W_{(R^2)}$
(see (\ref{stuWR2})). 

\medskip

To obtain the $R^2F^2$ effective action, we keep the $f_i$ and $\bar f_i$
fixed and take the limit as $k_i \to 0$. The second term on the rhs
in the above formula clearly tends to 0 in this limit. The first term has
a finite limit whose integral against the remaining anti-holomorphic
coefficient of $s$ in $\bar \Y_S$ vanishes. The resulting non-renormalization 
theorem thus holds pointwise in moduli but requires a cancellation 
of the integration over vertex operator insertion points.

\subsubsection{Effective action $R^4$ terms in the Heterotic string}

The calculation of the $R^4$ corrections are analogous to the 
ones for the $R^2F^2$ corrections, but 
technically more involved. The bosonic chiral half now
involves the chiral amplitude $\W_{(R^4)}$, in which all chiral
polarization vector $\epsilon_i $ must be re-expressed 
in terms of $f_i$ before the limit $k_i \to 0$ can be safely taken, (see \S 12.6).
By contrast with $\W_{(R^2)}$, even when combining with 
the superstring factor $\bar \Y_S$, the resulting
$s\W_{(R^4)}$, $t\W_{(R^4)}$, and $u\W_{(R^4)}$ cannot
be directly expressed in terms of $f_i$ without introducing 
simple poles in $s$, $t$ or $u$. However, it can be shown 
that the residues of these poles actually integrate to zero
against the anti-holomorphic differentials in $\bar \Y_S$.
For example, one such term is of the form,
\bea
2(f_1f_2)\p_1\p_2 G(1,2)\bigg(s\W_{(R^2)}(3,4)\bigg)
\eea
The remaining parts are obtained by expanding in powers
of $s$, $t$ and $u$ the exponential of the Green function
$\exp \{ \half\sum_{i<j} s_{ij} G(z_i,z_j) \}$; they are regular 
in the limit $k_i \to 0$, and integrate to 0 against the anti-holomorphic 
differentials in $\bar \Y_S$. This completes the proof 
that no two-loop $R^4$ terms arise in the Heterotic string.

\medskip

The implications of the two-loop non-renormalization of $R^4$ in the 
Heterotic string remain to be fully understood, and we plan to return
this this problem later. In particular, the Heterotic $Spin(32)/Z_2$
-- Type I duality \cite{Witten:1995ex,Bachas:1997xn,Bachas:1998rg} 
has been used to argue that non-vanishing two-loop corrections
to $R^4$ should arise on the Heterotic side \cite{Tseytlin:1995fy}.

\medskip
We also stress that our re-arrangement formulas for the chiral half of the bosonic string are valid for any genus.
Thus, as long as the superstring amplitude from the other chiral
sector contributes anti-holomorphic amplitudes at least
linear in $s,t$ and $u$, the preceding arguments should apply, and the non-renormalization theorems should hold to all orders of
string perturbation theory.

\subsection{Organization}

This paper is organized as follows.
In section \S 2, we provide the precise definitions
for all the ingredients making up the chiral amplitudes
${\cal B}[\delta]$. In section \S 3, a first group of identities involving 
sums over spin structures is summarized. These are
pointwise identities, by contrast with a more difficult group
of identities which will be encountered later, and which
involve integrals against the Beltrami differential $\mu(z)$.
Since the proofs of these pointwise identities are not required 
for the rest of the evaluation of the $N$-point function, they have been been 
relegated to  Appendices C and D.
In section \S 4, we establish the existence of the key scalar 
function $\Lambda(z)$. The same argument gives also the
integrals against $\mu(z)$ of  two of the simpler sums over spin structures,
namely $I_{13}$ and a symmetrized version of $I_{14}$.
In section \S 5, we prove the vanishing of the $N$-point 
function for $N\leq 3$. Even in the relatively simple
case of the $3$-point function, we need the previous identity
involving $I_{14}$, which is non-trivial since it relies implicitly 
on the existence of $\Lambda(z)$.
In section \S 6, we give an outline of the several steps in the 
evaluation of the 4-point function. The main formulas 
for sums over spin structures for the $4$-point function
are spelled out. The Dolbeault $\bar\p$-exact terms 
as well as three key identities which will be needed are identified.
In section \S 7, we derive the formulas for sums over spin 
structures announced in \S 6. The derivation relies on
some important relations between integrals against $\mu(z)$
of the more complicated sums $I_{15}$ and $I_{16}$.
These relations are proved in section \S 8.
The first key identity described in the outline given in
section \S 6 is proved in section \S 9. It results in the
cancellation of the kinematic factor $C_T$ from the final
amplitude. The second identity described in the outline given in
section \S 6 is proved in section \S 10. It involves the derivatives 
of the function $\Lambda(z)$, and results in the
cancellation of the contribution to the kinematic factor $K$
from the fermionic stress tensor. Finally, the third identity
is established in section \S 11. This identity is of particular importance, 
since it shows how all the effects of gauge choices,
namely $\varpi(z)$, $\mu(z)$, and $\Lambda(z)$, cancel out to
leave us with a gauge slice independent holomorphic amplitude ${\cal H}$. 
This completes the derivation of the 4-point function for the 
superstring chiral amplitudes and of the full amplitude for the
Type II superstring.
The 4-point function for the Heterotic string is evaluated in
section \S 12. The main step is the evaluation of the correlators of the internal fermions and of the massless bosons in the bosonic string. 
The formulas obtained involve matter correlators only and 
actually hold in any  genus.
The derivation of the amplitudes in terms of $k_i$ and $\epsilon_i$ 
is straightforward. However, their re-expression
in terms of the gauge invariant field strengths $f_i^{\mu\nu}$
is more difficult. Once these expressions are available, we can
readily derive their consequences for the low energy effective actions 
of the Heterotic strings. This is done in section \S 13.

\medskip
In Appendix A, we have collected the main formulas from the theories of Riemann surfaces and $\tet$-functions which we need. In Appendix B, we describe some particular geometric properties of hyperelliptic Riemann surfaces and of the unitary
gauge which play a role in the paper. 
The identities involving sums over spin structures can be divided into two groups, depending on whether they involve the factor
$\Z[\delta]$ or the factor $\Xi_6[\delta]$. The identities involving $\Z[\delta]$ are proved in Appendix C, while the identities involving $\Xi_6[\delta]$ are proved in Appendix D.

\vskip .5in

\noindent
{\bf \large Acknowledgments}

\smallskip

We are happy to acknowledge discussions with Per Kraus, John Schwarz,
Tomasz Taylor, Pierre Vanhove,  Chuan-Ji Zhu and especially Costas Bachas, 
Michael Gutperle and Boris Pioline.
One of us (E.D.) would like to thank the Aspen Center for Physics
where part of this work was completed.
The other (D.H.P.) would like to acknowledge the warm hospitality
of the Centro di Ricerca Matematica Ennio De Georgi in Pisa and
the Institute of Mathematical Sciences at the Chinese University of Hong Kong.

\newpage

\section{The Gauge-fixing Procedure with Vertex Operators}
\setcounter{equation}{0}

Our set up is the following \cite{V}. Let $\Sigma$ be the worldsheet, 
and assume
that its genus is $h=2$. We fix a canonical homology basis $A_I,B_I$,
$\#(A_I\cap A_J)=\#(B_I\cap B_J)=0$, $\#(A_I\cap B_J)=\delta_{IJ}$. Let
$\omega_I$ be the basis of holomorphic $(1,0)$-forms
dual to $A_I$. Let $\delta$ be an even spin structure. The chiral amplitude
${\cal B}[\delta]$ is obtained
by chiral splitting from the chirally symmetric amplitude
${\bf A}[\delta]$ defined by
\bea
\label{symmetric}
{\bf A} [\delta ] =\int DE_M{}^AD\Omega_M\delta(T)
\int\prod_{i=1}^N\, d^{2|2}{\bf z}_i E ({\bf z}_i)\
\<\prod_{i=1}^NV({\bf z}_i,\bar{\bf z}_i;  \e_i, \bar \e_i, k_i )\>_X,
\eea
where $(E_M{}^A, \Omega_M)$ describes the two-dimensional supergeometry, 
$\delta(T)$ denotes the Wess-Zumino torsion constraints,
${\bf z}=(z,\theta)$ are coordinates on the super worldsheet,
$E({\bf z})$ is the super worldsheet volume form,
and $V({\bf z},\bar{\bf z};  \e, \bar \e, k )$ is the superfield vertex
operator for the emission of the graviton multiplet
with momentum $k=(k^\mu)$ and polarization tensor
$\epsilon^\mu \, \bar \epsilon ^\nu$, $k^2=0$, $k\cdot\epsilon=0$ \cite{dpvertex}.
In Wess-Zumino gauge, the supergeometry $(E_M{}^A, \Omega _M)$ 
can be identified with a pair $(g_{mn},\chi_{m}{}^\alpha)$,
where $g_{mn}$ is a metric on $\Sigma$, and $\chi_m{}^\alpha$
is a gravitino field \cite{superg}.
In \cite{V}, it was shown that, after gauge fixing by the
procedure of projecting onto the super period matrix
introduced in \cite{I,II,III,IV} (following \cite{dp88,dp89Rome}), 
we can write
\bea
{\bf A}[\delta]
=
\int dp_I^\mu\,\bigg|{\cal B}[\delta](z_i;\epsilon_i,k_i,p_I)\bigg|^2,
\eea
with the chiral amplitude ${\cal B}[\delta]$ given by
\bea
\label{calb}
{\cal B}[\delta] (z_i; \epsilon_i, k_i,p_I)
&=&\prod_{I\leq J}d\Omega_{IJ}\,\int\! d^2 \zeta
\ {\prod_{a=1}^3b(p_a)\prod_{\alpha=1}^2\delta(\beta(q_{\alpha}))
\over
{\rm det}\,\Phi_{IJ+}(p_a)\,{\rm det}\,\<H_{\alpha}|\Phi_\beta^*\>}
\\
&&
\hskip .6in
\times
\bigg \< Q(p_I)\,
\exp \bigg  \{ {1\over 2\pi}\int \big (\chi S +\hat\mu T \big ) \bigg \} \,
\prod_{j=1}^N\V_j  \bigg \>.
\no
\eea
Here, $\zeta^\alpha$ are the supermoduli \cite{friedan, supermoduli}, $p_I^\mu$ 
are the internal loop momenta required for chiral splitting
\cite{dp89}, $S(z)$ and $T(z)$ are the supercurrent and the stress tensor
respectively, $\Phi_{IJ+}$ and $\Phi_\alpha^*$ are
superholomorphic $3/2$ differentials, $H_\alpha$ are dual super Beltrami
differentials. 
The term $Q(p_I)$ has been defined in (\ref{Q}).
The terms $\V_j$ are abbreviations for $\V(z_j,\epsilon_j,k_j)$, which are now
the full vertices described in (\ref{V}), incorporating both the naive vertex
$\V^{(0)}$ and the corrections $\V^{(1)}$ and $\V^{(2)}$ of \cite{V}. The term
$\hat\mu_{\bar z}{}^z$ is a
Beltrami differential for the deformation of complex structures
from the period matrix to the super period matrix. It is characterized in genus
$h=2$ by the following equation
\bea
\label{muequation}
\int_\Sigma d^2z \,\hat\mu\,\omega_I\omega_J
=
{1\over 8\pi}
\int_\Sigma d^2u\int_\Sigma d^2v\,
\omega_I(u)\,\chi_{\bar u}{}^+S_\delta(u,v)\chi_{\bar v}{}^+\omega_J(v),
\eea
where $S_\delta(u,v)$ is the Szeg\"o kernel. It is important to note that this
equation defines $\hat\mu$ only up to
a vector field $v^z$
\bea
\muhat \to \muhat +\p_{\bar z}v^z.
\eea
A choice of $\muhat $ is a choice of gauge, reflecting
the invariance of the theory under diffeomorphisms. Thus
the gauge fixed expression ${\cal B}[\delta]$ incorporates {\it two} choices of
gauge, both of which must cancel out in physical amplitudes: the choice of
gravitino slice $\chi$, and the choice
of Beltrami differential $\hat\mu$.

\medskip
It is convenient to expand ${\cal B}[\delta]$ into its connected
and disconnected parts ${\cal B}[\delta]^{(c)}$ and ${\cal B}[\delta]^{(d)}$.
The disconnected part ${\cal B}[\delta]^{(d)}$ consists by definition of
the contributions where the Wick contractions of the
supercurrent $S(z)$ and the stress tensor $T(z)$ are disconnected
from those of the vertex operators $\V_j$. The connected part
${\cal B}[\delta]^{(c)}={\cal B}[\delta]-
{\cal B}[\delta]^{(d)}$ consists of the rest. We arrive in this
way at the equation (\ref{BcBd}) given in the Introduction,
with the terms $\Y_1,\cdots,\Y_5$ given by\footnote{To simplify
notations throughout, the integration over the worldsheet $\Sigma$ will be 
abbreviated by $\int _\Sigma d^2 z \to \int$ when no confusion 
is expected to arise. In particular, we shall use the convenient notation
$\int \chi S = \int _\Sigma d^2 z \chiz S(z)$ and $\int \hat \mu T = 
\int _\Sigma d^2 z \muhat  T(z)$.}
\bea
\label{Ys}
\Y _1
& = &
{1 \over 8 \pi ^2} \left \< Q(p_I)\,  \int \! \chi S ~ \int \! \chi S ~
\prod _{i=1}^N \V_i ^{(0)} \right \> _{(c)}
\no \\
\Y _2
& = &
{1 \over 2 \pi } \left \< Q(p_I)\, \int \! \hat \mu T ~
\prod _{i=1}^N \V_i ^{(0)} \right \> _{(c)}
\no \\
\Y _3
& = &
{1 \over 2 \pi } \sum _{i=1} ^N \left \< Q(p_I)\, \int \! \chi S ~ \V ^{(1)} _i
~ \prod _{j \not= i}^N  \V_j ^{(0)}  \right \>
\no \\
\Y _4
& = &
\half \sum _{i \not= j} \left \< Q(p_I) ~
 \V^{(1)} _i  ~ \V^{(1)} _j  ~ \prod _{l \not= i,j}^{N}  \V_l ^{(0)} \right \>
\no \\
\Y _5
& = &
\sum _{i =1}^N \left \< Q(p_I) ~  \V ^{(2)} _i ~
\prod _{j \not= i}^{N}  \V_j ^{(0)} \right \>
\eea
The subindex (c) in $\Y_1$ and $\Y_2$ indicates that only contributions with
some vertex operators contracted with a current or a stress tensor are kept. It
is convenient to introduce the gauge invariant field strength $f_i^{\mu\nu}$ for
each particle by
\bea
\label{f}
f_i^{\mu\nu}=\e_i^\mu k_i^\nu-\e_i^\nu k_i^\mu,
\eea
and rewrite the $\V_i^{(0)}$ component of the corresponding vertex operator as
\bea
\label{Vanti}
\V_i^{(0)}=\bigg(\e_i^\mu\p_{z_i}x_+^\mu-{i\over 2}f_i^{\mu\nu}
\psi_+^\mu\psi_+^\nu\bigg)\, e^{ik_i\cdot x_+(z_i)}.
\eea

In \cite{V}, it was already shown that, upon pairing left and right movers and
integrating over the worldsheet $\Sigma$,
the physical amplitudes are gauge-slice independent. In this paper, we shall
make the particular choice of gravitino slice
$\chi=\sum_{\alpha}\zeta^{\alpha}\delta(z,q_{\alpha})$,
where $q_{\alpha}$ are the zeroes of a holomorphic differential
$\varpi(z)$. The differential $\varpi(z)$ is
arbitrary, and its ultimate cancellation out of the physical amplitudes will
serve as a check of the gauge slice independence.
Once $\chi$ is chosen,
the Beltrami differential $\muhat $
is constrained by the equation (\ref{muequation}),
but it is otherwise arbitrary. With $\chi$ having Dirac point support at $q_1$
and $q_2$,
as mentioned earlier in the Introduction,
it is convenient to introduce the Beltrami differential
$\mu(z)$ by
\bea
\label{mudef}
\muhat = S_\delta(q_1,q_2) \mu(z).
\eea
The factor $S_\delta(q_1,q_2)$ cancels then out
of the equation (\ref{muequation}) for $\muhat $, and the defining equation
for $\mu(z)$ becomes
\bea
\label{mutildedef}
\int \,\mu\,\omega_I\omega_J
=
{\zeta^1\zeta^2\over 8\pi} \bigg (
\omega_I (q_1)\omega_J (q_2) 
+ \omega_J (q_1)\omega_I (q_2) \bigg )
\eea
In particular, the Beltrami differential $\mu(z)$ can be taken to be
independent of the spin structure $\delta$.
Even though the gravitino slice $\chi$ consisted of Dirac point masses, the
Beltrami differential $\mu(z)$ can be taken to be smooth. If one wishes,
one can choose $\mu(z)$ to consist of point masses also. However, as
discussed in \cite{V}
and as will be seen explicitly in section \S 9, this would have to be done with
considerable care, especially if one takes a limit where the point supports of
$\mu(z)$ tend to $q_1$ or $q_2$.

\medskip

In terms of the Abel map from the surface $\Sigma$ into its Jacobian, 
the condition that $q_1, q_2$ are the zeroes of a holomorphic $(1,0)$-form
is equivalent to the condition,
\bea
\label{unitarygauge}
q_1+q_2-2\Delta=2\kappa ~ \in ~ {\bf Z}^2\oplus \Omega\,{\bf Z}^2
\eea
where $\Delta$ is the vector of Riemann constants \cite{fay,dp88},
and $2\kappa$ is an arbitrary full period. The superstring amplitudes 
will be shown to be independent of the independent variables $q_1$
and $\kappa$. The holomorphic $(1,0)$-forms which vanish at 
$q_1,q_2$ are determined up to a constant multiple (see Appendix \S D
for a proof). We choose such a form $\varpi(z)$ to be
\bea
\label{varpi}
\varpi(z)
& \equiv &
+ \omega_I(z)\p_I\tet(q_1-\Delta)e^{2\pi i\kappa'(q_1-\Delta)}
\no \\
& = &
-\omega_I(z)\p_I\tet(q_2-\Delta)e^{2\pi i\kappa'(q_2-\Delta)}.
\eea

\medskip

In the formula (\ref{BcBd}), the measures $d\mu_0[\delta]$ and
the measure $d\mu_2[\delta]$ appear, which are
defined in (\ref{dmu0dmu2}). The term
$\Z[\delta]$ can also be calculated explicitly \cite{vv2}, and we have,
\bea
\label{zee}
{\cal Z}  [\delta]
=
{\tet [\delta ](0)^5 \tet (p_1 + p_2 + p_3 - 3 \Delta )
\prod _{a<b} E(p_a, p_b) \prod _a \sigma (p_a)^3
\over
Z^{15}  \tet [\delta ](q_1 + q_2 - 2 \Delta ) E(q_1,q_2) \prod _\alpha
\sigma (q_\alpha )^2  \det \omega _I \omega _J (p_a)}.
\eea
Here, the prime form $E$, and the 1-form $\sigma (z)$ are defined
in Appendix \S A, specifically (\ref{prime}) and (\ref{sigma}), while 
$Z$ is the partition function for a single chiral boson, given by
\bea
\label{zeecube}
Z^3=
{\tet(r_1+r_2-r_3-\Delta)E(r_1,r_2)\sigma(r_1)\sigma(r_2)
\over
E(r_1,r_3)E(r_2,r_3)\sigma(r_3){\det}\,\omega_I(r_j)}
\eea
where $r_1,r_2,r_3$ are arbitrary generic points and $r_j$
in the determinant runs  over $r_1$ and $r_2$.
It was shown in \cite{IV} that $\Z [\delta]$ may be simply 
re-expressed in terms of the chiral partition function $\Z_B$ for the 
bosonic string, 
\bea
\Z [\delta] = {\Z _B  ~ Z^{12} ~ \tet [\delta ](0)^5
\over
\tet [\delta ](q_1 + q_2 - 2 \Delta ) E(q_1,q_2)
\sigma (q_1 )^2 \sigma (q_2 )^2}
\eea
In the unitary gauge, with $q_1 + q_2 - 2 \Delta = 2 \kappa$, 
the $\tet$-function in the denominator simplifies considerably and,
with the help of (\ref{periodshift}), may be recast in the following form
\bea
\label{zeedelta}
\Z [\delta ] = \Z_0 \, E(q_1,q_2) \, e^{4 \pi i \kappa ' \Omega \kappa'} ~
\< \kappa |\delta \> ~ \tet [\delta ](0)^4
\eea
where $\Z_0$ is $\delta$-independent and given by
\bea
\label{zeedelta1}
\Z _0 \equiv{ \Z _B ~ Z^{12} \over E(q_1, q_2)^2 \sigma (q_1)^2 \sigma (q_2)^2}
\hskip .5in 
\Z _\B =
{ 1 \over \pi ^{12} \Psi _{10}(\Omega)}
\eea
Note that an extra factor of $E(q_1,q_2)^{-1}$ has been included in the
definition of $\Z_0$ for later convenience. This makes $\Z_0$ a $(-1,0)$
form (with non-trivial monodromy) in both $q_1$ and $q_2$ with a double pole at
$q_1=q_2$ and no zeros. Note that (\ref{zeedelta}) clearly exposes
the $\delta$-dependence of $\Z[\delta]$.

\newpage

\section{Identities for Sums over Spin Structures}
\setcounter{equation}{0}

We shall evaluate directly the sum $\sum_\delta{\cal B}[\delta]$.
In the papers \cite{I,II,III,IV}, for the evaluation of the chiral superstring
measure for each fixed spin structure $\delta$, we had used a split gauge,
where the period matrix and the super period matrix coincide.
For the present evaluation of the $N$-point function with $N\leq 4$, where we
consider only a sum over even spin structures $\delta$,
the unitary gauge turns out to be more convenient, as
it turns out that a greater
number of terms cancels upon summing over $\delta$.
The unitary gauge is closely related to the Mandelstam representation of string
diagrams in the light-cone gauge. It has been used by many authors in the
evaluation of string amplitudes \cite{old0,GW,adp90}, and particularly in
\cite{lech, iz, zwzI, zwzII, zwzIII}
where the existence of many identities in this gauge was discovered.

\medskip

In this section, we list all the summation identities which will be needed.
Their detailed derivation is given in Appendices C and  D.
The summation identities can be divided into two groups,
those which involve the factor $\Z[\delta]$ and those which involve 
the factor $\Xi_6[\delta]$. The first group
splits itself into two subgroups, depending on whether
there is or is not an insertion of the fermion stress energy tensor.
From the form of the terms $\Y_1,\cdots,\Y_5$ given in
(\ref{Ys}), it is clear that the sums below
will appear in the $N$-point function with $N\leq 4$.

\subsection{Identities without the Fermion Stress Tensor}

For the $N$-point function when $N\leq 3$, only sums
with 4 or less Szeg\"o kernels can occur, when there is no fermion stress tensor
insertion. They are
\bea
I_1 & = & \sum _\delta \Z [\delta] ~ S_\delta (q_1, q_2)
\no \\
I_2 & = & \sum _\delta \Z [\delta] ~ S_\delta (q_1, q_2) S_\delta (z_1,z_2)^2
\no \\
I_3 & = & \sum _\delta \Z [\delta] ~ S_\delta (q_1, q_2) S_\delta (z_1,z_2)
S_\delta (z_2,z_3) S_\delta (z_3,z_1)
\no \\
I_4 & = & \sum _\delta \Z [\delta] ~ S_\delta (q_1, z_1) S_\delta (z_1,q_2)
\no \\
I_5 & = & \sum _\delta \Z [\delta] ~ S_\delta (q_1, z_1) S_\delta (z_1,q_2)
S_\delta (z_2,z_3) S_\delta (z_3,z_2)
\no \\
I_6 & = & \sum _\delta \Z [\delta] ~ S_\delta (q_1, z_1) S_\delta (z_1,z_2)
S_\delta (z_2,q_2)
\no \\
I_7 & = & \sum _\delta \Z [\delta] ~ S_\delta (q_1, z_1) S_\delta (z_1,z_2)
S_\delta (z_2,z_3) S_\delta (z_3,q_2)
\eea
For the $4$-point function, sums over $\delta$ with products of 5 Szeg\"o
kernels can occur. They are
\bea
I_8 & = & \sum _\delta \Z [\delta] ~ S_\delta (q_1, z_1) S_\delta (z_1, z_2)
S_\delta (z_2, z_3) S_\delta (z_3, z_4) S_\delta (z_4, q_2)
\no \\
I_9 & = & \sum _\delta \Z [\delta] ~ S_\delta (q_1, z_1) S_\delta (z_1, z_2)
S_\delta (z_2, q_2) S_\delta (z_3, z_4)^2 
\no \\
I_{10} & = & \sum _\delta \Z [\delta] ~ S_\delta (q_1, z_1) S_\delta (z_1, q_2)
S_\delta (z_2, z_3) S_\delta (z_3, z_4) S_\delta (z_4, z_2)
\no \\
I_{11} (z_1,z_2,z_3,z_4)
& = &
\sum _\delta \Z [\delta] S_\delta (q_1, q_2)
S_\delta (z_1, z_2)^2 S_\delta (z_3, z_4)^2
 \\
I_{12} (z_1,z_2,z_3,z_4)
& = &
\sum _\delta \Z [\delta] S_\delta (q_1, q_2)
S_\delta (z_1, z_2) S_\delta (z_2, z_3) S_\delta (z_3, z_4) S_\delta (z_4, z_1)
\no
\eea
The identities for these sums
that we need are:
\bea
\label{1-12}
&&
I_1=I_2=I_3=I_4=I_5=I_6=I_7=I_8=I_9=I_{10}=0
\nonumber\\
&&
I_{11}(z_i) = I_{12}(z_i)
=
-2\Z_0\prod_{i=1}^4\varpi(z_i).
\eea
In particular, $I_{11}(z_i)=I_{12}(z_i)$ are completely symmetric 
in all its variables, although this is not manifest from their definition.

\subsection{Identities with the Fermion Stress Tensor Insertion}

Next, we consider sums over spin structures involving the fermion
stress tensor. The expression $\varphi [\delta] $ accounting for the fermion 
stress tensor insertion can be defined by
\bea
\label{varphi}
\varphi [\delta] (w;z_1,z_2) \equiv
S_\delta (z_1, w)  \p_w S_\delta (w, z_2)
- S_\delta (z_2, w)  \p_w S_\delta (w, z_1)
\eea
In all evaluations, we shall make use of the Fay trisecant identities (see the
appendices, eq. (\ref{Fay3})) to recast $\varphi [\delta] $
as follows,
\bea
\varphi [\delta] (w;z_1,z_2)
=
- { \tet [\delta] (z_1 + z_2 - 2 w) E(z_1, z_2) \over
\tet [\delta ](0) E(z_1, w)^2 E(z_2, w)^2}
\eea
The sums over spin structures which arise at the level of the 2-, 3-, and
4-point functions are given by
\bea
I_{13} (w; z_1,z_2)
& = &
\sum _\delta \Z [\delta] S_\delta (q_1, q_2) \varphi [\delta] (w;z_1,z_2)
S_\delta (z_2, z_1)
\no \\
I_{14} (w; z_1,z_2,z_3)
& = &
\sum _\delta \Z [\delta] S_\delta (q_1, q_2) \varphi [\delta] (w;z_1,z_2)
S_\delta (z_2, z_3) S_\delta (z_3,z_1)
\no \\
I_{15} (w; z_1,z_2,z_3,z_4)
& = &
\sum _\delta \Z [\delta] S_\delta (q_1, q_2) \varphi [\delta] (w;z_1,z_2)
S_\delta (z_2, z_3) S_\delta (z_3, z_4) S_\delta (z_4, z_1)
\no \\
I_{16} (w; z_1,z_2; z_3, z_4)
& = &
\sum _\delta \Z [\delta] S_\delta (q_1, q_2) \varphi [\delta] (w;z_1,z_2)
S_\delta (z_2, z_1) S_\delta (z_3, z_4)^2
\eea
Actually, it will be more natural to work with following  (anti-) symmetrized
parts of $I_{15}$,
\bea
2 I_{15}^S (w;z_1,z_2,z_3,z_4)
& = &
  I_{15} (w;z_1,z_2,z_3,z_4) +  I_{15} (w;z_2,z_1,z_3,z_4)
\no \\
2 I_{15}^A (w;z_1,z_2,z_3,z_4)
& = &
 I_{15} (w;z_1,z_2,z_3,z_4) - I_{15} (w;z_2,z_1,z_3,z_4).
\eea
which will arise naturally in the amplitudes.

\medskip

Of these, the simplest one, $I_{13}$, is holomorphic and satisfies 
the following identity
\bea
\label{I13}
I_{13}(w;z_1,z_2)=4\Z_0\varpi(z_1)\varpi(z_2)\varpi(w)^2.
\eea
The remaining sums actually have poles in some of the $z_i$
and $w$, and will involve the Green's function $G(z;z_1,z_2;p_1,p_2)$ 
which is a $(1,0)$-form in $z$, and a scalar in $z_1$ and $z_2$,
with simple poles in $z$ at $z_1$ and $z_2$, and zeroes at $p_1$ and $p_2$.
Explicit forms for $G(z;z_1,z_2;p_1,p_2)$
are given in the appendix (\ref{green1}) and (\ref{green2})
\footnote{The Green's function $G(z;z_1,z_2;p_1,p_2)$ should not
be confused with the Green's gunction $G(z,w)$
introduced earlier in (\ref{scalarprop}), which depends only on two points $z$ and $w$. Which Green's function applies is usually clear from the context.}.
The expression $I_{14}$ is then given by
\bea
\label{I14}
I_{14} (w;z_1,z_2,z_3) =
2 \Z_0 \varpi (z_1) \varpi (z_2) \varpi (w)^2 \sum_{\alpha =1,2}
G(z_3;z_1,z_2;q_\alpha ,w) 
\eea
For the more complicated ones, $I_{15}$ and $I_{16}$,
it is most convenient to list the symmetrized and anti-symmetrized 
$I_{15}^S$ and $I_{15}^A$ parts respectively,  
\bea
\label{I15SA}
I_{15}^S (w;z_1,z_2,z_3,z_4)
& = &
 \Z_0 \varpi (z_1) \varpi (z_2) \varpi (w)^2
\big \{ G(z_3;z_1,z_2;q_1,w)  G(z_4;z_1,z_2;q_2,w)
\no \\ && \hskip 1.4in
+ G(z_3;z_1,z_2;q_2,w)  G(z_4;z_1,z_2;q_1,w)   \big \}
\no \\ && \no \\
I_{15}^A (w;z_1,z_2,z_3,z_4)
& = &
\Z_0 \varpi (z_1) \varpi (z_2) \varpi (w)^2
\big \{ G(z_3;z_4,z_1;q_1,w)  G(z_4;z_3,z_2;q_1,w)
\no \\ && \hskip 1.45in
- G(z_3;z_4,z_2;q_1,w)  G(z_4;z_3,z_1;q_1,w)
\no \\ && \hskip 1.45in
+ G(z_3;z_4,z_1;q_2,w)  G(z_4;z_3,z_2;q_2,w)
\no \\ && \hskip 1.45in
- G(z_3;z_4,z_2;q_2,w)  G(z_4;z_3,z_1;q_2,w)
 \big \}
 \no \\
\eea
Finally, $I_{16}$ is related to $I_{15}$ by
\bea
\label{I16}
I_{16} (w;z_1,z_2;z_3,z_4)
& = &
- I_{15}^S (w;z_1,z_2,z_3,z_4)
\no \\ &&
- \Z_0  \varpi (z_1) \varpi (z_2) \varpi (w)^2 \big \{
G(z_3;z_4,z_1;q_1,w)  G(z_4;z_3,z_2;q_1,w)
\no \\ && \hskip 1.55in
+ G(z_3;z_4,z_2;q_1,w)  G(z_4;z_3,z_1;q_1,w)
\no \\ && \hskip 1.55in
+ G(z_3;z_4,z_1;q_2,w)  G(z_4;z_3,z_2;q_2,w)
\no \\ && \hskip 1.55in
+ G(z_3;z_4,z_2;q_2,w)  G(z_4;z_3,z_1;q_2,w) \big \}
\no \\ &&
\eea

\subsection{Identities involving $\Xi_6[\delta](\Omega)$}

For the $1,2,3$-point functions, we need the following identities:
\bea
\label{vanishing1}
I_{17}  & = &
\sum _\delta \Xi _6 [\delta ] \ \tet [\delta ](0)^4  
\\
I_{18} & = &
\sum _\delta \Xi _6 [\delta ] \ \tet [\delta ](0)^4 S_\delta (z_1,z_2)^2
\no \\
I_{19}  & = &
\sum _\delta \Xi _6 [\delta ] \ \tet [\delta ](0)^4
S_\delta (z_1,z_2) S_\delta (z_2,z_3) S_\delta (z_3,z_1) 
\no \\
I_{20} (z_1,z_2;z_3,z_4) & = & 
\sum_\delta \Xi_6[\delta] \tet[\delta](0)^4
S_\delta(z_1,z_2)^2S_\delta(z_3,z_4)^2
\no \\
I_{21} (z_1,z_2,z_3,z_4) & = & 
\sum_\delta \Xi_6[\delta] \tet[\delta](0)^4
S_\delta(z_1,z_2) S_\delta(z_2,z_3) S_\delta(z_3,z_4) S_\delta(z_4,z_1)
\no
\eea
In Appendix D, it is shown that these sums take the following values,
\bea
I_{17}=I_{18}=I_{19} 
& = & 0
 \\
I_{20} (z_1,z_2;z_3,z_4) 
& = & 
- 4 \pi ^4 \Psi _{10} \bigg ( 
\Delta (z_1,z_3) \Delta (z_2,z_4) + \Delta (z_1,z_4) \Delta (z_2,z_3) \bigg )
\no \\
I_{21} (z_1,z_2,z_3,z_4) 
& = & 
+ 2  \pi ^4 \Psi _{10} \bigg ( 
\Delta (z_1,z_2) \Delta (z_3,z_4) - \Delta (z_1,z_4) \Delta (z_2,z_3) \bigg )
\no
\eea
where we have expressed the result in terms of the antisymmetric
biholomorphic 1-form $\Delta(x,y)$, defined in (\ref{Delta}); we repeat it
here for convenience,
\bea
\Delta(x,y)=
\omega_1(x)\omega_2(y)-\omega_2(x)\omega_1(y).
\eea
(The bilinear form $\Delta(x,y)$ should not be confused with the vector $\Delta$ of Riemann constants.) Clearly, the totally symmetrized parts of both $I_{20}$ and $I_{21}$
vanish; as a result, the quantity ${\cal S}$, introduced in \cite{I} by
\bea
{\cal S} (1234)
=
 - {1 \over 192 \pi ^6 \Psi _{10}}
\omega _I (1) \omega _J (2) \omega _K (3) \omega _L (4)
\sum _\delta \Xi _6 [\delta]  \tet [\delta ]^3 \p_I \p_J \p_K \p_L
\tet [\delta](0) \quad
\eea
vanishes identically. The quantity ${\cal T}$ of \cite{I} is indeed
reproduced by the above formulas as the antisymmetric 
part of $I_{20}$, in the following way,
\bea
{\cal T} (i,j|k,l) 
& = & 
{1 \over 32 \pi ^6 \Psi _{10}} \bigg ( I_{20} (i,k;l,j) - I_{20}(i,l;k,j) \bigg )
\no \\
& = & - {1 \over 32 \pi^2} \Delta (i,j) \Delta (k,l)
\eea

\newpage

\section{The Scalar Function $\Lambda(z)$}
\setcounter{equation}{0}

From the expression for $\Y_2$ in (\ref{Ys}), it is evident that we shall
also need identities involving integrals against $\hat\mu _{\bar w} {}^w$
of sums over spin structures involving the stress
tensor. In this  section, we discuss two
of the simplest such integrals, involving $I_{13}$ and $I_{14}$.
In the process, we also establish the existence of a scalar function
$\Lambda(z)$, which plays a fundamental role in the
proof of the gauge slice independence of the $N$-point function.

\subsection{The total Derivative Formula for $\mu (z)\,\varpi(z)$}

The unitary gauge has a remarkable property, which is one of the 
most important preliminary results that we need:
the expression $\mu(w)\,\varpi(w)$ is a $(0,1)$-form which integrates to
$0$ when paired against an arbitrary holomorphic $(1,0)$ form, 
\bea
\label{muvarpi}
\int \mu(w)\,\varpi(w)\,\omega_I(w)=0,
\qquad\qquad I=1,2.
\eea
Indeed, from the definition of $\mu$ in (\ref{mudef}) and $\varpi(z)$
in (\ref{varpi}), it follows that
\bea
\label{integrated13}
\int  \mu (w) \, \varpi (w) \omega _J (w)
& \sim &
\omega _{\{ I} (q_1) \omega _{J\} } (q_2)
\p _I \tet (q_1 - \Delta )
 \\
& \sim &
\p _{q_1} \tet (q_1 - \Delta ) \omega _J (q_2)
- \p _{q_2} \tet (q_2 - \Delta ) \omega _J (q_1)
= 0
\no
\eea
Thus, there exists a single-valued
scalar function $\Lambda(z)$ such that
\bea
 \mu (z) \varpi (z) = \p_{\bar z} \Lambda(z)
\eea
The function $\Lambda$ may be solved for in terms of $\mu$,
\bea
\Lambda(z) = \Lambda_0 - {1 \over 2 \pi} \int  \mu (w) \varpi (w) \p_w
\ln E(z,w)
\eea
The monodromy in $z$ of the integrand cancels upon integration
in view of (\ref{muvarpi}) and the above expression is well-defined 
and single-valued.  Under a change of slice
$\delta _v  \mu _{\bar z} {}^z =  \p _{\bar z} v^z$,
generated by a smooth vector field $v^z$, the function $\Lambda$ 
transforms as follows
\bea
\delta \Lambda(z)  =  \delta \Lambda _0 + \varpi  (z) v^z (z)  
\eea
Since $\varpi(z)$ vanishes at $q_1$ and $q_2$, it follows
that $\Lambda(q_1) - \Lambda(q_2)$ is slice-independent.

\subsection{Integrals of $\mu$ with $I_{13}$ and $I_{14}$}

In view of the identity (\ref{I13}) for $I_{13}$,
the identity (\ref{integrated13}) implies at once
\bea
\label{muI13}
\int \mu(w)\,I_{13}(w;x,y)=0.
\eea
Next, we claim that the following identity holds
\bea
\label{muI14}
\int \mu(w)\varpi(w)\big\{
I_{14}(w;z_1,z_2,z_3)
+
I_{14}(w;z_2,z_3,z_1)
+
I_{14}(w;z_3,z_1,z_2)\big\}=0.
\eea
For this, we note that the Green's function $G(x;p_1,p_2;q_\alpha, w)$, 
for fixed $\alpha$, as a function of $w$, is a scalar and has simple  poles 
at $p_1,\, p_2$ and $q_\beta \equiv 2 \Delta - q_\alpha$. 
It does not have a pole at $x$ and at $q_\alpha$. 
The residues at $p_1$ and $p_2$ are well-known,
the third was deduced using the fact that
\bea
\int \omega _I(w) \p _{\bar w} G(x;p_1,p_2;q_\alpha, w) =0
\eea
and we have\footnote{Note that we have
$ \varpi (p_1) \Delta (q_\beta , p_2) = \varpi (p_2) \Delta (q_\beta , p_1)$.}
\bea
\label{resG1}
 \p _{\bar w} G(x;p_1,p_2;q_\alpha, w)
& = &
2 \pi \delta (w,p_2) {\varpi (x) \over \varpi (p_2)}
-  2 \pi \delta (w,p_1) {\varpi (x) \over \varpi (p_1)}
\no \\ &&
+ 2 \pi \delta (w,q_\beta ) {\varpi (x) \, \Delta (p_1,p_2) \over \varpi (p_1)
\Delta (q_\beta , p_2)} \quad
\eea
or, in a form which often will also be useful,
\bea
\label{resG2}
 \p _{\bar w} \left \{ \varpi (w) G(x;p_1,p_2;q_\alpha, w) \right \}
=
2 \pi \varpi (x) \{ \delta (w,p_2) - \delta (w,p_1) \}
\eea
The role of the factor $\varpi(w)$ is to cancel the
pole in $w$ at $q_\beta$, since $\varpi(w)$ vanishes there.
Returning to the proof of the desired identity, we can apply
(\ref{Lambda}) and integrate by parts to rewrite the left hand side of
(\ref{muI14}) as
\bea
&&
-2\Z_0
\int\Lambda(w)\p_{\bar w}
\big\{\varpi(z_1)\varpi(z_2)\varpi(w)G(z_3;z_1,z_2;q_1,w)
\\
&&
\hskip 1.2in +
\varpi(z_2)\varpi(z_3)\varpi(w)G(z_1;z_2,z_3;q_1,w)
\nonumber\\
&&
\hskip 1.2in 
+
\varpi(z_1)\varpi(z_2)\varpi(w)G(z_2;z_3,z_1;q_1,w)+(q_1\leftrightarrow
q_2)\big\}
\no
\eea
Using the  identity (\ref{resG2}),  we see that all the terms are  proportional to
$\varpi(z_1)\varpi(z_2)\varpi(z_3)$, and that their coefficients sum to $0$.

\newpage

\section{The $N$-point Function for $N\leq 3$}
\setcounter{equation}{0}

With the identities for sums over spin structures given in the previous section,
we can now return to the evaluation of the $N$-point function for $N\leq 3$. 
It is immediately seen from the identities involving $\Xi_6[\delta](\Omega)$ 
listed in \S 3.3 that, for $N\leq 3$, all
contributions from the disconnected component ${\cal B}[\delta]^{(d)}$ cancel
upon summing over $\delta$. Thus, in this section, we concentrate on the
connected component ${\cal B}[\delta]^{(c)}$.

\subsection{The 1-point Function}

For the 1-point function, the $\Y_1$ contribution vanishes by the $I_4=0$ identity;
the $\Y_2, \Y_3$, and $\Y_5$ contributions vanish by $I_1=0$;
there is manifestly no $\Y_4$ contribution;
thus the 1-point function vanishes.

\subsection{The 2-point Function}

For the 2-point function, the $\Y_1$ and $\Y_5$ contributions vanish by
the identities $I_2=I_6=0$, the $\Y_3$ contribution vanishes by
the identity $I_4=0$, and the $\Y_4$ contribution vanishes by the identity
$I_1=0$. In the $\Y_2$ contribution, there is no
contribution from the bosonic stress tensor in view of the identity $I_2=0$.
Thus the 2-point function reduces to the
contributions to $\Y_2$ from the
fermionic stress tensor. These are
proportional to
\bea
\int \mu(w)\,\sum_\delta
\Z[\delta]\,S_\delta(q_1,q_2)\varphi[\delta](w;z_1,z_2)\,
S_\delta(z_2,z_1)
=
\int \mu(w)\,I_{13}(w;z_1,z_2).
\eea
But this vanishes by the identity (\ref{muI13}).

\subsection{The 3-point Function}

For the 3-point function,
the contributions from $\Y_1$ compute as follows. The single-link string of
contractions
between the super current insertions $S(q_1)$ and $S(q_2)$ cancels by $I_1 = I_2
= I_3=0$.
The double-link string between $S(q_1)$ and $S(q_2)$ cancels by
$I_4 = I_5=0$. The triple-link string between $S(q_1)$ and $S(q_2)$
cancels by $I_6 =0$, and the quadruple link string cancels by $I_7=0$.
Thus, all contributions to $\Y_1$ vanish.

\medskip

Contributions from $\Y_3$ compute as follows. The single-link string
between $S(q_1)$ and $S(q_2)$ cancels by $I_1 = I_2 =0$.
The double-link string between $S(q_1)$ and $S(q_2)$ cancels by
$I_4 = 0$. The triple-link string between $S(q_1)$ and $S(q_2)$
cancels by $I_6 =0$, and there is no quadruple link string.
Thus, all contributions to $\Y_3$ vanish.
Contributions to $\Y_4$ arise only through the single-link string
between $S(q_1)$ and $S(q_2)$ and cancels by $I_1 =0$.

\medskip

To evaluate $\Y_5$, we use the fact that all the $\delta$-dependence of
$\hat \mu$ is through $S_\delta (q_1,q_2)$, so that the $\delta$-resummed
$\Z[\delta] \Y_5$ is proportional to,
\bea
\sum _\delta \Z[\delta ] S_\delta (q_1,q_2) \left \{
a_1 + a_2 S_\delta (z_i,z_j)^2 +
a_3 S_\delta (z_1,z_2) S_\delta (z_2,z_3) S_\delta (z_3,z_1) \right \}
\eea
where $a_1,a_2,a_3$ are $\delta$-independent. The terms proportional to
$a_1,a_2,a_3$ respectively cancel by using $I_1=0$, $I_2=0$ and $I_3=0$.
Hence the contribution of $\Y_5$ vanishes.

\medskip

Only the contributions to $\Y_2$ remain.
The bosonic stress tensor gives no contributions by $I_1=I_2=I_3=0$. By
definition of $\Y_2$, we need consider only the connected parts in
the contributions from the fermionic stress tensor.
Since the fermionic stress tensor insertion is normal ordered, at least two
vertex operators of the three must contribute a fermion
bilinear in $\Y_2$. Thus, we get two contributions, one from
two insertions of fermion bilinears and one from 3 insertions
(i.e. at each vertex) of fermion bilinears.
The correlator with two insertions is proportional to
$\varphi [\delta] (w, z_2, z_3) S_\delta (z_2,z_3)$. Upon summation over
$\delta$ against $\Z [\delta]$, this yields $I_{13}(w;z_1,z_2)$, whose
integration against $\mu$ vanishes in view of (\ref{muI13}). Hence,
the contribution of the term with two insertions to the GSO projected amplitude
vanishes.

\medskip

The contribution with three insertions of vertex fermion bilinears 
is proportional to (we omit constant factors and the correlator of the $x_+$ field),
\bea
f_1 ^{\mu _1 \nu _1}   f_2 ^{\mu _2 \nu _2}  f_3 ^{\mu _3 \nu _3}
 \left \< \int \hat \mu T _\psi \,
\psi _+ ^{\mu_1} \psi _+ ^{\nu_1} (z_1)
\psi _+ ^{\mu_2} \psi _+ ^{\nu_2} (z_2)
\psi _+ ^{\mu_3} \psi _+ ^{\nu_3} (z_3) \right \>
\eea
Upon carrying out the contractions in all possible ways, the polarization
coefficient is proportional to the kinematic factor for the 3-point function $C_3$,
given by
\bea
C_3=f_1^{\mu\nu}f_2^{\nu\rho}f_3^{\rho\mu}
=
(\epsilon_1\cdot k_3)
(\epsilon_2\cdot k_1)
(\epsilon_3\cdot k_2)
-
(\epsilon_1\cdot k_2)
(\epsilon_2\cdot k_3)
(\epsilon_3\cdot k_1).
\eea
The Green function contributions are proportional to,
\bea
& & 
C_3 \int \hat \mu _{\bar w}{}^w \bigg \{
\varphi [\delta ] (w;z_1,z_2) S_\delta (z_2,z_3) S_\delta (z_3,z_1)
  +
\varphi [\delta ] (w;z_2,z_3) S_\delta (z_3,z_1) S_\delta (z_1,z_2)
\no \\ && \hskip 1.5in   +
\varphi [\delta ] (w;z_3,z_1) S_\delta (z_1,z_2) S_\delta (z_2,z_3) \bigg \}
\eea
Upon summation over $\delta$ against $\Z[\delta]$, this yields the symmetrized
combination of $I_{14}$, whose integral against $\hat \mu$ vanishes by
(\ref{muI14}).
Thus the 3-point function vanishes.

\newpage

\section{The Chiral 4-point Function: Outline}
\setcounter{equation}{0}

We turn now to the evaluation of the 4-point function.
Since the calculations are lengthy, it may be helpful to list here
the main steps, with details left to subsequent sections.

\subsection{Kinematic Invariants for the 4-point Function}

Recall that $f_i^{\mu\nu}=\e_i^\mu k_i^\nu-\e_i^\nu k_i^\mu$
denotes the ``gauge-invariant field strength" of the $i$-th particle.
The kinematical factors for the 4-point function are given by
\bea
\label{kin4}
K (1,2,3,4)
& \equiv &
(f_1f_2) (f_3 f_4) + (f_1f_3) (f_2 f_4) + (f_1f_4) (f_2 f_3)
\no \\ &&
- 4 (f_1 f_2 f_3 f_4)  - 4 (f_1 f_3 f_2 f_4)  - 4 (f_1 f_2 f_4 f_3)
\no \\
C_T (i,j|k,l)
& \equiv &
(f_i f_k) (f_j f_l) - (f_i f_l) (f_j f_k)
\no \\ &&
+ 2 (f_i f_j f_k f_l)  - 2 (f_i f_j f_l f_k)
\eea
Here, $K$ is the familiar factor totally symmetric in its arguments, 
while $C_T$ is antisymmetric
in $i\leftrightarrow j$ and antisymmetric in $k\leftrightarrow l$.
Here and later, we  use the notation,
\bea
(f_i f_j) & \equiv & f_i ^{\mu \nu } f_j ^{\nu \mu}
\no \\
(f_i f_j f_k ) & \equiv & f_i ^{\mu \nu } f_j ^{\nu \rho} f_k ^{\rho \mu}
\no \\
(f_i f_j f_k f_l) & \equiv & f_i ^{\mu \nu } f_j ^{\nu \rho} f_k ^{\rho \sigma}
f_l ^{\sigma \mu}.
\eea

\subsection{Chiral Amplitudes after Summing over Spin Structures}

The GSO summation over $\delta$ considerably simplifies the 
chiral amplitude, and produces the following expressions.  The
contributions of the disconnected components are
\bea
\label{calBd}
\sum_\delta {\cal B}[\delta]^{(d)}
&=&
- { 1 \over 16\pi^2} \bigg(C_T(1,2|3,4)\Delta(1,2)\Delta(3,4)
+ C_T(1,3|2,4)\Delta(1,3)\Delta(2,4)
\nonumber\\
&&
\qquad\qquad
+ C_T(1,4|2,3)\Delta(1,4)\Delta(2,3)\bigg)
\,\<Q(p_I)\prod_{i=1}^4e^{ik_i\cdot x_+(z_i)}\>.
\eea
This formula reproduces formula (9.10) of \cite{I}, and shows that the 
totally symmetric function ${\cal S}$ in \cite{I} vanishes.

\medskip

Next, we shall list the contributions from the connected components,
which arise from the $\delta$-summation of $\Y_1$, $\Y_2$, $\Y_3$, 
$\Y_4$, and $\Y_5$. The contribution $\Y_2$ arises from the stress 
tensor insertion. It will be convenient to decompose $\Y_2$ further
according to contribution from the bosonic
stress tensor $T_x$ or from the fermionic stress tensor $T_\psi$,
\bea
\label{stress}
T_x=-\half\p_zx_+^\mu\p_zx_+^\mu,
\qquad
T_\psi=\half\psi_+^\mu\p_z\psi_+^\mu,
\eea
The contribution from $ T_\psi$ is further split according to whether
the kinematic factor is proportional to the symmetric $K$ or the 
antisymmetric $C_T$. In a straightforward notation, we thus have,
as follows,
\bea
\Y_2 = \Y_{2x} + \Y_{2\psi} ^S + \Y _{2\psi }^A
\eea
The contributions of the connected components are then given as follows,
\bea
\label{Y1}
\sum_{\delta}\Z[\delta]\Y_1&=&
-{\zeta^1\zeta^2\over 32\pi^2}
\Z_0\,K \, \prod_{i=1}^4\varpi(z_i) \,
\left \<Q(p_I)\p x_+(q_1)\p x_+(q_2)
\prod_{j=1}^4e^{ik_j\cdot x(z_j)} \right \>_{(c)}
\\
\label{Y2x}
\sum_\delta\Z[\delta]\Y_{2x} & = &
{1\over 8\pi}\Z_0\, K 
\prod_{i=1}^4\varpi(z_i)
\int \mu(w)
\left \<Q(p_I)\p x_+^\mu\p x_+^\mu(w)
\prod_{j=1}^4e^{ik_j\cdot x_+(z_j)}\right \>_{(c)}
\\
\label{Y2psiS}
\sum_\delta\Z[\delta]\Y_{2\psi} ^S & = &
 {1\over 2}\, K \Z_0 \sum _{i=1}^4 \p \Lambda (z_i) \prod _{j\not= i} \varpi (z_j)
\, \left \<Q(p_I)\prod_{i=1}^4e^{ik_i\cdot x_+(z_i)} \right \>
\\
\label{Y2psiA}
\sum_\delta\Z[\delta]\Y_{2\psi} ^A & = &
{\zeta^1\zeta^2\over 16\pi^2}
\bigg(C_T(1,2|3,4)\I_{16}^A(1,2|3,4)
+
C_T(1,3|2,4)\I_{16}^A(1,3 |2,4)
\nonumber\\
&&
\qquad\qquad
+
C_T(1,4|2,3)\I_{16}^A(1,4|2,3)\bigg)
\,\left \<Q(p_I)\prod_{i=1}^4e^{ik_i\cdot x_+(z_i)} \right \>.
\\
\label{Y3}
\sum_{\delta}\Z[\delta]\Y_3&=&0
\\
\label{Y4}
\sum_{\delta}\Z[\delta]\Y_4&=&0
\\
\label{Y5}
\sum_{\delta}\Z[\delta]\Y_5&=&
\half\Z_0 K \sum_{i=1}^4
d\bar z_i\p_{\bar z_i}\bigg(\Lambda(z_i) \prod_{j\not=i}\varpi (z_j)
\,\left \<Q(p_I)\prod_{j=1}^4e^{ik_j\cdot x_+(z_j)} \right \>\bigg)
\eea
Here $\Lambda(z)$ is the single valued scalar function defined by
(\ref{Lambda}). The expressions $\I_{16}^S$ and $\I_{16}^A$ are symmetric and 
antisymmetric versions of the integral  $\int \mu I_{16}$, and are defined by
\bea
\I_{16}^S(z_1,z_2,z_3,z_4)
&=&
{1\over 12}\sum_{\sigma\in S^4}\I_{16}(z_{\sigma
1},z_{\sigma2},z_{\sigma3},z_{\sigma4})
\\
\I_{16}^A(z_1,z_4|z_2,z_3)
&=&
{1\over 3}\bigg(\I_{16}(z_1,z_2,z_3,z_4)
+\I_{16}(z_3,z_4,z_1,z_2)-(z_2\leftrightarrow z_3)\bigg),
\eea
with $\I_{16}(z_1,z_2,z_3,z_4)$ itself being defined by
\bea
\I_{16}(z_1,z_2,z_3,z_4)={1\over 2\pi}\int
\mu(w)I_{16}(w;z_1,z_2,z_3,z_4).
\eea

\subsection{Elimination of the Dolbeault $\bar\p$-exact Components}

It is now clear that all the terms in $\sum_\delta{\cal B}[\delta]$ which are
not tensor products of pure $(1,0)$-forms are given by the expression
\bea
\label{dolexact}
\half\Z_0\, K \sum_{i=1}^4
d\bar z_i\p_{\bar z_i}\bigg(\int \! d^2\zeta \, \Lambda(z_i) \, \prod_{j\not=i}\varpi (z_j)
\,\left \<Q(p_I)\prod_{j=1}^4e^{ik_j\cdot x_+(z_j)} \right \>\bigg)
\eea
arising from $\sum_\delta\Z[\delta]\Y_5$. Note that for each $i$, the
contribution to this expression is manifestly a Dolbeault $\bar\p$-exact 
$(0,1)$-form in $z_i$, and a holomorphic (and hence closed) 
$(1,0)$-form in the remaining variables. The Dolbeault $\bar\p$-exactness 
allows us to eliminate such expressions by
de~Rham $d$-exact differentials, that is,
\bea
\sum_\delta\Z[\delta]\Y_5
&=&
-\half\Z_0\, K 
\sum_{i=1}^4(dz_i\p_{z_i}  \, \Lambda(z_i))\, \prod_{j\not=i}\varpi (z_j) 
\,\left \<Q(p_I)\prod_{j=1}^4e^{ik_j\cdot x_+(z_j)} \right \>
\nonumber\\
&&
-
\half\Z_0\, K \sum_{i=1}^4
\, \Lambda(z_i) \, \prod_{j\not=i}\varpi (z_j) 
\, \left \<Q(p_I)ik_i\cdot \p_{z_i}x_+\,\prod_{j=1}^4e^{ik_j\cdot
x_+(z_j)} \right \>
\nonumber\\
&&
+\half\Z_0\, K 
\sum_{i=1}^4d_{z_i}\bigg( \, \Lambda(z_i) \, \prod_{j\not=i}\varpi (z_j)
\, \left \<Q(p_I)\prod_{j=1}^4e^{ik_j\cdot x_+(z_j)} \right \>\bigg)
\eea
Thus we can write
\bea
\sum_\delta {\cal B}[\delta]
={\cal H}+\half\Z_0\, K 
\sum_{i=1}^4d_{z_i}\bigg(\int \! d^2\zeta \, \Lambda(z_i) \, \prod_{j\not=i}\varpi (z_j)
\, \left \<Q(p_I)\prod_{j=1}^4e^{ik_j\cdot x_+(z_j)} \right \>\bigg)
\eea
with the  form ${\cal H}$ given by
\bea
\label{Hsum}
{\cal H}
&=&
\sum_\delta {\cal B}[\delta]^{(d)}
+\int\,d^2\zeta\, \sum _\delta \Z[\delta ] \left (
\Y_1 + \Y_{2x} + \Y_{2\psi}^S + \Y_{2\psi}^A \right )
\nonumber\\
&&
-\half\Z_0\, K 
\sum_{i=1}^4(dz_i\p_{z_i} \int \! d^2\zeta \, \Lambda(z_i))\, \prod_{j\not=i}\varpi (z_j)
\, \left \<Q(p_I)\prod_{j=1}^4e^{ik_j\cdot x_+(z_j)} \right \>
\nonumber\\
&&
-\half\Z_0\, K \sum_{i=1}^4
\int \! d^2\zeta \, \Lambda(z_i) \, \prod_{j\not=i}\varpi (z_j)
\, \left \<Q(p_I)ik_i ^\mu \p x_+ ^\mu (z_i) \,\prod_{j=1}^4e^{ik_j\cdot
x_+(z_j)} \right \>
\eea
From the expressions for $\Y_1$, $\Y_{2x}$, $\Y_{2\psi}^S$ and $\Y_{2 \psi }^A$,
and the form of $\B [\delta] ^{(d)}$, it is manifest that ${\cal H}$ is purely 
a $(1,0)$ form; this form is holomorphic away from $z_i = z_j$ for $i \not= j$.

\subsection{Explicit Evaluation of ${\cal H}$}

It remains only to evaluate ${\cal H}$. The following three types of 
cancellations and recombining mechanism take place. First, we have,
\bea
\label{cancellation1}
\sum_\delta{\cal B}[\delta]^{(d)}+ \sum _\delta \Z[\delta ] \int \! d^2\zeta \,
\Y_{2\psi}^A
=0.
\eea
so that all the contributions proportional to the kinematic invariant 
$C_T$ cancel out of the full amplitude. The terms above turn out, 
however, to be individually independent of gauge choices. 

\medskip

The next  cancellation and recombining mechanism are more remarkable 
since, starting from terms in the left hand side which depend individually 
on both gauge choices of gravitino slice $\chiz$ and Beltrami differential 
$\muhat $, the combined outcome is a gauge-slice independent amplitude. 
This can only occur thanks to the presence of the  function $\Lambda(z)$ 
which arises from the Dolbeault $\bar\p$-exact terms.
The first is a  cancellation,
\bea
\label{cancellation2}
\sum _\delta \Z[\delta] \Y_{2\psi}^S 
-\half\Z_0\, K 
\sum_{i=1}^4(dz_i\p_{z_i}\Lambda(z_i))\,
\,\<Q(p_I)\prod_{j=1}^4e^{ik_j\cdot x_+(z_j)}\>\prod_{j\not=i}\varpi (z_j)
= 0
\eea
and the second is a recombining mechanism, given by,
\bea
\label{cancellation3}
&&
\sum_\delta\Z[\delta] \left ( \Y_1+ \Y_{2x} \right )
-\half\Z_0 \, K \sum_{i=1}^4
\Lambda(z_i) \, \prod_{j\not=i}\varpi (z_j)
\, \left \<Q(p_I)ik_i ^\mu \p x_+ ^\mu (z_i)\,\prod_{j=1}^4e^{ik_j\cdot
x_+(z_j)} \right \>
\nonumber\\
&&
\qquad
=
{\zeta ^1 \zeta ^2 \over 64\pi^2}\, K 
\, \Y_S \,
\exp \left \{ i\pi p_I^\mu\Omega_{IJ}p_J^\mu
+
2\pi i\sum_jp_I^\mu k_j^\mu\int_{z_0}^{z_j}\omega_I \right \}
\prod_{i<j}E(z_i,z_j)^{k_i\cdot k_j}.
\eea
Here $\Y_S$ is the holomorphic factor defined in
(\ref{Ystar}). This completes the proof of the formulas
(\ref{H}) and (\ref{explicitH}) announced in the Introduction.

\medskip

In the formulas (\ref{cancellation2})
and (\ref{cancellation3}), we see once again the mechanism for gauge slice
independence of gauge-fixed superstring amplitudes at work: as in the case of
the chiral superstring measure \cite{II}, the gauge choices of gravitino slice
$\chiz$ and Beltrami differential $\muhat $
are closely entertwined. They can only cancel when combined with one another.

\newpage

\section{Sums over Spin Structures for the 4-point Function}
\setcounter{equation}{0}

In this section, we combine the results of summations over spin
structures in \S 3 with the structure of kinematical invariants.

\subsection{The Disconnected Component $\sum_\delta{\cal B}[\delta]^{(d)}$}

As we had already noticed in the process of calculating the $N$-point 
function for $N\leq 3$, all the terms in the disconnected component 
$\sum_\delta {\cal B}[\delta]^{(d)}$ with fewer than 4 fermion bilinears 
$\psi_+^\mu\psi_+^\nu(z_i)$ cancel because of the identities 
$I_{17}=I_{18}=I_{19}=0$. This cancellation persists in the calculation
of the 4-point function.
Thus we need only consider the expression,
\bea
\<Q(p_I)\prod_{i=1}^4e^{ik_i\cdot x_+(z_i)}\>\,
\sum_\delta \Xi_6[\delta]\, \tet[\delta]^4 \, \W_0[\delta]
\eea
where $\W_0[\delta]$ is defined by
\bea
\label{W0}
\W_0[\delta]=\left ( - { i \over 2} \right ) ^4
\left \< \prod _{j=1}^4 f_j ^{\mu _j \nu _j}
\psi _+ ^{\mu _j} \psi _+ ^{\nu _j} (z_j)  \right \>.
\eea
The Wick contractions work out as follows
\bea
 \W_0 [\delta] & = & +
{1 \over 4 } (f_1f_2) (f_3 f_4) ~ S_\delta (z_1,z_2) ^2 S_\delta (z_3,z_4)^2
\no \\ && +
{1 \over 4 } (f_1f_3) (f_2 f_4) ~ S_\delta (z_1,z_3) ^2 S_\delta (z_2,z_4)^2
\no \\ && +
{1 \over 4 } (f_1f_4) (f_2 f_3) ~ S_\delta (z_1,z_4) ^2 S_\delta (z_2,z_3)^2
\no \\ && -
(f_1 f_2 f_3 f_4) ~
S_\delta (z_1,z_2)  S_\delta (z_2,z_3)  S_\delta (z_3,z_4) S_\delta (z_4,z_1)
\no \\ && -
(f_1 f_3 f_2 f_4) ~
S_\delta (z_1,z_3)  S_\delta (z_3,z_2)  S_\delta (z_2,z_4) S_\delta (z_4,z_1)
\no \\ && -
(f_1 f_2 f_4 f_3) ~
S_\delta (z_1,z_2)  S_\delta (z_2,z_4)  S_\delta (z_4,z_3) S_\delta (z_3,z_1)
\eea
We can now combine this formula for $\W_0[\delta]$ and the formulas for
$I_{20}$, and $I_{21}$ and their symmetrized ${\cal S}$ and antisymmetrized
${\cal T}$ versions, which multiply the kinematic invariants $K$ and $C_T$
respectively. Since ${\cal S}=0$, no contribution proportional to the 
symmetric invariant $K$ remains. Using the expression for ${\cal T}$
produces the announced result (\ref{calBd}).

\subsection{The Contributions from $\Y_1$}

The insertions $\psi _+ (q_1) $ and $\psi _+(q_2)$ must be contracted
with the fermion bilinears from the vertex insertions in the form of a
linear string. We refer to the number of fermion propagators in this string
as the {\sl  length} of the string. The string may be multiplied by
non-intersecting closed fermion $n$-cycle, with $n$ fermion propagators.

\medskip

To $\Y_1$, only length 1 contributes. Indeed, at length 2, we may have
a 0-cycle, a 2-cycle or a 3-cycle, which vanish by $I_4=I_5=I_{10}=0$
respectively.  At length 3, the 0-cycle and 2-cycle vanish by $I_6=I_9=0$;
at length 4, the 0-cycle vanishes by $I_7=0$; while at length 5, the
0-cycle vanishes by $I_8=0$.

\medskip

At length 1, the contributions from the 0-cycle, the 2-cycle and the 3-cycle
cancel by $I_1=I_2=I_3=0$ respectively. But the contributions from
two 2-cycles, and a single 4-cycle are non-vanishing and are governed by
the contraction of the fermion bilinears only in each of the 4 vertex operators.
This results in the term $W_0[\delta]$
which had already been evaluated in the previous
subsection.

\medskip

The contraction of the two supercurrent insertions produces a factor
of $S_\delta (q_1,q_2)$.
Summing this combined result against $\Z [\delta ] $, we obtain,
\bea
\sum _\delta \Z [\delta ] S_\delta (q_1,q_2)  \W_0 [\delta]
& = & +
{1 \over 4 } (f_1f_2) (f_3 f_4) ~ I_{11} (z_1,z_2;z_3,z_4)
 -
(f_1 f_2 f_3 f_4) ~ I_{12} (z_1,z_2,z_3,z_4)
\no \\ && +
{1 \over 4 } (f_1f_3) (f_2 f_4) ~ I_{11} (z_1,z_3;z_2,z_4)
 -
(f_1 f_3 f_2 f_4) ~ I_{12} (z_1,z_3,z_2,z_4)
\no \\ && +
{1 \over 4 } (f_1f_4) (f_2 f_3) ~ I_{11} (z_1,z_4;z_2,z_3)
-
(f_1 f_2 f_4 f_3)~ I_{12} (z_1,z_2,z_4,z_3)
\no \\
\eea
Now, $I_{11}$ and $I_{12}$ are equal and are totally symmetric
functions of their 4 vertex points $z_i$. Therefore, all terms are
proportional to  $I_{11}$ and only the combined kinematical factor $K$ enters.
We then end up with
\bea
\sum _\delta \Z [\delta ] S_\delta (q_1,q_2)  \W_0 [\delta]
=  - \half \Z_0 \, K \, \prod _{i=1} ^4 \varpi (z_i)
\eea
Assembling all contributions to $\Y_1 [\delta]$ and carrying out the summation
over $\delta$, we have
\bea
\sum _\delta \Z [\delta ]   \Y_1 [\delta]
 =
- {\zeta ^1 \zeta ^2 \over 32 \pi ^2}  \, \Z_0 \, K \, \prod _{i=1} ^4 \varpi
(z_i)
\left \< Q(p_I) \, \p x_+ (q_1) \p x_+ (q_2)
\prod _{j=1} ^4 e^{i k_j \cdot x_+ (z_j)} \right \> _{(c)}.
\eea
The $\p x_+ (q_1) \p x_+ (q_2)$ operator is normal ordered since the fermion
operators in the supercurrent have already been contracted. This prescription
is indicated on the correlator with the subscript $(c)$. This is the formula
(\ref{Y1}).

\subsection{Contribution to $\Y_{2x}$ (bosonic stress tensor)}

The contribution to $\Y_2$ from the bosonic stress tensor only
was denoted by $\Y_{2x}$. Its only non-vanishing contribution  
arises from the contraction of the fermion bilinears in all 4 vertex operators.
These contractions are as in $\W_0$. To sum over spin structures,
we write $\muhat = S_\delta(q_1,q_2)\, \mu(z)$, so the $\delta$-dependence 
of the Beltrami differential $\hat \mu$ is isolated. It is now straightforward to
express the contribution from $\Y_{2x}$ and apply the
identities (\ref{1-12}),
\bea
\sum _\delta \Z [\delta ] \Y_{2x} 
=  {1 \over 8 \pi} \Z_0 \, K \prod _{i=1}^4 \varpi (z_i)
\int  \mu (w) \left \< Q(p_I) \p x_+ ^\mu \p x_+ ^\mu (w) \prod _{j=1}
^4
e^{ik _j \cdot x_+ (z_j)} \right \>_{(c)}
\eea
Note that the bosonic correlator must be connected.
This is the formula (\ref{Y2x}).

\subsection{Contribution from $\Y_{2\psi}$ (fermionic stress tensor)}

The insertion of the fermionic stress tensor $T_\psi$ may be formulated
in terms of the insertion of a modified Szego kernel $S_\delta '$,
to be defined below.
Since we are computing the connected part only, the lowest contribution
is that of a 2-cycle, which gives $I_{13}$. This quantity integrates to
zero against $\hat \mu$ however. For the 3-cycle, the contribution
yields $I_{14}$. Its integration against $\hat \mu$ vanishes upon cyclic
symmetrization of its arguments, and this is precisely the combination
that enters into the 4-point amplitude (just as it was the one that
entered into the 3-point function). The fact that in the 4-point function
further bosonic contractions must be carried out is immaterial.
Thus, all these contributions vanish in the 4-point function.

\subsubsection{The integrated fermionic propagator $S_\delta'(x,y)$}

The only remaining correlator is when the fermionic stress tensor is inserted in
a correlator where each vertex operator contribution is limited to its fermion
bilinear prefactor. Since the stress tensor is integrated against $\muhat $,
it is useful to introduce the following correlator
$S_\delta'(x,y)$ in order to work out carefully the combinatorics,
\bea
\label{Sprime}
S_\delta'(x,y)
=
{1\over 2\pi}\left\<\int\hat \mu T_\psi\ \psi_+(x)\psi_+(y)\right\>.
\eea
The integrand can be evaluated using the definition of $T_\psi$
in (\ref{stress}). One finds
\bea
\left\<\int\hat \mu T_\psi\ \psi_+(x)\psi_+(y)\right\>
=\half
\varphi[\delta](w;x,y),
\eea
where $\varphi[\delta](w;x,y)$ was defined in (\ref{varphi}).
This gives the following formula for $S_\delta'(x,y)$
\bea
S_\delta'(x,y)
=
{1\over 4\pi}
\int \hat\mu _{\bar w} {}^w \,\varphi[\delta](w;x,y).
\eea

Returning now to the evaluation of the sums involving $\Y_2^\psi$,
the object of interest is
\bea
 \W_1 [\delta]  \equiv
\left ( - { i \over 2} \right ) ^4 {1 \over 2 \pi}
\left \< \int \hat \mu T ~ \prod _{j=1}^4 f_j ^{\mu _j \nu _j}
\psi _+ ^{\mu _j} \psi _+ ^{\nu _j} (z_j)  \right \>
\eea
It is convenient to work this out in terms of the modified
Szeg\"o kernel $S_\delta '$ first. This may be done by
successively replacing, in each term  in  $\W_0$, one
Szeg\"o kernel $S_\delta$ factor  by $S_\delta '$, and summing
up all contributions,
\bea
\W_1 [\delta]
& = & -
{1 \over 2 } (f_1f_2) (f_3 f_4)  \left  \{
S_\delta (z_1,z_2) S_\delta ' (z_1,z_2) S_\delta (z_3,z_4)^2
+
S_\delta (z_1,z_2)^2 S_\delta ' (z_3,z_4) S_\delta (z_3,z_4) \right \}
\no \\ && -
{1 \over 2 } (f_1f_3) (f_2 f_4)   \left  \{
S_\delta (z_1,z_3) S_\delta ' (z_1,z_3) S_\delta (z_2,z_4)^2
+
S_\delta (z_1,z_3)^2 S_\delta ' (z_2,z_4) S_\delta (z_2,z_4) \right \}
\no \\ && -
{1 \over 2 } (f_1f_4) (f_2 f_3)   \left  \{
S_\delta (z_1,z_4) S_\delta ' (z_1,z_4) S_\delta (z_2,z_3)^2
+
S_\delta (z_1,z_4)^2 S_\delta ' (z_2,z_3) S_\delta (z_2,z_3) \right \}
\no \\ && +
(f_1 f_2 f_3 f_4) ~   \{
S_\delta ' (z_1,z_2)  S_\delta (z_2,z_3)  S_\delta (z_3,z_4) S_\delta (z_4,z_1)
\no \\ && \hskip 1in +
S_\delta (z_1,z_2)  S_\delta ' (z_2,z_3)  S_\delta (z_3,z_4) S_\delta (z_4,z_1)
\no \\ && \hskip 1in +
S_\delta (z_1,z_2)  S_\delta (z_2,z_3)  S_\delta ' (z_3,z_4) S_\delta (z_4,z_1)
\no \\ && \hskip 1in +
S_\delta (z_1,z_2)  S_\delta (z_2,z_3)  S_\delta (z_3,z_4) S_\delta ' (z_4,z_1) 
\}
\no \\ && +
(f_1 f_3 f_2 f_4) ~ \{
S_\delta ' (z_1,z_3)  S_\delta (z_3,z_2)  S_\delta (z_2,z_4) S_\delta (z_4,z_1)
\no \\ && \hskip 1in +
S_\delta (z_1,z_3)  S_\delta ' (z_3,z_2)  S_\delta (z_2,z_4) S_\delta (z_4,z_1)
\no \\ && \hskip 1in +
S_\delta (z_1,z_3)  S_\delta (z_3,z_2)  S_\delta ' (z_2,z_4) S_\delta (z_4,z_1)
\no \\ && \hskip 1in +
S_\delta (z_1,z_3)  S_\delta (z_3,z_2)  S_\delta (z_2,z_4) S_\delta ' (z_4,z_1)
\}
\no \\ && +
(f_1 f_2 f_4 f_3) ~ \{
S_\delta ' (z_1,z_2)  S_\delta (z_2,z_4)  S_\delta (z_4,z_3) S_\delta (z_3,z_1)
\no \\ && \hskip 1in +
S_\delta (z_1,z_2)  S_\delta ' (z_2,z_4)  S_\delta (z_4,z_3) S_\delta (z_3,z_1)
\no \\ && \hskip 1in +
S_\delta (z_1,z_2)  S_\delta (z_2,z_4)  S_\delta ' (z_4,z_3) S_\delta (z_3,z_1)
\no \\ && \hskip 1in +
S_\delta (z_1,z_2)  S_\delta (z_2,z_4)  S_\delta (z_4,z_3) S_\delta ' (z_3,z_1)
\}
\eea

\subsubsection{Formulas in terms of the integrated ${\cal I}_{15}$
and ${\cal I}_{16}$}

Next, we carry out the sum over spin structures against $\Z[\delta]$.
For this we need the following integrated versions of
the sums $I_{15}$ and $I_{16}$
introduced in section \S 3.2,
\bea
\label{calI15I16}
{\cal I}_{15}(z_1,z_2,z_3,z_4)&=&
{1\over 2\pi}
\int \mu (w) \,I_{15}(w;z_1,z_2,z_3,z_4)
\nonumber\\
{\cal I}_{16}(z_1,z_2,z_3,z_4)&=&
{1\over 2\pi}
\int \mu (w) \,I_{16}(w;z_1,z_2,z_3,z_4),
\eea
and their cyclically permuted versions ${\cal I}_{15}^C$ and ${\cal I}_{16}^C$
\bea
\label{calI15C}
{\cal I}_{15}^C(1,2,3,4)
&=&
{\cal I}_{15}(1,2,3,4)
+
{\cal I}_{15}(2,3,4,1)
+
{\cal I}_{15}(3,4,1,2)
+
{\cal I}_{15}(4,1,2,3) \quad
\\
\label{calI16C}
{\cal I}_{16}^C(1,2,3,4)
&=&
{\cal I}_{16}(1,2,3,4)
+
{\cal I}_{16}(3,4,1,2).
\eea
Here we have abbreviated $z_i$ by $i$.
Using the definitions of $\I_{15}^C$ and $\I_{16}^C$, the previous expressions
for $\sum _\delta \Z[\delta] \W_1 [\delta]$
reduce to\footnote{Care is needed in obtaining the correct
sign upon identifying the corresponding $\I_{15}$ or $\I_{16}$ factors.}
\bea
\sum _\delta \Z[\delta] \W_1 [\delta]
& = &
+ {1 \over 4 } (f_1f_2) (f_3 f_4)  \, \I_{16} ^C (1,2;3,4) +
\half (f_1 f_2 f_3 f_4) \,  \I_{15} ^C (1,2,3,4)
\no \\ && +
{1 \over 4 } (f_1f_3) (f_2 f_4) \, \I_{16} ^C (1,3;2,4) +
\half (f_1 f_3 f_2 f_4) \, \I_{15} ^C (1,3,2,4)
\no \\ && +
{1 \over 4 } (f_1f_4) (f_2 f_3) \,  \I_{16} ^C (1,4;2,3) +
\half (f_1 f_2 f_4 f_3) \, \I_{15} ^C (1,2,4,3) \quad
\eea
To arrive at an expression in terms of the kinematic factors $K$ and $C_T$, we
introduce the symmetrized versions
of $\I_{15}$ and $\I_{16}$
\bea
\label{calI15S}
3\I_{15}^S(1,2,3,4)
&=&\I_{15}^C(1,2,3,4)
+
\I_{15}^C(1,3,4,2)
+
\I_{15}^C(1,4,2,3)
\\
\label{calI16S}
3\I_{16}^S(1,2,3,4)
&=&
\I_{16}^C(1,2;3,4)
+
\I_{16}^C(1,3;4,2)
+
\I_{16}^C(1,4;2,3).
\eea
as well as the following antisymmetrized versions
\bea
\label{calI15A}
3\I_{15}^A(1,4|2,3)
&=&
\I_{15}^C(1,2,3,4)-\I_{15}^C(1,3,2,4)
\\
3\I_{16}^A(1,4|2,3)
&=&
\I_{16}^C(1,2;3,4)
-
\I_{16}^C(1,3;2,4).
\eea
The quantities $\I_{15}^C$, $\I_{15}^S$, $\I_{15}^A$,
$\I_{16}^C$, $\I_{16}^S$, $\I_{16}^A$
satisfy many important identities. Leaving temporarily their full description
and derivation to the next section, we note that these identities imply that
$\I_{15}^S$ and $I_{16}^S$ are
invariant under all permutations of the 4 points,
that the following inversion formulas hold,
\bea
\I_{15}^C (1,2,3,4) & = &
\I_{15} ^S (1,2,3,4) + \I_{15} ^A (2,1|3,4) + \I_{15}^A (1,4|2,3)
\no \\
\I_{16}^C (1,2;3,4) & = &
\I_{16} ^S (1,2,3,4) + \I_{16} ^A (1,4|2,3) + \I_{16}^A (1,3|2,4)
\eea
and that we also have
\bea
\I _{15} ^S (1,2,3,4) & = & - 2 \, \I _{16} ^S (1,2,3,4)
\no \\
\I _{15} ^A (1,4|2,3) & = & -  \, \I _{16} ^A (1,4|2,3).
\eea
Combining all, we obtain the following
formula,
\bea
\label{kine}
\sum _\delta \Z [\delta ] \W _1 [\delta]
& = &
+ {1 \over 4} K \, \I _{16}^S
+ {1 \over 4} C_T (1,2|3,4) \I _{16}^A (1,2|3,4)
\no \\ && \hskip .7in
+ {1 \over 4} C_T (1,3|2,4) \I _{16}^A (1,3|2,4)
\no \\ && \hskip .7in
+ {1 \over 4} C_T (1,4|2,3) \I _{16}^A (1,4|2,3) \quad
\eea
This gives formulas  (\ref{Y2psiS}), and (\ref{Y2psiA}).

\subsection{Contributions from $\Y_3$ and $\Y_4$}

To $\Y_3$, the contributions of length 1 are with a 0-cycle, a 2-cycle
and a 3-cycle, which vanish by $I_1=I_2=I_3=0$. The contributions of length 2
are with a 0-cycle and a 2-cycle and vanish by $I_4=I_5=0$.
Finally, the contributions of lengths 3 and 4 are with 0-cycles only and
vanish by $I_6=I_7=0$. Thus, the full $\Y_3$ vanishes.

\medskip

The arguments for $\Y_4$ are analogous. At length 1, we have
a 0-cycle and a 2-cycle, which vanish by $I_1=I_2=0$. The contributions
of lengths 2 and 3 are with 0-cycles only and vanish by $I_4=I_5=0$.
Thus, the full $\Y_4$ vanishes.

\subsection{Contributions from $\Y_5$}

The entire contribution to $\Y_5$ is proportional to $\hat \mu$,
which is proportional to
$S_\delta (q_1,q_2)$. As  a result, upon summation over $\delta$,
the contributions with a 0-cycle, a 2-cycle and a 3-cycle cancel
by $I_1=I_2=I_3=0$ respectively. There only remain the contributions
from two 2-cycles and one 4-cycle, which yield $I_{11}$
and $I_{12}$ respectively. But these quantities were evaluated in $\W_0$
and yield a kinematical factor proportional to $K$,
\bea
\sum _\delta \Z [\delta ] \Y_5 [\delta ]
=
\half \Z_0 \, K
\sum _{j=1} ^4 \mu (z_j) \left \<
Q(p_I) \prod _{j=1}^4 e^{i k_j \cdot x_+ (z_j)} \right \>
\, \prod _{i=1}^4 \varpi (z_i)
\eea
This object is a $(1,0)$ form in 3 of its $z_i$ assignments but a $(0,1)$-form
in the remaining fourth one.

\medskip

It is at this stage that we can isolate the Dolbeault
$\bar\p$-exact form. Since each contribution above contains
the combination
$\hat \mu {\bar z_i} {}^{z_i} \varpi (z_i)$,
we can rewrite this factor as $\p_{\bar z_i} \Lambda(z_i)$
in view of the equation (\ref{Lambda})
and obtain
\bea
\sum _\delta \Z [\delta ] \Y_5 [\delta ]
=
\half  \Z_0 \, K
\sum _{j=1} ^4  \, d\bar z_i \p_{\bar z_i}\bigg( \Lambda (z_i)
\<  Q(p_I) \prod _{j=1}^4 e^{i k_j \cdot x_+ (z_j)} \>
\prod _{j \not= i} \varpi (z_j)\bigg).
\eea
This is the desired formula (\ref{Y5}).

\newpage

\section{Integrals of $I_{15}$ and $I_{16}$ against $\hat\mu $}
\setcounter{equation}{0}

In the last section, we made use of certain symmetry properties
and relations between $\I_{15}^S$, $\I_{15}^A$, $\I_{16}^S$, to derive an intermediate formula for the contribution $\Y_2^\psi$. In this section, 
we establish these relations. To do so, we shall express these integrals 
in terms of the following simpler integrals $\I$ and $\J_\alpha$, with
$\alpha =1,2$, defined by,
\bea
\I  (z_1,z_2,z_3,z_4)
& = &
{1 \over 2\pi}   \int  \hat\mu _{\bar w} {}^w  \varpi(w)^2 I_{15}^S(w;z_1,z_2,z_3,z_4)
\\
\J _\alpha (z_1,z_2,z_3,z_4) & = &
{1 \over 2 \pi} \Z_0 \varpi (z_1) \varpi (z_2)
\int  \hat \mu _{\bar w} {}^w \varpi (w)^2
G(z_3 ;z_4 ,z_1 ;q_\alpha,w) G(z_4 ; z_3, z_2 ;q_\alpha,w) \no
\eea
Recall that $G(z;z_1,z_2;q,w)$ is the Green's function in
$z$, with poles at $z=z_1$ and $z=z_2$, and zeroes at $z=q$
and $z=w$, appearing in the evaluation of $I_{15}$ and $I_{16}$
in section \S 3.  Clearly, the functions $\I$ and $\J_\alpha$ satisfy the following symmetry properties
\bea
\label{symmetryIJ}
\I(1,2,3,4)=\I(2,1,3,4)&=&\I(1,2,4,3)=\I(2,1,4,3)
\no \\
\J_\alpha(1,2,3,4)&=&\J_\alpha(2,1,4,3).
\eea

\subsection{Evaluation of the Integral $\I$}

We express $\I$ in terms of the function $\Lambda(w)$,
using the equation (\ref{Lambda}). Since $\Lambda (w)$ is 
a single-valued and smooth function for smooth $\hat \mu$,
we can integrate by parts, and we have
\bea
\I  (z_1,z_2,z_3,z_4)
&=&
- {1 \over 2\pi}  \Z_0 \varpi (z_1) \varpi (z_2) \int \Lambda(w) \p _{\bar w} 
\bigg \{
\varpi (w) G(z_3;z_1,z_2;q_1,w)
\nonumber\\
&& \hskip 1.6in \times G(z_4;z_1,z_2;q_2,w)
 + (q_1
\leftrightarrow q_2) \bigg \}
\eea
Inside the braces, the first $G$-factor has a simple pole in $w$ at $q_2$, but
is regular at $q_1$, while the second $G$-factor is regular at $q_2$,
but has a simple pole at $q_1$. The prefactor $\varpi (w)$ kills both poles.
Thus, only the double poles remain at $z_1$ and $z_2$.

\medskip

To compute the integral, we need the  asymptotics of $G$,
near $z_{1,2}$, which is given by
\bea
G(x;z_1,z_2;q_\alpha, w)
& = &
- \left ( {1 \over w-z_1}  - \half \p \ln \varpi(z_1) \right )
{\varpi (x) \over \varpi (z_1)}
+ \gamma (x;z_1;z_2;q_\alpha) 
\no \\
G(x;z_1,z_2;q_\alpha, w)
& = &
+ \left ( {1 \over w-z_2} - \half \p \ln \varpi(z_2) \right )
 {\varpi (x) \over \varpi (z_2)}
- \gamma (x;z_2;z_1;q_\alpha)
\eea
up to terms which are $\O (w-z_1)$ and $\O (w-z_2)$.
The finite part $\gamma$ may be evaluated for example by
representing $G$ in terms of the prime form and its derivatives.
\bea
G(z;z_i,z_j;q_\alpha,w) 
= \tau_{ij} (z) - {\Delta (z,q_\alpha) \over \Delta (w,q_\alpha)} \tau_{ij}(w)
- {\Delta (z,w) \over \Delta (q_\alpha,w)} \tau_{ij} (q_\alpha)
\eea
where the Abelian differential of the third kind is defined by
\bea
\tau _{ij} (z) = \p _z \ln { E(z,z_i) \over E(z,z_j)}
\eea
The finite part is now given by
\bea
\label{gamma1}
\gamma (x;z_1;z_2;q_\alpha)
=
\tau_{12}(x) -{\Delta (x,z_1) \over \Delta (q_\alpha,z_1)} \tau_{12}(q_\alpha)
+ {\varpi (x) \over \varpi (z_1)} \p_{z_1} \ln \bigg ( \varpi (z_1)^\half  E(z_1,z_2)
\bigg )
\eea
To evaluate the integral giving $\I$, we use the asymptotic 
behaviors at the poles $z_1$ and $z_2$ and we find,
We then have a fairly economical way of recasting the result,
\bea
\label{eye}
\I  (z_1,z_2,z_3,z_4)
& = &
- 2 \rho_1 \p \Lambda(z_1) -2 \rho_2  \p \Lambda(z_2)
\\ &&  +  \bigg \{
\Lambda(z_1) \, \rho_4 \, \gamma (z_4;z_1;z_2;q_2)  +
\Lambda(z_1) \, \rho_3 \, \gamma (z_3;z_1;z_2;q_1)
\no \\ && \quad +
\Lambda(z_2) \, \rho_4 \, \gamma (z_4;z_2;z_1;q_2) +
\Lambda(z_2) \, \rho_3 \, \gamma (z_3;z_2;z_1;q_1)   + ( q_1 \leftrightarrow
q_2).
\bigg \}
\no
\eea
where we have introduced the following abbreviation,
\bea
\rho _i \equiv \Z_0 \prod _{j \not= i} \varpi (z_j)
\eea

\subsection{Evaluation of the integrals $\J_\alpha$}

To evaluate $\J_\alpha$, we express $\hat \mu$ in terms 
of $\Lambda$, using (\ref{Lambda}), and integrate by parts in $w$,
\bea
\J _\alpha (z_1,z_2,z_3,z_4) & = &
- {1 \over 2 \pi} \Z_0 \varpi (z_1) \varpi (z_2)
\int  \Lambda (w) \p_{\bar w} \bigg \{ \varpi (w)
G(z_3 ;z_4 ,z_1 ;q_\alpha,w) 
\no \\ &&
\hskip 2.1in \times
G(z_4 ; z_3, z_2 ;q_\alpha,w) \bigg \}
\eea
This time, simple poles arise in $w$ at $z_1,z_2$.
The first $G$ factor also has a pole in $w$ at $z_4$,
but the second factor has a zero there. Conversely,
the second $G$ factor has a pole at $w=z_3$ but
the first $G$ factor has a zero there.
There is also a simple pole at $q_\beta = - q_\alpha + 2 \Delta + 2 \kappa$.
The residue at $q_\beta$ is given by (\ref{resG1}) and (\ref{resG2}), 
and the fact that,
\bea
\lim _{w \to q_\beta} \varpi (w) G(z;p_1,p_2;q_\alpha,w)
= \p \varpi (q_\beta) {\varpi (z) \Delta (p_1,p_2)
\over \varpi (p_2) \Delta (q_\beta , p_1)}
\eea
Putting all together, we have 
\bea
\label{jay}
\J _\alpha (z_1,z_2,z_3,z_4) & = &
- \Lambda(z_1) \, \rho _4 \, G(z_4;z_3,z_2;q_\alpha,z_1)
- \Lambda(z_2) \, \rho _3 \, G(z_3;z_4,z_1;q_\alpha,z_2)
\no \\ &&
- \Z_0 \Lambda(q_\beta) \, \p \varpi (q_\beta) \, c_\beta ^2 \,
\Delta (z_1,z_4) \Delta (z_2,z_3)
\eea
where the following combinations are defined by,
\bea
c_\beta  ^2 = {\varpi (u) \varpi (v)
\over \Delta (q_\beta ,u) \Delta (q_\beta ,v)}
\eea
and are independent of $u$ and $v$.

\subsection{The Integrals $\I_{15}$ and $\I_{16}$
in terms of $\I$ and $\J_\alpha$}

In section \S 7, equation (\ref{calI15I16}), we introduced 
$\I_{15}$ and $\I_{16}$, which are integrals of $\hat \mu$
against $I_{15}$ and $I_{16}$. Using the definitions of 
$\I$ and $\J_\alpha$, they take the form,
\bea
\label{IJform}
\I_{15} (1,2,3,4)
& = &
+ \I (1,2,3,4) 
+ \sum _{\alpha =1,2} \bigg ( \J_\alpha  (1,2,3,4) - \J_\alpha (2,1,3,4) \bigg )
\no \\
\I_{16} (1,2,3,4)
& = &
- \I (1,2,3,4) 
- \sum _{\alpha =1,2} \bigg ( \J_\alpha  (1,2,3,4) + \J_\alpha  (2,1,3,4) \bigg )
\eea

\subsubsection{Evaluation of the Symmetrized Integrals
$\I_{15}^S$ and $\I_{16}^S$}

The symmetrized forms $\I_{15}^S$ and $\I_{16}^S$ are easily
evaluated using (\ref{IJform}) and  the fact that
\bea
\label{calJsum}
\sum_{\sigma \in S_4}
\J _\alpha \left  (\sigma (1), \sigma (2), \sigma (3), \sigma (4) \right )=0
\eea
and we obtain,
\bea
\label{box2}
 \I_{15} ^S (1,2,3,4) =  - 2 \I_{16}^S (1,2,3,4) 
\eea
as well as the following explicit form,
\bea
\label{box4}
\I_{16} ^S (1,2,3,4)
& = &
- {1 \over 3} \bigg  (
\I(1,2;3,4) + \I(1,3;2,4) + \I(1,4;2,3)
\no \\ &&  \quad +
\I(3,4;1,2) + \I(2,4;1,3) + \I(2,3;1,4) \bigg).
\eea

\subsubsection{Evaluating the Antisymmetrized Integrals
$\I_{15}^A$ and $\I_{16}^A$}

The antisymmetrized combinations are somewhat more involved.
After some simplifications, using the symmetries of $\I$ and $\J_\alpha$,
and the identity (\ref{calJsum}), we derive the relation,
\bea
\label{box3}
 \I ^A _{15} (1,4|2,3) =-   \I ^A _{16} (1,4|2,3) 
\eea
and
\bea
\label{asymm1}
3 \I ^A _{16} (1,4|2,3) & = &
- \I (1,2;3,4) - \I(3,4;1,2) + \I (1,3;2,4) + \I(2,4;1,3)
\no \\ &&
- \sum _{\alpha =1,2} \bigg (
\J_\alpha (1,2,3,4) + \J_\alpha (2,1,3,4) + \J_\alpha (3,4,1,2) 
\no \\ && \hskip .5in 
+ \J_\alpha (4,3,1,2) - \J_\alpha (1,3,2,4) - \J_\alpha (3,1,2,4)
\no \\ && \hskip .5in
- \J_\alpha (2,4,1,3)  - \J_\alpha (4,2,1,3) \bigg )
\eea

\newpage

\section{First Cancellation: Terms Involving $C_T$}
\setcounter{equation}{0}

We begin now the evaluation of the holomorphic amplitude ${\cal H}$. 
As outlined in section \S 6.4, the evaluation will involve two specific 
cancellations and one recombination mechanism. The first is conceptually the
easiest, since it will turn out to involve only gauge slice independent quantities. 
We treat it in this section. The key ingredient is the evaluation of the integral
$\I_{16}^A$, which, in view of (\ref{box3}), also determines $\I_{15}^A$.
In (\ref{asymm1}), $\I_{16}^A$ is given in terms of the basic integrals
$\I$, calculated in (\ref{eye}) and $\J_\alpha$, calculated in (\ref{jay}).
It will turn out that a key ingredient is the gauge slice independent
combination $\Lambda (q_1) - \Lambda (q_2)$, which we evaluate first below.

\subsection{Evaluation of $\Lambda(q_1)-\Lambda(q_2)$}

It has been pointed out earlier that, under a change of Beltrami differential 
by a vector field $v^z$, the function $\Lambda(z)$ defined up to an additive 
constant by the equation (\ref{Lambda}) changes by $\varpi(z)v^z(z)$. 
Since $\varpi(z)$ vanishes at $q_1$ and $q_2$, it follows that the quantity 
$\Lambda(q_1)-\Lambda(q_2)$ does not depend on the choice of
Beltrami differential within its equivalence class. 
we can write
\bea
\Lambda(q_1) - \Lambda(q_2) =
- {1 \over 2 \pi} \int _w \mu (w) \varpi (w) G(w;q_1,q_2;p_1',p_2')
\eea
where the Beltrami differential $\mu$ is defined by (\ref{mutildedef}),
and the points $p_1', p_2'$ are arbitrary generic points, upon which the
integral is independent.\footnote{Choosing $p_1', p_2' \in \{ q_1,q_2\}$ would
not constitute a generic choice and results in a singularity in $G$. Thus,
one cannot simply choose $\mu$ to be supported at $q_1,q_2$.}

\medskip

It is manifest that $\mu$ is integrated against a single-valued
holomorphic
2-form in $w$, which confirms the independence of $\Lambda(q_1) - \Lambda(q_2)$
of the choice of slice
for $\mu$.
We can evaluate it by expressing the holomorphic 2-form in terms
of $\omega_I\omega_J$, and then making use of the defining equation
(\ref{mutildedef}) for $\mu$. However, since
the answer is known to be gauge-independent, we can proceed faster by taking
$\mu$ concentrated at 3 generic points $p_a$,
\bea
\mu (w) = {\zeta ^1 \zeta ^2 \over 4 \pi }
\sum _{a=1}^3 \mu _a \delta (w,p_a)
\eea
Inverting the matrix $\omega _I \omega _J(p_a)$,
we get that $\mu _a = \varpi _a (q_1,q_2)$ in the notation of
\cite{II,IV}. Their explicit form, for arbitrary points $q_1$, and $q_2$
is helpful and given as follows,
\bea
\label{mutilde}
\mu_1  & = &
\half {\Delta (q_1,p_2) \Delta (q_2,p_3) + \Delta (q_1,p_3) \Delta (q_2,p_2)
\over
\Delta (p_1,p_2) \Delta (p_1,p_3)}
\no \\
\mu_2  & = &
\half {\Delta (q_1,p_1) \Delta (q_2,p_3) + \Delta (q_1,p_3) \Delta (q_2,p_1)
\over
\Delta (p_2,p_1) \Delta (p_2,p_3)}
\no \\
\mu_3  & = &
\half {\Delta (q_1,p_1) \Delta (q_2,p_2) + \Delta (q_1,p_2) \Delta (q_2,p_1)
\over
\Delta (p_3,p_1) \Delta (p_3,p_2)}
\eea
With this choice, the general expression for $\Lambda(q_1)-\Lambda(q_2)$ is
given by
\bea
\Lambda(q_1) - \Lambda(q_2) =
- {\zeta ^1 \zeta ^2 \over 8 \pi^2} \sum _a \mu _a \varpi (p_a)
G(p_a;q_1,q_2;p_1',p_2')
\eea
The arbitrary points in the Green function may be chosen conveniently  as
follows, $p_1' = p_1$ and $ p_2'=p_2$,  so that in the sum over $a$ only the 
value $a=3$ is non-vanishing. The result is
\bea
\Lambda(q_1) - \Lambda(q_2) =
- {\zeta ^1 \zeta ^2 \over 8 \pi^2}  \mu _3 \varpi (p_3)
G(p_3;q_1,q_2;p_1,p_2)
\eea
It remains to calculate the combination
\bea
\mu _3 \varpi (p_3)  G(p_3;q_1,q_2;p_1,p_2)
& = &
{\Delta (q_1,p_1) \Delta (q_2,p_2) + \Delta (q_1,p_2) \Delta (q_2,p_1)
\over
2 \Delta (p_1,p_3) \Delta (p_2,p_3)}
\varpi (p_3)
\no \\
&& \hskip - 1.5in \times
{\tet (p_3-q_1-q_2+p_1+p_2- \Delta) E(p_3,p_1) E(p_3,p_2) E(q_1,q_2)
\sigma (p_3)
\over
\tet (-q_2+p_1+p_2- \Delta) E(p_3,q_1) E(p_3,q_2) E(q_1,p_1)E(q_1,p_2)
\sigma (q_1)}
\eea
It is readily checked that this expression is a well-defined,
single-valued and holomorphic scalar in each $p_a$, and must thus be
independent of all $p_a$. Independence of $p_3$ allows us to choose
alternatively $p_3 = q_1$ and $p_3=q_2$, which yield in turn,
\bea
\lim _{p_3 \to q_1} \mu _3 \varpi (p_3)  G(p_3;q_1,q_2;p_1,p_2)
& = &
+ { c_1 \over c_2} \p \varpi (q_1)
\no \\
\lim _{p_3 \to q_2} \mu _3 \varpi (p_3)  G(p_3;q_1,q_2;p_1,p_2)
& = &
- { c_2 \over c_1} \p \varpi (q_2)
\eea
Here, we define $c_1$ and $c_2$ by
\bea
\label{cees}
\varpi (z) = c_\alpha \Delta (q_\alpha, z)
\eea
In particular, we obtain the following useful identity,
\bea
\label{cid}
c_1^2\p\varpi(q_1)+c_2^2\p\varpi(q_2)=0.
\eea
Assembling all pieces, we obtain the formula
\bea
\label{Lambdadiff}
\Lambda(q_1) - \Lambda(q_2) =
- {\zeta ^1 \zeta ^2 \over 8 \pi^2} { c_1 \over c_2} \p \varpi (q_1) =
+ {\zeta ^1 \zeta ^2 \over 8 \pi^2} { c_2 \over c_1} \p \varpi (q_2).
\eea

\subsection{Remarks on Singular Choices of $\mu(z)$}

As an aside, we discuss pitfalls associated with choices of $\mu(z)$
supported at points $p_a$, with $p_a$ tending to the $q_\alpha$. These choices can lead to
apparently contradictory statements,
which would have to be sorted out very carefully in order to arrive at a reliable answer.

\medskip

For example, a first choice of the points $p_a$ is the following.
Let $q_1$ and $q_2$ at first be
arbitrary points (not the zeros of a holomorphic (1,0)-form) and let $p_1\to
q_1$
and $p_2 \to q_2$. Using the formulas (\ref{mutilde}), we then have
\bea
\mu_1   =
\half {\Delta (q_2,p_3) \over \Delta (q_1,p_3)},
\quad
\mu_2   =
\half {\Delta (q_1,p_3) \over \Delta (q_2,p_3)},
\quad 
\mu_3   =
\half { \Delta (q_1,q_2) \Delta (q_2,q_1)
\over
\Delta (p_3,q_1) \Delta (p_3,q_2)}.
\eea
Taking now the limit where $q_1$ and $q_2$ become the distinct
zeros of a holomorphic 1-form, and using again the definitions of $c_\beta$,
we obtain,
\bea
\mu _1 = \half {c_1 \over c_2}
\hskip .7in
\mu _2 = \half {c_2 \over c_1}
\hskip .7in
\mu _3 = 0.
\eea

\medskip

However, we can also consider the following second choice. Let $q_1$ and $q_2$
from the outset be
the zeros of a holomorphic 1-form (i.e. unitary gauge). Then, $\hat \Omega _{IJ}
- \Omega _{IJ}$ is rank 1 only. Now, in (\ref{mutilde}), let $p_1 \to q_1$,
then we get instead
\bea
\mu _1 = {c_1 \over c_2}
\hskip .7in
\mu _2 = 0
\hskip .7in
\mu _3 = 0
\eea
Clearly, this results in a very different formula for the amplitudes $\Y_i$'s.

\medskip
Another manifestation of subtleties with special choices of
points $p_a$ for the support of $\mu$ is perhaps even easier to see: if
one sets $p_1=q_1$ and $p_2=q_2$, then
$\mu\varpi$ vanishes identically. Thus $\Lambda(z)$ also vanishes
identically, contradicting the formula which we just
obtained for $\Lambda(q_1)-\Lambda(q_2)$.

\medskip
Thus a reliable outcome is guaranteed only if a smooth $\mu(z)$ is used,
as long as there are terms either involving an un-integrated $\mu(z)$
appears (such as the corrections ${\cal V}^{(2)}(z)$ to the vertex operators),
or involving $\mu(z)$ integrated against meromorphic
correlation functions (such as in  $\Y_2$).
Only after all such terms cancel, and when only terms involving
integrals of $\mu(z)$ against holomorphic 2-forms remain
(as in the calculation of $\Lambda(q_1)-\Lambda(q_2)$
in the previous subsection),
can one choose $\mu(z)$ to be supported at special points.

\subsection{Evaluation of the Coefficient $\Z_0c_1c_2\p\varpi(q_1)\varpi(q_2)$}

First, we observe that, if we view the surface $\Sigma$ as a double cover of the sphere, then the factor $\Z_0c_1c_2\p\varpi(q_1)\varpi(q_2)$
is actually a well-defined scalar function
on the sphere. This is because it is symmetric in $q_1$ and $q_2$, has no monodromy when $q_1$ and $q_2$ go around the same cycle with opposite directions, and its tensor weight
works out to be 0, when the tensor weights of all
factors are taken into account.
Alternatively, we shall see shortly from the relation
(\ref{I16Z0cc}) below that it is also a coefficient of proportionality between $\I_{16}^A(1,4|2,3)$
and $\Delta(1,4)\Delta(2,3)$. Now $\I_{16}$ is a scalar in the
underlying parameters $q_1$ and $q_2$, since the components
$\Z[\delta]$ and $S_\delta(q_1,q_2)$ of $I_{16}$ are forms of weights $-3/2$ 
and $1/2$ respectively in each of these variables,
and the Beltrami differential $\mu(z)$ should be viewed as valued
in the tensor product $T_{1,0}^{q_1}(\Sigma)\otimes T_{1,0}^{q_2}(\Sigma)$. Integrating $I_{16}$ against $\mu(z)$ results then in a scalar function, which descends on the sphere,
since it is symmetric in $q_1$ and $q_2$.
From the expression of $\Z_0c_1c_2\p\varpi(q_1)\varpi(q_2)$,
it is holomorphic, and thus it suffices to evaluate it at a single point.

\medskip

We begin by deriving a more convenient formula for $\varpi(w)$, using (\ref{sigma})
in which we set $r_1=q_1$ and then take the limit $p_2 \to r_3=w$. One gets,
\bea
\varpi (w) = - e^{2 \pi i \kappa ' (q_1-\Delta)}
{\tet (q_1+w-z-\Delta) E(w,q_1) \sigma (w) \over E(z,q_1) E(z,w) \sigma  (z)}
\eea
It is now straightforward to take the derivative at $q_1$ and we obtain,
\bea
\p \varpi (q_1) & = & - e^{2 \pi i \kappa ' (q_1-\Delta)}
{\sigma (q_1) \, \tet (2 q_1 - z_1 - \Delta) \over \sigma (z_1) E(z_1,q_1)^2}
\no \\
\p \varpi (q_2) & = & + e^{2 \pi i \kappa ' (q_2-\Delta)}
{\sigma (q_2) \, \tet (2 q_2 - z_2 - \Delta) \over \sigma (z_2) E(z_2,q_2)^2}
\eea
To evaluate the coefficients $c_\alpha$, we proceed from (\ref{zeecube})
and set $p_1=z$, $p_2=q_1$ and let $w_0 \to z$, so that we get
\bea
Z^3 \Delta (z,q_1)
& = &
-  \varpi (z) \sigma (q_1)  e^{- 2 \pi i \kappa ' (q_1-\Delta)}
\no \\
Z^3 \Delta (z,q_2)
& = &
+  \varpi (z) \sigma (q_2)  e^{- 2 \pi i \kappa ' (q_2-\Delta)}
\eea
which yields the following equations for the coefficients $c_\alpha$,
\bea
c_1  & = &  + Z^3   \sigma (q_1 ) ^{-1} e^{2 \pi i \kappa ' (q_1 -\Delta)}
\no \\
c_2  & = &  - Z^3  \sigma (q_2 ) ^{-1} e^{2 \pi i \kappa ' (q_2 -\Delta)}
\eea
Assembling the following products, we get
\bea
c_1 c_2 \p \varpi (q_1)  \p \varpi (q_2)
=
Z^6 e^{8 \pi i \kappa ' \kappa } \,
{ \tet (2 q_1 - z_1 - \Delta) \tet (2 q_2 - z_2 - \Delta)
\over
\sigma (z_1) \sigma (z_2) E(z_1,q_1)^2 E(z_2,q_2)^2}
\eea
The $\tet$-functions may be evaluated using
(\ref{zeecube}) where $r_1=q_2$, $r_2 = z_1$, and $r_3 = q_1$,
and then using $q_1 + q_2 = 2 \Delta + 2 \kappa$, and finally by
interchanging $q_1$ and $q_2$. We obtain,
\bea
\tet (q_2-q_1+z_1-\Delta)  = \tet (2 q_1-z_1-\Delta -2 \kappa )
=
Z^3 {\sigma (q_1) E(q_2,q_1) E(z_1,q_1) \Delta (q_2,z_1)
\over
\sigma (q_2) \sigma (z_1) E(q_2,z_1) }
\no \\
\tet (q_1-q_2+z_2-\Delta)  = \tet (2 q_2-z_2-\Delta -2 \kappa )
=
Z^3 {\sigma (q_2) E(q_1,q_2) E(z_2,q_2) \Delta (q_1,z_2)
\over
\sigma (q_1) \sigma (z_2) E(q_1,z_2) }
\no
\eea
Since $2 \kappa$ is a full period, we may use the periodicity
of the $\tet$-function,
\bea
\tet (\zeta - 2 \kappa) = \tet (\zeta) e^{- 4 \pi \kappa ' \Omega \kappa'
+ 4 \pi \kappa ' \zeta }
\eea
Multiplying  both, one finds,
\bea
&&
 \tet (2 q_1 - z_1-\Delta) \tet (2 q_2 - z_2-\Delta)
 e^{ + 8 \pi i \kappa ' \Omega \kappa ' - 4 \pi i \kappa ' (z_1+z_2-2 \Delta) }
\no \\
&& \hskip .5in =
-  Z^6 { E(q_1,q_2)^2 E(z_1,q_1) E(z_2,q_2) \Delta (q_1,z_2) \Delta (q_2,z_1)
\over
\sigma (z_1) \sigma (z_2) E(z_1,q_2) E(z_2,q_1)}
\eea
Combining all these results, we get
\bea
\label{combo}
\Z_0 c_1 c_2 \p \varpi (q_1)  \p \varpi (q_2)
=
- {\Z_B  \, e^{4 \pi i \kappa ' (z_1+z_2-2 \Delta) }
Z^{24} \Delta (q_1,z_2) \Delta (q_2,z_1)
\over
\prod _{\alpha =1,2} \bigg (
\sigma (z_\alpha )^2  \sigma (q_\alpha )^2
E(z_\alpha ,q_1) E(z_\alpha ,q_2) \bigg )}
\eea
As a check, one may use the monodromy transformations of the
prime form in (\ref{primemonod}) and of $\sigma$ in (\ref{sigmamonod})
to check that this expression is indeed independent of $z_1$, $z_2$
$q_1$ and $q_2$.

\medskip

Given that (\ref{combo}) is independent of $z_1$, $z_2$ $q_1$ and $q_2$, 
we may evaluate it by taking the limit $q_2 \to q_1$ which is smooth. 
Since $q_1 + q_2 = 2 \Delta + 2 \kappa$, we take $\kappa = \nu _1$
to be any odd spin structure, so that $q_1, q_2 \to p_1=\Delta + \nu _1$.
The above result simplifies further and we have,
\bea
\Z_0 c_1 c_2 \p \varpi (q_1)  \p \varpi (q_2)
=
- {\Z_B \,  e^{4 \pi i \nu_1 ' (z_1 + z_2 - 2 \Delta) }
Z^{24} \Delta (p_1,z_1) \Delta (p_1,z_2)
\over
\sigma (z_1)^2 \sigma (z_2)^2  \sigma (p_1 )^4
E(z_1 ,p_1)^2 E(z_2 ,p_1)^2}
\eea
To evaluate this combination, we let $z_1$ and $z_2$ also tend
to branch points $z_1 \to p_2$, and $z_2 \to p_3$, with $p_1,p_2,p_3$
all mutually distinct. All ingredients may now be expressed in terms
of the following quantities, which were calculated in eqs. (3.9) and (3.14) 
of \cite{IV},
\bea
\Delta (p_i,p_j)
& = &
- \M _{\nu _i \nu _j} ^{-1} \omega _{\nu_i} (p_j) \omega _{\nu_j} (p_i)
\no \\
\omega _{\nu_i} (p_j) \sigma (p_j)^{-1}
& = &
C_j Z^{-3} \M_{\nu _i \nu _j}
\no \\
E(p_1,p_2) ^{-2}
& = &
\omega _{\nu_3} (p_1) \omega _{\nu_3} (p_2) \, \tet [\nu _3 ] (p_1-p_2)^{-2}
\no \\
E(p_1,p_3) ^{-2}
& = &
\omega _{\nu_2} (p_1) \omega _{\nu_2} (p_3) \, \tet [\nu _2] (p_1-p_3)^{-2}
\eea
Putting all together, we have
\bea
\label{combo1}
\Z_0 c_1 c_2 \p \varpi (q_1)  \p \varpi (q_2)
=
-  C_0 \, C_1 ^4 C_2 ^2 C_3 ^2 e^{4 \pi i \nu_1 ' (\nu_2+\nu_3)}
{\Z_B \,  \M _{12}^2 \M_{13}^2 \M_{23}^2 \over \tet [\nu_1+\nu_2+\nu_3](0)^4}
\eea
Using the expression for $\Z_B$ and the product relation for $\M_{ij}$'s,
derived in \S 4.2 of  \cite{IV},
\bea
\M _{12}^2 \M_{13}^2 \M_{23}^2
= \pi ^{12} \tet [\nu_1+\nu_2+\nu_3](0)^4  \Psi _{10}
\eea
we find that the ratio factor on the rhs of (\ref{combo1}) equals to 1. 
The remaining factors are
\bea
C_i
& = &
\exp \{ -i \pi \nu _i ' \Omega \nu _i ' - 2 \pi i \nu _i ' \nu _i ''\}
\hskip 1in i=1,2,3
\no \\
C_0
& = &
{\tet [\nu_1+\nu_2+\nu_3](0)^4
\over
\tet [\nu _3 ] (p_1-p_2)^2 \tet [\nu _2] (p_1-p_3)^2 }
\eea
The  calculation of $C_0$ is routine. The result is
\bea
C_0 & =  &
\exp  \{
+  4 \pi i \nu _1 ' \Omega \nu _1 '
+ 2 \pi i \nu_2 ' \Omega  \nu _2  '
+ 2 \pi i \nu _3 ' \Omega \nu _3 '
- 4 \pi i  \nu _1 ' \Omega (\nu_2' + \nu_3')
\no \\ && \hskip .5in
- 4 \pi i \nu _3 '(\nu_1 '' - \nu _2 '') - 4 \pi i \nu_2 ' (\nu _1 '' - \nu
_3'')
\}
\eea
Putting all the exponential factors together, we get
\bea
-  C_0 \, C_1 ^4 C_2 ^2 C_3 ^2 e^{4 \pi i \nu_1 ' (\nu_2+\nu_3)} =1
\eea
and we finally obtain,
\bea
\label{qdependence}
 \Z_0 c_1 c_2 \p \varpi (q_1)  \p \varpi (q_2) =+ 1.
\eea

\subsection{Evaluation of the Antisymmetric Integral $\I^A _{16}$}

We start from the expression (\ref{asymm1}) for $\I^A _{16}$ and make
use of the explicit expressions for $\I$ and $\J_\alpha$ in (\ref{eye}) and
(\ref{jay}). The derivative terms $\p \Lambda$ of (\ref{eye}) cancel out
in the combination of $\I$ that enters (\ref{asymm1}). The remaining terms
are proportional to $\Lambda(i)$ with $i=1,2,3,4$ and to $\Lambda(q_\alpha)$. 
The result is
\bea
3 \I ^A _{16} (1,4|2,3) =  \sum _{i=1}^4 \tilde a_i \Lambda(i)
- 6 \Delta (1,4)  \Delta (2,3) \Z_0
\sum _\beta   \Lambda(q_\beta ) \p \varpi (q_\beta) c_\beta ^2
 \eea
 We first concentrate on the terms in $\Lambda(i)$.
Given the symmetries under the interchange of vertex points in
$\I^A _{16}$, it suffices  to compute the coefficient of a single $\Lambda(i)$
term, say $\Lambda(1)$. The coefficient for $\Lambda(1)$ is given by
\bea
\tilde a_1 & = &
\sum _\alpha \bigg \{
- \rho _4  \gamma (4;1,2;q_\alpha)
- \rho _3  \gamma (3;1,2;q_\alpha)
+ \rho _4  \gamma (4;1,3;q_\alpha)
\no \\ && \hskip .35in
+ \rho _2  \gamma (2;1,3;q_\alpha)
- \rho _4 G(4;3,2;q_\alpha,1)
- \rho _3 G(3;4,2;q_\alpha,1)
\no \\ && \hskip .35in
 + \rho _4 G(4;2,3;q_\alpha,1)
+ \rho _2 G(2;4,3;q_\alpha,1) \bigg \}
\eea
Working out these expression in terms of Abelian differentials, using (\ref{gamma1}), we have
\bea
\tilde a_1
& = &
2 \rho _1 \tau_{23}(1) + 2 \rho _2 \tau_{14} (2)
- 2 \rho _3 \tau_{14} (3) - 2 \rho _4 \tau_{23}(4)
\no \\ &&
+ \sum _\alpha \bigg [
\rho _4 { \Delta (4,1) \over \Delta (q_\alpha,1)} \tau_{23} (q_\alpha)
+
\rho _3 { \Delta (3,2) \over \Delta (q_\alpha,2)} \tau_{14} (q_\alpha) \bigg ]
\eea
It is readily checked that $\tilde a_1$ is a well-defined, single-valued
holomorphic 1-form in each $z_i$. For example, the simple pole in
$z_1-z_2$ from the first two terms has residue $2 \rho_1 - 2 \rho_2$
which vanishes at $z_1=z_2$. Also, the apparent poles when $z_i \to q_\alpha$
are cancelled by the $\rho _4$ and $\rho_3$ factors, each of
which contains a factor of $\varpi (z_i)$ which vanishes as $z_i \to q_\alpha$.
The absence of monodromy in the $z_i$ should of course follow
since the starting point was a single-valued expression, but it
may also be checked explicitly on the above formula.

\medskip

Using the above result that $\tilde a_1$ is a single-valued holomorphic
1-form in each $z_i$, and using antisymmetry of $\tilde a_1$ under
the interchange of 1 and 4 as well as under the interchange of 2 and 3,
it is clear that the entire $z_i$-dependence must be proportional to
$\Delta (1,4) \Delta (2,3)$. The same holds for each factor $\tilde a_i$.
\bea
\tilde a_i & = & a_i \Delta (1,4) \Delta (2,3)
\eea
Here, $a_i$ is now $z_j$-independent and, for $i=1$ is given by
\bea
a_1 \Delta (1,4) \Delta (2,3)
& = &
2 \rho _1 \tau_{23}(1) + 2 \rho _2 \tau_{14} (2)
- 2 \rho _3 \tau_{14} (3) - 2 \rho _4 \tau_{23}(4)
\no \\ &&
+ \sum _\alpha \bigg [
\rho _4 { \Delta (4,1) \over \Delta (q_\alpha,1)} \tau_{23} (q_\alpha)
+
\rho _3 { \Delta (3,2) \over \Delta (q_\alpha,2)} \tau_{14} (q_\alpha) \bigg ]
\eea
The points $z_i$ are  arbitrary, so the $z_j$-independent
coefficient may be determined by choosing convenient special values
for which $\Delta (1,4) \Delta (2,3) \not=0$. We choose $z_4 = q_1$,
so that
\bea
a_1 \Delta (1,q_1) \Delta (2,3)
& =  &
-  \rho_4  \tau_{23}(q_1)
+ \rho_4  {\Delta (q_1,1) \over \Delta (q_2,1)} \tau_{23} (q_2)
\no \\ &&
+ \varpi (1) \varpi(2) {\Delta (3,2) \over \Delta (q_1,2)} \lim _{z_4 \to q_1}
\varpi(z_4) \tau_{14} (q_1)
\eea
The limit is readily evaluated and equals $\p\varpi(q_1)$. Using
$\rho_4 = \varpi(1) \varpi(2) \varpi(3)$ and (\ref{cees}), 
as well as a simplification by an overall factor of $\Delta (1,q_1)$, we obtain
\bea
a_1  \Delta (2,3)
=
\varpi(2) \varpi (3) \bigg (c_1   \tau_{23}(q_1) - c_2  \tau_{23} (q_2) \bigg )
+ c_1^2 \Delta (2,3) \p \varpi (q_1)
\eea
The points 2 and 3 being still arbitrary, we make the choice $z_3=q_1$,
\bea
a_1  \Delta (2,q_1)
& = &
c_1  \varpi(2) \lim _{z_3 \to q_1} \varpi (z_3) \tau_{23}(q_1)
+ c_1^2 \Delta (2,q_1) \p \varpi (q_1)
\no \\
& = &
c_1 \varpi (2) \p\varpi(q_1) + c_1^2 \Delta (2,q_1) \p \varpi (q_1)
=0
\eea
Hence the coefficients $a_i$ and thus $\tilde a_i$ all vanish. As a result, we
obtain
\bea
\I ^A _{16} (1,4|2,3) =  - 2 \Delta (1,4)  \Delta (2,3)  \Z_0
\sum _{\beta =1} ^2  \Lambda(q_\beta ) c_\beta ^2 \p \varpi (q_\beta)
\eea
Finally, we assemble this with the relation (\ref{cid}) between
the $c_\alpha^2\omega(q_{\alpha})$ for $\alpha =1,2$ and the formula (\ref{Lambdadiff}) for $\Lambda(q_1)-\Lambda(q_2)$
to obtain,
\bea
\label{I16Z0cc}
\I_{16}^A(1,4|2,3)
=
{\zeta^1\zeta^2\over 4\pi^2}
\Delta(1,4)\Delta(2,3)
\Z_0c_1c_2\p\varpi(q_1)\varpi(q_2).
\eea
Using now the value of the overall normalization factor (\ref{qdependence}), we find
\bea
\label{I16Aform}
 \I ^A _{16} (1,4|2,3)
 =    {\zeta ^1 \zeta ^2 \over 4 \pi^2} \Delta (1,4)  \Delta (2,3).
\eea

\subsection{Cancellation of the kinematic invariant $C_T$}

With the formula for $\I_{16}^A$ which we just obtained,
it now clear that the result (\ref{Y2psiA}) holds, and that, 
using also (\ref{calBd}), the equation
(\ref{cancellation1}) is verified, and thus
all the terms involving the kinematic invariant $C_T$ cancel.\footnote{That this
cancellation should occur was suggested to us by John Schwarz.}

\newpage

\section{Second Cancellation: Terms Involving $\p\Lambda(z)$}
\setcounter{equation}{0}

In this section, we show that the contribution 
arising from the fermionic stress tensor which is proportional to
the symmetric kinematical invariant $K$, namely 
$\sum _\delta \Z[\delta ] \Y_{2\psi}^S$, 
is cancelled by the de~Rham $d$-exact counterterms
from $\sum_\delta\Z[\delta]\Y_5$.

\medskip

From (\ref{kine}), it is clear that the entire contribution 
$\sum _\delta \Z[\delta ] \Y_{2\psi}^S$ is proportional to $\I_{16}^S$,
which in turn is proportional to the symmetrized version of the integral $\I$,
as given in (\ref{box4}). We recall these formulas here for convenience,
\bea
\I_{16}^S (1,2,3,4)
& = &
-{1 \over 3} \bigg  (
\I(1,2;3,4) + \I(1,3;2,4) + \I(1,4;2,3)
\no \\ &&  \quad +
\I(3,4;1,2) + \I(2,4;1,3) + \I(2,3;1,4) \bigg)
\eea
where the function ${\cal I}$ has been calculated in (\ref{eye}),
and was found to be given by
\bea
\label{eye1}
\I  (z_1,z_2,z_3,z_4)
& = &
- 2 \rho_1 \p \Lambda(z_1) -2 \rho_2  \p \Lambda(z_2)
\\ &&  +  \bigg \{
\Lambda(z_1) \rho_4  \gamma (z_4;z_1;z_2;q_1)  +
\Lambda(z_1) \rho_3  \gamma (z_3;z_1;z_2;q_1)
\no \\ && \quad +
\Lambda(z_2)\rho_4  \gamma (z_4;z_2;z_1;q_1) +
\Lambda(z_2) \rho_3  \gamma (z_3;z_2;z_1;q_1)   + ( q_1 \leftrightarrow
q_2)
\bigg \}
\no
\eea
where we use of the following notation,
\bea
\gamma (x;p_1;p_2;q_\alpha)
=
\tau_{12} (x) - {\Delta (x,p_1) \over \Delta (q_\alpha, p_1)} \tau_{12}
(q_\alpha)
+ {\varpi (x) \over \varpi (p_1)} \p_{p_1}
\ln \left ( \varpi (p_1)^\half E(p_1,p_2) \right )
\eea
The effect of the symmetrization is given as follows,
\bea
\I_{16} ^S (1,2,3,4)
& = &
\sum _{i=1}^4 \bigg (
2 \rho _i \p \Lambda(z_i) - {1 \over 3} \Lambda(z_i) F_i (z_1,z_2,z_3,z_4)
\bigg ).
\eea
The $F_i$ are $(1,0)$-forms in each $z_j$. It suffices to examine $F_1$,
as the expression for the remaining $F_i$ may be obtained as the
3 cyclic permutations of $F_1$. We find,
\bea
F_1
& = &
+ \rho _2  \gamma (2;1,3;q_1) + \rho _2  \gamma (2;1,4;q_1)
\no \\ &&
+ \rho _3 \gamma (3;1,2;q_1) + \rho _3 \gamma (3;1,4;q_1)
\no \\ &&
+ \rho _4 \gamma (4;1,2;q_1) + \rho _4  \gamma (4;1,3;q_1)
+ (q_1 \leftrightarrow q_2)
\eea
in the notation introduced in section \S 8.1.

\subsection{$F_1$ is a single-valued holomorphic 1-form}

It will be useful to write out this object in detail by representing
$\gamma$ by the above formula,
\bea
F_1 & = &
2 \rho _2 \tau _{13}(2) + 2 \rho _2 \tau_{14}(2) +
2 \rho _3 \tau _{12}(3) + 2 \rho _3 \tau_{14}(3) +
2 \rho _4 \tau _{12}(4) + 2 \rho _4 \tau_{13}(4)
\no \\ && +
2 \rho _1 \p _1 \ln \left ( \varpi (1)^3 E(1,2) E(1,3) E(1,4) \right )
\no \\ && +
\Delta (1,2) \varpi (3) \varpi (4)
    \bigg ( c_1 \tau_{13} (q_1) + c_2 \tau_{13}(q_2)
    + c_1 \tau_{14} (q_1) + c_2 \tau_{14}(q_2)\bigg )
\no \\ && +
\Delta (1,3) \varpi (2) \varpi (4)
    \bigg ( c_1 \tau_{12} (q_1) + c_2 \tau_{12}(q_2)
    + c_1 \tau_{14} (q_1) + c_2 \tau_{14}(q_2)\bigg )
\no \\ && +
\Delta (1,4) \varpi (2) \varpi (3)
    \bigg ( c_1 \tau_{12} (q_1) + c_2 \tau_{12}(q_2)
    + c_1 \tau_{13} (q_1) + c_2 \tau_{13}(q_2)\bigg )
\eea
We claim that $F_1$ is a single-valued, holomorphic 1-form
in each $z_j$. It is easy to show that all poles in $F_1$
cancel in each of the following limits :
 $z_2 \to z_1$, $z_2 \to z_3$, $z_1 \to q_1$, and $z_2 \to q_1$.
 By symmetry in the arguments, all other cases are equivalent to
 one of these. Checking cancellation of the monodromies is
 slightly more involved. Monodromies around homology cycles $A_I$
 are manifestly zero. Under $z_1 \to z_1 + B_I$, we have
 \bea
F_1 & \to &
F_1 + 8 \pi i \rho _2 \omega _I (2) + 8 \pi i \rho _3 \omega _I (3)
+  8 \pi i \rho _4 \omega _I (4) - 24 \pi i \rho _1 \omega _I (1)
\no \\ &&
+4  \pi i \bigg ( c_1 \omega _I (q_1) + c_2 \omega _I (q_2) \bigg )
\bigg ( \varpi (3) \varpi (4) \Delta (1,2) + \varpi (2) \varpi (4) \Delta (1,3)
\no \\ && \hskip 2in
+ \varpi (2) \varpi (3) \Delta (1,4) \bigg )
\eea
To show vanishing, we use the fact that $c_1 \omega _I (q_1)
= c_2 \omega _I (q_2)$ and the following relation,
\bea
\rho _4 \omega _I(1) - \rho _1 \omega _I(4)
+ c_1 \omega _I (q_1) \Delta (1,4) \varpi (2) \varpi (3) =0
\eea
The vanishing  of the monodromy under $z_2 \to z_2 + B_I$, is shown 
with the same tools as for the $z_1$ monodromy.

\subsection{Vanishing of $F_1$ at Two Generic Points}

Since $F_1$ is a well-defined single-valued
holomorphic 1-form in each of its arguments $z_j$, it is easy
to show that it actually vanishes. To do so, it suffices to
show that $F_1$, as a form in $z_4$, vanishes at $q_1$
(and thus automatically at $q_2$) and at $z_1$. The first
is shown as follows,
\bea
\lim _{z_4 \to q_1} F_1
& = &
 \rho _4 \tau _{12} (q_1) +  \rho _4 \tau_{13} (q_1)
- {c_2 \over c_1} \rho _4 \tau _{12} (q_2)
- {c_2 \over c_1} \rho _4 \tau _{13} (q_2)
\no \\ &&
+ c_1 \p \varpi (q_1) \varpi (2) \Delta (1,3)
+ c_1 \p \varpi (q_1) \varpi (3) \Delta (1,2)
\eea
This combination vanishes with the help of the following identity,
\bea
\varpi (z_1) \varpi (z_2) \bigg (
c_1 \tau_{12}(q_1) - c_2 \tau _{12} (q_2) \bigg )
=
- c_1 ^2 \p \varpi (q_1) \Delta (1,2)
\eea
Next, we take the limit $z_4 \to z_1$, which takes the following form,
\bea
\hat F_1 = { 1 \over \varpi (z_1)} \lim _{z_4 \to z_1} F_1
& = &
2 \varpi (2) \varpi (3) \p _1 \ln \bigg (\varpi (1) E(1,2) E(1,3) \bigg )
\no \\ &&
+ 2 \varpi (1) \varpi (2) \tau_{12}(3)
+ 2 \varpi (1) \varpi (3) \tau_{13}(2)
\no \\ &&
+ \varpi (2) \Delta (1,3) (c_1 \tau_{12}(q_1) + c_2 \tau_{12}(q_2) )
\no \\ &&
+ \varpi (3) \Delta (1,2) (c_1 \tau_{13}(q_1) + c_2 \tau_{13}(q_2) )
\eea
In a manner completely analogous to the one used for
the original $F_1$, one easily shows that $\hat F_1$ is
a well-defined, single-valued holomorphic 1-form in each of
its arguments $z_1,z_2,z_3$.
Finally, we show that $\hat F_1$ vanishes by evaluating it
as $z_3 \to q_1$ and $z_3 \to z_1$. The latter is immediate,
while the former holds by
\bea
\lim _{z_3 \to q_1} \hat F_1
=
\varpi (1) \varpi (2) (\tau_{12}(q_1) - {c_2 \over c_1} \tau_{12}(q_2) )
+ \Delta (1,3) c_1 \p \varpi (q_1) =0
\eea
This completes the proof of the fact that $F_1$ is a single-valued,
holomorphic 1-form in $z_4$, which vanishes at $q_1$ and $z_1$,
and must therefore vanish identically, since the points $q_1$ and
$z_1$ are generic and independent. As a result, we have
$F_i=0$ for $i=1,2,3,4$ and we are left with the following simple result,
\bea
\I_{16}^S(1,2,3,4)
=2 \sum_{i=1}^4\rho_i\p\Lambda(z_i),
\eea
or, altogether after replacing $\rho_i$ by its
definition $\rho_i=\Z_0\prod_{j\not=i}\varpi(z_j)$,
\bea
\I_{15} ^S (1,2,3,4)   = -2 \I_{16}^S (1,2,3,4) =
  -4 \,\Z_0\sum _{i=1}^4  \p \Lambda(z_i)\,
  \prod_{j\not=i}\varpi(z_j) 
\eea
which, together with (\ref{kine}) yields (\ref{Y2psiS}).

\subsection{Cancellation of $\Y_{2S}^\psi$ and the
$\partial\Lambda(z)$ counterterm}

Recalling the formula (\ref{Y2psiS}), which was just established,
it follows immediately $\sum _\delta \Z[\delta] \Y_{2\psi}^S$ cancels the
counterterm proportional to $\partial\Lambda(z_i)$ in
the formula (\ref{Hsum}) for the holomorphic amplitude ${\cal H}$.
This is the second cancellation announced earlier
in (\ref{cancellation2}).

\newpage

\section{Recombining Mechanism of the $\varpi$, $\mu$, and $\Lambda$}
\setcounter{equation}{0}

We come now to the remaining three terms in the expression (\ref{Hsum}) 
for the holomorphic amplitude ${\cal H}$, whose origins are respectively 
the insertions of the two supercurrents from $\Y_1$ when the two 
fermions in the supercurrents are contracted with one another, 
the deformation of  complex structures resulting from the bosonic stress 
tensor $T_x$ in $\Y_{2x}$, and the $\Lambda(z)$ part of the counterterms 
in $\Y_5$ arising from completing the de~Rham $d$-exact differentials. The explicit expressions are as follows,
\bea
\Y_1 & \to &
- { \zeta ^1 \zeta ^2 \over 32 \pi^2} \Z_0 K
\left \< Q(p_I) \p x_+ ^\mu (q_1) \p x_+ ^\mu (q_2) \prod _l
e^{ik_l \cdot x_+ (z_l)} \right \> _{(c)} \prod _i \varpi (z_i)
\\
\Y _2 & \to &
{ 1 \over 8 \pi} \Z_0 \, K  \, \int _w \mu (w)
\left \< Q(p_I) \p x_+ ^\mu (w) \p x_+ ^\mu (w) \prod _l
e^{ik_l \cdot x_+ (z_l)} \right \> _{(c)} \prod _i \varpi (z_i)
\no \\
\Y_5 & \to &
- \half \Z_0 \, K  \sum _j \Lambda(z_j)  \,
\left \< Q(p_I) \, i k _j ^\mu  \p x_+ ^\mu (z_j ) \prod _l
e^{ik_l \cdot x_+ (z_l)} \right \> \prod _{i \not= j} \varpi (z_i)
\no
\eea
The first term depends on the gauge choice $\varpi(z)$, while the 
other two terms depend on both the gauge choices $\varpi(z)$
and $\muhat $. Note that the points $q_1$ and $q_2$
depend on $\varpi(z)$, while $\Lambda(z)$ depends on both
$\varpi(z)$ and $\muhat $. We shall show now how all these gauge 
choices cancel, upon summing the above three terms, leading to a 
gauge-independent formula for the holomorphic amplitude ${\cal H}$.

\subsection{Terms bilinear in $p_I$}

These arise in the above contributions $\Y_1$ and $\Y_{2x}$,
from the contractions of $\p x_+$ with $Q(p_I)$. We use the contraction
formula
\bea
Q(p_I) \, \p x_+ ^\mu (z) = 2 \pi p_I ^\mu \omega _I (z) Q(p_I)
\eea
and obtain the following contributions,
\bea
\Y_1 & \to &
- { \zeta ^1 \zeta ^2 \over 32 \pi^2} \Z_0 \, K 
4 \pi ^2 p_I ^\mu p_J ^\mu \omega _I (q_1) \omega _J (q_2)
\left \< Q(p_I)  \prod _l  e^{ik_l \cdot x_+ (z_l)} \right \> \prod _i \varpi
(z_i)
\\
\Y _{2x} & \to &
{ 1 \over 8 \pi} \Z_0 \, K   \, \int \mu (w)
4 \pi ^2 p_I ^\mu p_J ^\mu \omega _I (w) \omega _J (w)
\left \< Q(p_I)  \prod _l  e^{ik_l \cdot x_+ (z_l)} \right \>  \prod _i \varpi (z_i)
\no
\eea
Using the defining relation (\ref{mutildedef}) for $\mu$, the
sum of these two terms vanishes.

\subsection{Terms linear in $p_I$}

These terms arise from $\Y_1$, $\Y_{2x}$, and $\Y_5$ and have the
following factor in common,
\bea
\label{commonfactor}
\Z_0 \, K \left  \< Q(p_I)  \prod _l  e^{ik_l \cdot x_+ (z_l)} \right  \>
\eea
which we omit to write below. The terms are as follows,
\bea
\Y_1 & \to &
i { \zeta ^1 \zeta ^2 \over 16 \pi} \sum _j p_I \cdot k_j
\bigg ( \omega _I (q_1) \p_{q_2} \ln E(q_2,z_j)
+  \omega _I (q_2) \p_{q_1} \ln E(q_1,z_j) \bigg )  \prod _i \varpi (z_i)
\no \\
\Y _{2x} & \to &
- {i \over 2} \sum _j p_I \cdot k_j   \, \int  \mu (w)
\omega _I (w) \p _w \ln E(w,z_j) \prod _i \varpi (z_i)
\no \\
\Y_5 & \to &
- i \pi  \sum _j p_I \cdot k_j \Lambda(z_j)  \, \omega _I (z_j)
 \prod _{i \not= j} \varpi (z_i)
\eea
Note that the $w$-integral in $\Y_{2x}$ is over a single-valued
integrand in view of the summation over $k_j$ with $\sum _j k_j=0$.
On the other hand, the function $\Lambda(z_j)$ is defined only up to an
additive constant. 
For our present purposes, it is convenient to write it as
\bea
\Lambda(z) = \Lambda(w_0) - { 1 \over 2 \pi} \int  \mu (w) \varpi (w)
\p _w \ln {E(w,z) \over E(w,w_0)},
\eea
where $w_0$ is an auxiliary point. Under a
change of slice, $ \delta \mu (w) = \p _{\bar w} v(w)$, we have
\bea
\delta \Lambda (z) = v(z) \varpi (z) + \delta \Lambda(w_0) - v(w_0) \varpi (w_0)
\eea
To recover the naive transformation law, one would have to
require $\delta \Lambda(w_0) = v(w_0) \varpi (w_0)$.

\medskip

In $\Y_{2x}$, the integral of $\mu(w)$ is against a 2-form
which is single-valued in $w$, but which is not holomorphic,
since there are poles at the insertion points $z_j$. These
are cancelled by the contribution from $\Y_5$. Therefore,
it is advantageous to combine $\Y_{2x}$ and $\Y_5$. We shall
seek to carry out the integration over $w$ explicitly by
recasting the integrand as $\mu (w)$ times a single-valued
holomorphic 2-from in $w$. The starting point is
\bea
\Y _{2x} + \Y_5
\to
- {i \over 2} \int \mu (w) \phi _I (w;z_j,w_0)
-i \pi  \Lambda(w_0) \sum _j p_I \cdot k_j   \,  
\prod _{i \not= j} \varpi (z_i) \omega _I (z_j)
\eea
where
\bea
\phi _I (w;z_j,w_0) & =  & \sum _j p_I \cdot k_j \prod _{i \not= j} \varpi (z_i)
\bigg (
\varpi (z_j) \omega _I (w)  \p _w \ln E(w,z_j)
\no \\ && \hskip 1.2in
-  \omega _I(z_j)  \varpi (w) \p_w \ln {E(w,z_j) \over E(w,w_0)} \bigg )
\eea
The $w$-integral still depends on the slice choice of $\mu(w)$
since the single-valued 2-form $\phi _I (w;z_j,w_0)$ is not holomorphic,
but has a single simple pole at $w=w_0$, with residue
\bea
 \sum _j p_I \cdot k_j \prod _{i \not= j} \varpi (z_i)
\omega _I(z_j)  \varpi (w_0)
\eea
We define the Green function $G_2(w,w_0)$ to be of type $(2,0)$
in $w$ and $(-1,0)$ in $w_0$, with a single pole at $w=w_0$,
and unit residue. It satisfies,
\bea
\p_{\bar w} G_2 (w,w_0) = 2 \pi \delta (w,w_0)
\eea
This Green function is not unique since one may always add
a holomorphic 2 form in $w$ without changing the defining
equations, but these differences will be immaterial for our purposes.
We rearrange the above combination as follows,
\bea
\phi _I (w;z_j,w_0)
=
\phi _I ^{(h)} (w;z_j,w_0) +
G_2 (w,w_0) \varpi (w_0)
 \sum _j p_I \cdot k_j \omega _I(z_j)   \prod _{i \not= j} \varpi (z_i)
\eea
By construction,  $\phi _I ^{(h)} (w;z_j,w_0)$ is
a single-valued holomorphic 2-form in $w$,
\bea
\phi _I ^{(h)} (w;z_j,w_0)
& =  &
\sum _j p_I \cdot k_j \prod _{i \not= j} \varpi (z_i)
\bigg (
\varpi (z_j) \omega _I (w)  \p _w \ln E(w,z_j)
\no \\ && \hskip 1.3in
-  \omega _I(z_j)  \varpi (w) \p_w \ln {E(w,z_j) \over E(w,w_0)}
\no \\ && \hskip 1.3in
- \omega _I(z_j)    G_2 (w,w_0) \varpi (w_0)
\bigg )
\eea
Thus, its integral against $\mu$ is independent of the choice for $\mu$.
Once again, when this is the case, it can be correctly evaluated
using point insertions for $\mu$. A convenient choice is
\bea
\mu (w) =
{\zeta ^1 \zeta ^2 \over 8 \pi} {c_1 \over c_2} \delta (w,q_1) +
{\zeta ^1 \zeta ^2 \over 8 \pi} {c_2 \over c_1} \delta (w,q_2)
\eea
As a result, and using $c_1 \omega _I(q_1) = c_2 \omega _I(q_2)$, we have,
\bea
&&
\int  \mu (w) \phi _I ^{(h)} (w;z_j,w_0)
 \\ && \hskip .5in =
{\zeta ^1 \zeta ^2 \over 8 \pi} \sum _j p_I \cdot k_j \prod _{i \not= j} \varpi
(z_i)
\bigg \{
- \omega _I (z_j)
\left (  {c_1 \over c_2} G_2(q_1,w_0) + {c_2 \over c_1} G_2(q_2,w_0) \right )
\varpi (w_0)
 \no \\ && \hskip 1.2in
+\varpi (z_j) \omega _I(q_1) \p_{q_2} \ln E(q_2,z_j)
+ \varpi (z_j) \omega _I(q_2) \p_{q_1} \ln E(q_1,z_j)
\bigg \}
\no
\eea
The combination in the last line above is cancelled by $\Y_1$.
Thus we are left with
\bea
\Y_1 + \Y _{2x} + \Y_5
& \rightarrow & \!
{i \over 2} \sum _j p_I \cdot k_j   \,  \prod _{i \not= j} \varpi (z_i) \omega
_I (z_j)
\bigg ( {\cal C} (w_0,q_1,q_2) - \int  \mu (w) G_2 (w,w_0) \varpi (w_0)
\bigg )
\no \\
{\cal C} (w_0,q_1,q_2)
& = &
- 2 \pi \Lambda(w_0)  +
{\zeta ^1 \zeta ^2 \over 8 \pi}
\left (  {c_1 \over c_2} G_2(q_1,w_0) + {c_2 \over c_1} G_2(q_2,w_0) \right )
\varpi (w_0)
\no
\eea
Now the choice of $\Lambda(w_0)$ is at our disposal,
and the simplest procedure is to choose it to cancel $\C$, namely,
\bea
\label{fchoice}
 2 \pi { \Lambda(w_0) \over \varpi(w_0)} & = &
{\zeta ^1 \zeta ^2 \over 8 \pi}
\left (  {c_1 \over c_2} G_2(q_1,w_0) + {c_2 \over c_1} G_2(q_2,w_0) \right )
- \int \mu (w) G_2 (w,w_0)
\eea
Notice that this choice yields the simple transformation law 
$\delta \Lambda (w_0)= v (w_0) \varpi (w_0)$.
With it, the terms linear in $p_I$ cancel.
(Alternatively, $\Lambda(w_0)$ can be kept arbitrary.
In that case, it produces a de~Rham $d$-exact differential, when combined with 
similar contributions from the terms of order 0 in the internal momenta 
$p_I$ to be derived later. This de~Rham $d$-exact differential can be absorbed
in the other de~Rham $d$-exact differentials
in the expression (\ref{Dolbeault}) for the full chiral amplitude $\sum_\delta
{\cal B}[\delta]$.)

\subsection{Terms independent of $p_I$}

We again omit the common factor (\ref{commonfactor}),
and denote the corresponding equalities by $\to$.
The contractions then yield the following expressions,
\bea
\Y_1 & \to &
{ \zeta ^1 \zeta ^2 \over 32 \pi ^2} \prod _l \varpi (z_l)
\sum _{i,j} k_i \cdot k_j \,
\p _{q_1} \ln E(q_1,z_i) \, \p _{q_2} \ln E(q_2,z_i)
\no \\
\Y_{2x} & \to &
- {1 \over 8 \pi} \prod _l \varpi (z_l)
\sum _{i,j} k_i \cdot k_j \, \int  \mu (w)
\p _w \ln E(w,z_i) \, \p _w \ln E(w,z_i)
\no \\
\Y_5 & \to &
- \half \sum _{i,j} k_i \cdot k_j \, \prod _{l\not= j} \varpi (z_l)
\Lambda(z_j)  \p _{z_j} \ln E(z_j,z_i)
\eea
The combination $\Y_2 + \Y_5$ is gauge slice-independent.
We regroup terms in terms of the 2-form in $w$ occurring in the
integral over $w$,
\bea
\phi (w;z_i,w_0) & = &
- {1 \over 8 \pi} \prod _l \varpi (z_l)
\sum _{i,j} k_i \cdot k_j \,
\p _w \ln E(w,z_i) \, \p _w \ln E(w,z_i)
\\ &&
+ {1 \over 4 \pi} \sum _{i,j} k_i \cdot k_j \,  \prod _{l\not= j} \varpi (z_l)
\p _{z_j} \ln E(z_j ,z_i) \, \varpi (w) \p _w \ln {E(w,z_i) \over E(w,w_0)}
\no
\eea
Because of the conservation of momenta $\sum _i k_i=0$, $\phi$
is a well-defined, single-valued  2-form in $w$. The poles
at $w=z_i$ cancel between the first and the second terms,
and only a single pole remains at $w=w_0$ with residue
\bea
- {1 \over 4 \pi} \sum _{i,j} k_i \cdot k_j \,  \prod _{l\not= j} \varpi (z_l)
\p _{z_j} \ln E(z_j ,z_i) \, \varpi (w_0)
\eea
In the same spirit as the manipulations for the terms linear
in $p_I$, we introduce the holomorphic 2-form,
\bea
\label{phihol}
\phi ^{(h)} (w;z_i,w_0)
=
 \phi (w;z_i,w_0)
+ {1 \over 4 \pi} G_2 (w,w_0) \varpi (w_0) \prod _{l\not= j} \varpi (z_l)
 \sum _{i,j} k_i \cdot k_j \,   \p _{z_j} \ln E(z_j ,z_i)
\no \\
\eea
Hence, we have
\bea
\Y_{2x} + \Y_5 & \sim &
\int  \mu (w) \phi ^{(h)} (w;z_i,w_0)
\\ &&
-  \bigg ( \half  \Lambda(w_0)
+ {1 \over 4 \pi} \varpi (w_0) \int  \mu (w) G_2 (w,w_0) \bigg )
 \sum _{i,j} k_i \cdot k_j \,  \prod _{l\not= j} \varpi (z_l)
\p _{z_j} \ln E(z_j ,z_i)
\no
\eea
Making again the choice (\ref{fchoice}) as we did for the terms
linear in $p_I$, the last term simplifies, and we have
\bea
\label{y2y5}
\Y_{2x} + \Y_5 & \to &
\int  \mu (w) \phi ^{(h)} (w;z_i,w_0)
-  {\zeta ^1 \zeta ^2 \over 32 \pi^2}
\bigg (  {c_1 \over c_2} G_2(q_1,w_0) 
\\ && \hskip .6in
+ {c_2 \over c_1} G_2(q_2,w_0) \bigg )
\varpi (w_0)
 \sum _{i,j} k_i \cdot k_j \,  \prod _{l\not= j} \varpi (z_l)
\p _{z_j} \ln E(z_j ,z_i)
\no
\eea
Since $\phi ^{(h)} $ is well-defined, single-valued and holomorphic in
$w$, we may choose $\mu$ as we did earlier, and compute
the $w$-integral explicitly. The term involving $G_2$ in (\ref{phihol})
cancels the entire last term in (\ref{y2y5}), and we are left with
\bea
\Y_{2x} + \Y_5
& \to &
- {\zeta ^1 \zeta ^2 \over 64 \pi^2}
 \prod _l \varpi (z_l)
\sum _{i,j} k_i \cdot k_j \,
\bigg (
{c_1 \over c_2} \p _{q_1} \ln E(q_1,z_i) \, \p _{q_1} \ln E(q_1,z_i)
\no \\ && \hskip 1.8in
+
{c_2 \over c_1} \p _{q_2} \ln E(q_2,z_i) \, \p _{q_2} \ln E(q_2,z_i)
\bigg ) \quad
\eea
This result nicely combines with $\Y_1$, and may be expressed as
follows,
\bea
\Y_1 + \Y_{2x} + \Y_5
& \to &
- {\zeta ^1 \zeta ^2 \over 64 \pi^2}  \prod _l \varpi (z_l)
\sum _{i,j} k_i \cdot k_j \, {1 \over c_1 c_2}
\\ && \hskip .5in \times
\prod _{\ell =i,j}
\bigg (  c_1  \p _{q_1} \ln E(q_1,z_\ell)  - c_2 \p _{q_2} \ln E(q_2,z_\ell)
\bigg )
\no
\eea
In view of the sum $\sum _i k_i=0$, we may let
\bea
\p _{q_\alpha} \ln E(q_\alpha ,z_\ell)
\to
\p _{q_\alpha } \ln {E(q_\alpha,z_\ell) \over E(q_\alpha,w_1)}
\qquad \alpha =1,2
\eea
for an arbitrary point $w_1$. We then use the relation,
\bea
c_1 \p_{q_1} \ln {E(q_1,z_\ell) \over E(q_1,w_1)}
-
c_2 \p_{q_2} \ln {E(q_2,z_\ell) \over E(q_2,w_1)}
=
- c_1 ^2 \p \varpi (q_1) { \Delta (z_\ell, w_1) \over \varpi (z_\ell) \varpi
(w_1)}
\eea
and, using $- c_1 ^2 \p \varpi (q_1)= + c_2 ^2 \p \varpi (q_2)$, we get
\bea
\Y_1 + \Y_{2x} + \Y_5
\to 
{\zeta ^1 \zeta ^2 \over 64 \pi^2}  \, c_1 c_2 \p \varpi (q_1) \p \varpi (q_2)
\, \Y_S
\eea
with 
\bea
\label{YSint}
\Y _S =
\prod _l \varpi (z_l) \sum _{i,j} k_i \cdot k_j
{ \Delta (z_i, w_1) \Delta (z_j, w_1)
 \over
 \varpi (z_i)  \varpi (z_j)  \varpi (w_1)^2} \qquad
\eea
Of course, this entire combination is in fact independent of $w_1$.
Restoring the factor (\ref{commonfactor}) and using (\ref{qdependence}),
we recover the formula of (\ref{explicitH}) and (\ref{H}).

\subsubsection{Alternative formulas for $\Y_S$}

We may evaluate $\Y_S$ in (\ref{YSint})  more explicitly in terms of the 
Mandelstam variables,
\bea
s & = & - (k_1+k_2)^2 = - 2 k_1 \cdot k_2 = - 2 k_3 \cdot k_4
\no \\
t & = & - (k_2+k_3)^2 = - 2 k_2 \cdot k_3 = - 2 k_1 \cdot k_4
\no \\
u & = & - (k_1+k_3)^2 = - 2 k_1 \cdot k_3 = - 2 k_2 \cdot k_4
\eea
Since $\Y_S$ in (\ref{YSint}) is independent of $w_1$ we may set $w_1=z_4$.
Using the following simple identity,
\bea
\varpi (3) \Delta (2,4) - \varpi (2) \Delta (3,4) = \varpi (4) \Delta (2,3)
\eea
and permutations thereof, as well as the relation $s+t+u=0$, we find
\bea
3 \Y_S & =  &
(t-u) \Delta (1,2) \Delta (3,4) +
(s-t) \Delta (1,3) \Delta (4,2) +
(u-s) \Delta (1,4) \Delta (2,3)
\no \\
& = &
+ (k_1-k_2) \cdot (k_3-k_4) \Delta (1,2) \Delta (3,4)
\no \\ &&
+ (k_1-k_3) \cdot (k_2-k_4) \Delta (1,3) \Delta (2,4)
\no \\ &&
+ (k_1-k_4) \cdot (k_2-k_3) \Delta (1,4) \Delta (2,3)
\eea
which are manifestly totally symmetric under the interchange of
any points. One may prefer the non-manifestly symmetric form,
\bea
\Y_S = - s \Delta (1,4) \Delta (2,3) + t \Delta (1,2) \Delta (3,4)
\eea
Thus we have obtained the desired formula
(\ref{cancellation3}).

\medskip
The derivation of the formulas (\ref{H}) and
(\ref{explicitH}) for the gauge-fixed chiral superstring amplitude is now
complete.

\newpage

\section{The 4-Point Function for the Heterotic String}
\setcounter{equation}{0}

We shall use a formulation of the Heterotic string in terms of internal chiral 
worldsheet fermions $\lambda ^I (z)$ for $I=1,\cdots ,32$ \cite{Gross:1985fr}.
For the case of $HO=Spin(32)/{\bf Z}_2$, all 32 fermions have the same spin structure $\kappa$, while for $HE= E_8 \times E_8$, the 32 fermions are 
split into two groups of 16, with the same spin
structures $\kappa _1$ and $\kappa _2$ within each group.

\subsection{Correlators of Internal Fermions}

For both $HO$ and $HE$, the gauge currents are 
\bea
j^a (z) = \half T^a _{IJ} \lambda ^I \lambda ^J (z)
\eea
where the $T^a _{IJ}$ are representation matrices of 
$SO(32)$ for $HO$ and of the $SO(16) \times SO(16)$
subgroup for $HE$. (Spin fields of $SO(16) \times SO(16)$
will be required to represent those roots of $E_8 \times E_8$
that do not lie within $SO(16) \times SO(16)$,
but we shall not need them here.)
The correlator required for the $R^2 F^2$ contributions is
\bea
\< j^{a_1} (z_1) j^{a_2} (z_2) \> _\kappa 
=
\half \tr (T^{a_1} T^{a_2}) S_\kappa (z_1,z_2)^2
\eea
The one required for the $F^4$  terms is more complicated and 
will depend upon whether we consider the $HO$ or $HE$ theory.
The first case is when all 4 external gauge particles 
lie within the same gauge group, which is always true
for the $HO$ theory. All fermions then have the same 
spin structure $\kappa$, and the correlator is given as follows,
\bea
\< \prod _{i=1}^4 j^{a_i} (z_i)  \> _\kappa 
& = &
+ {1 \over 4} \tr (T^{a_1} T^{a_2}) \tr (T^{a_3} T^{a_4})
    S_\kappa (z_1,z_2)^2 S_\kappa (z_3,z_4)^2
 \\ &&
+ {1 \over 4} \tr (T^{a_1} T^{a_3}) \tr (T^{a_2} T^{a_4})
    S_\kappa (z_1,z_3)^2 S_\kappa (z_2,z_4)^2
\no \\ &&
+ {1 \over 4} \tr (T^{a_1} T^{a_4}) \tr (T^{a_2} T^{a_3})
    S_\kappa (z_1,z_4)^2 S_\kappa (z_2,z_3)^2
\no \\ &&
- \tr (T^{a_1} T^{a_2} T^{a_3} T^{a_4}) 
    S_\kappa (z_1, z_2) S_\kappa (z_2,z_3)
    S_\kappa (z_3, z_4) S_\kappa (z_4,z_1)
\no \\ &&
- \tr (T^{a_1} T^{a_2} T^{a_4} T^{a_3}) 
    S_\kappa (z_1, z_2) S_\kappa (z_2,z_4)
    S_\kappa (z_4, z_3) S_\kappa (z_4,z_1)
\no \\ &&
- \tr (T^{a_1} T^{a_3} T^{a_2} T^{a_4}) 
    S_\kappa (z_1, z_3) S_\kappa (z_3,z_2)
    S_\kappa (z_2, z_4) S_\kappa (z_4,z_1)
\no
\eea
The second case occurs only for the $HE$ theory, when  two
external states (say 1,2) belong to the first $E_8$,
while the other two (3,4) belong to the second $E_8$.
The first group of fermions has spin structure $\kappa_1$, while 
the second will have $\kappa_2$, which are independent
of one another. Since mixed traces then vanish, we are left with 
\bea
\< \prod _{i=1}^4 j^{a_i} (z_i)  \> _{\kappa_1, \kappa _2} 
=
+ {1 \over 4} \tr (T^{a_1} T^{a_2}) \tr (T^{a_3} T^{a_4})
    S_{\kappa_1} (z_1,z_2)^2 S_{\kappa_2} (z_3,z_4)^2
\eea
The spin structure summations involve, for $k=1,2$, the following expressions,
\bea
\Psi _{4k} & = & \sum _\kappa \tet [\kappa](0)^{8k}
 \\
F_{4k} ^{(2)} (z_1,z_2) 
& = & 
\sum _\kappa \tet [\kappa ](0)^{8k} 
    S_\kappa (z_1,z_2)^2
\no \\
F_{4k} ^{(2,2)} (z_1,z_2;z_3,z_4) 
& = & 
\sum _\kappa \tet [\kappa ](0)^{8k} 
    S_\kappa (z_1,z_2)^2 S_\kappa (z_3,z_4)^2
\no \\
F_{4k} ^{(4)} (z_1,z_2,z_3,z_4) 
& = & 
\sum _\kappa \tet [\kappa ](0)^{8k} 
    S_\kappa (z_1,z_2) S_\kappa (z_2,z_3) 
    S_\kappa (z_3,z_4) S_\kappa (z_4,z_1)
\no
\eea
We recognize $\Psi_{4k}$ as the familiar modular forms 
of weight $4k$.  The quantity $F^{(2)}_{4k}$ is easily computed 
using the Fay identity,
\bea
S_\kappa (z,w)^2 = \p_z \p_w \ln E(z,w) + \omega _I (z) \omega _J (w)
\p_I \p_J \tet[\kappa ](0) /\tet [\kappa ](0)
\eea
Using the heat equation for $\tet$, we have $\p_I \p_J \tet[\kappa ](0)
= 4 \pi i \p_{IJ} \tet [\kappa](0)$, where $\p_{IJ}$ denotes the 
symmetrized derivative with respect to $\Omega _{IJ}$.
As a result, we have 
\bea
F_{4k} ^{(2)} (z_1,z_2) =
\Psi _{4k}  \p_z \p_w \ln E(z_1,z_2) + 
{4 \pi i \over 8k } \omega _I (z_1) \omega _J (z_2) \p_{IJ} \Psi _{4k}.
\eea

\subsection{The $F^4$ and $F^2 F^2$ Amplitudes}

The partition functions for the internal
fermions in the $HO$ and $HE$ theories are 
given respectively by 
\bea
\Z_{HO} =\tet [\kappa ](0)^{16}
\hskip 1in
\Z_{HE} =\tet [\kappa_1 ](0)^8 \tet [\kappa _2](0)^8.
\eea
The correlators of the internal fermions are derived
from the general formulas of the previous section.
The spin structure summed gauge current correlator are denoted by ${\cal W}$. 

\medskip

For the $O(32)$ Heterotic string,
where all 4 external particles are in the same gauge group,
and the chiral fermions have the same spin structure $\kappa$,
we have
\bea
\W_{(F^4)}^{HO} (z_1,z_2,z_3,z_4) & = & 
\sum _{ \kappa } \tet [\kappa ](0)^{16}
~
\< \prod _{i=1}^4 j^{a_i} (z_i)  \> _{ \kappa _1, \kappa_2 }
\eea
and hence, in terms of the functions $F_{4k}^{(2,2)}$ and
$F_{4k}^{(4)}$,
\bea
\label{WF4HO}
\W_{(F^4)}^{HO} (z_1,z_2,z_3,z_4)
& = &
+ {1 \over 4}\ \bigg\{\tr (T^{a_1} T^{a_2}) \tr (T^{a_3} T^{a_4}) \, 
    \, F^{(2,2)} _{8} (z_1,z_2;z_3,z_4)
    \no\\
    &&
    \qquad\quad
    +\, {\rm cyclic\ permutations\ of \ (234) }
    \ \bigg\}
\no \\ 
&&
- \ \bigg\{\tr (T^{a_1} T^{a_2} T^{a_3} T^{a_4}) \,
     \, F^{(4)} _{8} (z_1,z_2,z_3,z_4)
     \no\\
     &&
     \qquad\quad
    +(3\leftrightarrow 4)
    +(2\leftrightarrow 3)\bigg\}.
\eea
For the $E_8\times E_8$ Heterotic string, we have to consider two
possible cases ${\cal W}_{F^4}^{HE}(z_1,z_2,z_3,z_4)$ and
${\cal W}_{(F^2F^2)}^{HE}(z_1,z_2|z_3,z_4)$, corresponding respectively 
to the case where all 4 gauge particles lie in the same $E_8$, and the case 
where the gauge particles 1,2 lie in the first $E_8$, and the gauge 
particles 3,4 lie in the second.
In the first case, we have
\bea
\W_{(F^4)}^{HE}(z_1,z_2,z_3,z_4)
&=&
\sum_{\kappa_1,\kappa_2}\tet[\kappa_1](0)^8\tet[\kappa_2](0)^8
\<\prod_{i=1}^4j^{a_i}(z_i)\>_{\kappa_1}
\eea
resulting into
\bea
\label{WF4HE1}
\W_{(F^4)}^{HE} (z_1,z_2,z_3,z_4)
& = &
+ {1 \over 4}\ \bigg\{\tr (T^{a_1} T^{a_2}) \tr (T^{a_3} T^{a_4}) \, 
    \Psi_4\, F^{(2,2)} _{4} (z_1,z_2;z_3,z_4)
    \no\\
    &&
    \qquad\quad
    +\, {\rm cyclic\ permutations\ of \ (234) }
    \ \bigg\}
\no \\ 
&&
- \ \bigg\{\tr (T^{a_1} T^{a_2} T^{a_3} T^{a_4}) \,
     \Psi_4\, F^{(4)} _{4} (z_1,z_2,z_3,z_4)
     \no\\
     &&
     \qquad\quad
    +(3\leftrightarrow 4)
    +(2\leftrightarrow 3)\bigg\}.
\eea
In the second case, we have
\bea
\W_{(F^2F^2)}^{HE}(z_1,z_2|z_3,z_4)
=
\sum_{\kappa_1,\kappa_2}\tet[\kappa_1](0)^8\tet[\kappa_2](0)^8
\<\prod_{i=1}^4j^{a_i}(z_i)\>_{\kappa_1,\kappa_2},
\eea
resulting into
\bea
\label{WF4HE2}
\W_{(F^2F^2)}^{HE}(z_1,z_2|z_3,z_4)
=
{1\over 4}\,
tr(T^{a_1}T^{a_2})\,tr(T^{a_3}T^{a_4})
\
F_4^{(2)}(z_1,z_2)\, F_4^{(2)}(z_3,z_4).
\eea

\medskip

Now the contribution of the chiral half of the bosonic string
in 10 dimensions for the exponential part
$\prod_{i=1}^4e^{ik_i\cdot x(z_i)}$ of 
4 massless bosons of momenta $k_i$,
at fixed internal momentum $p_I^\mu$, is given by\footnote{
In assembling left and right chiral halfs, we shall follow the standard 
convention for Heterotic string nomenclature,  and identify 
{\sl left-movers  =  holomorphic = bosonic  string chiral half},   and 
{\sl right-movers  =  anti-holomorphic = superstring chiral half}.}
\bea
\label{bosonichalf}
\Z_{BOS}(z_i,k_i;p_I^\mu)
=
{\prod_{I\leq J}d\Omega_{IJ}
\over
\pi^{12}\Psi_{10}(\Omega)}
\exp \bigg \{ i\pi p_I^\mu\Omega_{IJ}p_J^\mu
+
ip_I^\mu\sum_{i=1}^4 k_i\int^{z_i}\omega_I \bigg \}
\prod_{i<j}E(z_i,z_j)^{k_i\cdot k_j}
\quad
\eea
With this convention, we obtain the Heterotic $F^4$ and $F^2F^2$ amplitudes by
combining (\ref{bosonichalf}) with the contributions of the internal fermions
as described in (\ref{WF4HO}-\ref{WF4HE2}), matching with the 
anti-holomorphic contribution from the  Type II superstrings at the 
same internal momenta $p_I^\mu$, and integrating out the $p_I^\mu$.
The net effect of integrating out the $p_I^\mu$ is to shift
the multi-valued expression $\ln\,|E(z_i,z_j)|^2$ to the single-valued expression $G(z_i,z_j)$ defined in (\ref{scalarprop}). The final expression for the scattering
of 4 gauge bosons is then given by the formula
(\ref{bfAF4}) quoted in the Introduction.

\subsection{The $R^2 F^2$ amplitude}

Next, we consider the scattering of two gravitons and two gauge bosons. Again, for the scattering of two gauge bosons,
we have two formulas for the correlator $\W_{(F^2)}(z_1,z_2)$
of internal fermions,
depending on whether we consider the $Spin(32)/{\bf Z}_2$ theory or the 
$E_8\times E_8$ theory. They are given respectively by,
\bea
\label{WF2HO}
\W_{(F^2)}^{HO} (z_1,z_2) 
& = & 
+ \half \tr (T^{a_1} T^{a_2}) 
    \, F^{(2)} _{8} (z_1,z_2)
\eea
and by
\bea
\label{WF2HE}
\W_{(F^2)}^{HE} (z_1,z_2) 
& = & 
+ \half \tr (T^{a_1} T^{a_2}) 
    \Psi _4 \, F^{(2)} _{4} (z_1,z_2).
\eea

\medskip

Putting together left and right moving parts of the Heterotic 
amplitude at fixed moduli and vertex insertion points, 
but integrating over the internal momenta, we have
\bea
\A_4 (1,2,3,4) & = &
\W_{(F^2)} (3,4) \, \overline{\Y_S (1,2,3,4)} 
\prod _{i<j} |E(i,j) | ^{2 k_i \cdot k_j} 
\\ && 
\times  \int d^{20} p \, \W _{(R^2)} ^{(p)} (1,2) \,  
\bigg|\exp \big \{ i \pi p_I\Omega_{IJ} p_J 
- 2i \pi p_I \sum _i k_i  \int ^{z_i} \! \! \omega _I \big \}\bigg|^2. 
\no
\eea
Here, we have abbreviated the insertion points by their label,
$i \equiv z_i$. The graviton part of the bosonic left moving side is 
given by
\bea
\W_{(R^2)} ^{(p)} (1,2) =
- \epsilon _1 ^\mu  \epsilon _2 ^\mu \p_1 \p_2 \ln E(1,2)
+ \prod _{\ell =1,2}
\bigg ( 2 \pi \epsilon _\ell ^\mu p_I ^\mu  \omega _I (\ell) 
- i \sum _i \epsilon _\ell ^\mu k_i ^\mu  \p_\ell \ln E(\ell,i) \bigg ) \quad
\eea
To carry out the integral over $p$, we first introduce the shifted 
variable $\tilde p$,
\bea
\label{ptilde}
\tilde p_I ^\mu = 
p_I ^\mu + 
\{ (\Im \Omega) ^{-1}\} _{IJ} \sum _i k^\mu _i \Im \int ^{z_i} \omega _J
\eea
and use the familiar Gaussian integral formula,
\bea
\int \! d^{20} \tilde p \, 
\exp \bigg \{ - 2\pi \tilde p_I ^\mu \tilde p_J ^\mu \Im \Omega _{IJ} 
+  b^\mu _I \tilde p^\mu _I \bigg \} 
= {1 \over [\det (2 \Im \Omega )]^5 }
\exp \left \{ {b^\mu _I b^\mu _J \over 8 \pi}   \{ (\Im \Omega) ^{-1}\} _{IJ}  \right \}
\quad
\eea
In terms of the single-valued scalar Green function
$G(z,w)$,
the above integral may be re-expressed as follows,
\bea
 \int d^{20} p \, \W _{(R^2)} ^{(p)} (1,2) \,  
\exp \bigg \{- 2 \pi p_I p_J \Im \Omega _{IJ} 
- 4 \pi p_I \sum _i k_i \Im \int ^i \! \! \omega _I \bigg \} 
\prod _{i<j} |E(i,j) | ^{2 k_i \cdot k_j} &&
\no \\
 =
({\rm det}\,{\rm Im}\,\Omega)^{-5}
\,\W_{(R^2)} (1,2)  \prod _{i<j} e ^{-k_i\cdot k_j G(i,j)} \quad && \quad
\eea
where the Heterotic graviton part is defined by
\bea
\label{WR2}
\W_{(R^2)} (1,2) = 
  \epsilon _1 ^\mu  \epsilon _2 ^\mu  \p_1 \p_2 G (1,2)
-  \sum _{i,j}  \epsilon _1 ^\mu k_i ^\mu   \epsilon _2 ^\nu k_j ^\nu
\p_1 G(1, i)  \p_2 G(2, j) 
\eea
As a result, the full amplitude becomes,
\bea
\A_4 (1,2,3,4) =
({\rm det}\,{\rm Im}\,\Omega)^{-5}\,\W_{(R^2)} (1,2) \, \W_{(F^2)} (3,4) \, \overline{\Y_S (1,2,3,4)} \,
\prod _{i<j} e ^{-k_i\cdot k_j G(i,j)},
\eea
and hence we obtain the formula
(\ref{bfAR2F2}), upon substituting in the suitable expression
(\ref{WF2HO}) or (\ref{WF2HE}) for the corresponding Heterotic 
theory and integrating over insertion points and moduli.

\subsection{The $R^2F^2$ amplitude in terms of $f_i$}

It is possible to recast the amplitude in terms of the $f^{\mu \nu}_{1,2}$
tensors, which makes gauge invariance explicit, up to exact differential terms,
which may be omitted. To do this, we start from the definition, 
$f_i ^{\mu \nu} = \epsilon _i ^\mu k_i ^\nu -  \epsilon _i ^\nu k_i ^\mu $ 
for $i=1,2$ and use the contraction with any generic momentum $\ell ^\nu$
to extract $\epsilon _i$ up to a gauge transformation,
\bea
(k_i \cdot \ell) \epsilon _i ^\mu =
f_i ^{\mu \nu} \ell ^\nu + (\epsilon _i \cdot \ell) k_i ^\mu
\eea
Since the bosonic side of the Heterotic string is essentially half
of the bosonic string, one cannot expect to recast it all by itself
in terms of $f_i$ without introducing poles in the kinematic 
variables $s,t,u$. Physically, the reason that such poles are 
required finds its origin in the fact that 2- and 3-point functions 
in the purely bosonic string do have non-zero loop corrections.
The kinematic poles represent 1-particle reducible parts 
that correspond to such corrections.

\medskip

For the full Heterotic string, however, the superstring side produces
a factor linear in $s,t,u$ and these factors have the potential of 
canceling the poles from the bosonic side. We now show how
this works out. Since the $R^2F^2$ amplitude has a preferred
channel (here taken to be the $s$-channel), we shall 
work out the combinations $s \W_{(R^2)} (1,2)$ and 
$(t-u) \W_{(R^2)} (1,2)$, which suffice to reconstruct 
the full amplitude, using the conservation of momentum and 
the second form for $\Y_S$ in 
(\ref{Ystar}).

\subsubsection{$s\, \W_{(R^2)} (1,2)$ in terms of $f_i$}

From now on in this section, we change notations slightly and 
denote by the same symbol $\W_{(R^2)}(1,2)$
the chiral amplitude found in (\ref{WR2}) together with the factor
\bea
\label{calE}
{\cal E}={\rm exp}\bigg(-\sum_{i<j}k_i\cdot k_j\,G(z_i,z_j)\bigg).
\eea
This should cause no confusion, and it avoids a proliferation
of new symbols.

\medskip

The starting point for the derivation of a formula for $\W_{(R^2)}(1,2)$ in terms of $f_i$ is
\bea
s \W_{(R^2)}  (1,2) & = & 
-2 (k_1 \cdot  k_2 ) (  \epsilon _1 \cdot  \epsilon _2 )  \, \p_1 \p_2 G (1,2) \, \E
 \no \\ && 
+2   \sum _{i,j}  (k_1 \cdot k_2 ) (  \epsilon _1\cdot  k_i )   
( \epsilon _2 \cdot  k_j ) \,
\p_1 G(1, i)  \p_2 G(2, j) \, \E
\eea
We re-express the kinematic combination under the sum as follows,
\bea
2 (k_1 \cdot k_2 ) (  \epsilon _1\cdot  k_i )    ( \epsilon _2 \cdot  k_j )
& = & 
- 2 k_i ^\mu f_1 ^{\mu \nu} f_2 ^{\nu \rho} k_j ^\rho
+ (k_2 \cdot k_j) X^{(1)} _i + (k_1 \cdot k_i) X^{(2)}_j
\no \\
X^{(1)} _i  
& = & 
2 (\epsilon _1 \cdot k_i ) (\epsilon _2 \cdot k_1) - (k_1 \cdot k_i)
(\epsilon _1 \cdot \epsilon _2)
\no \\
X^{(2)} _j  
& = & 
2 (\epsilon _2 \cdot k_j ) (\epsilon _1 \cdot k_2) - (k_2 \cdot k_j)
(\epsilon _1 \cdot \epsilon _2)
\eea
The combination of the $X$-terms may be re-arranged by adding a
total derivative form. 
We then find,
\bea
\sum _{i,j} (k_1 \cdot k_i) X_j ^{(2)} \p_1 G(1,i) \p_2 G(2,j) \E
& = & 
X_1 ^{(2)} \p_1\p_2 G(1,2) \E
 - \p_1 \bigg ( \sum _j  X^{(2)}_j \p_2 G(2,j) \E \bigg )
\no \\
\sum _{i,j} (k_2 \cdot k_j) X_i ^{(1)} \p_1 G(1,i) \p_2 G(2,j) \E
& = & 
X_2 ^{(1)} \p_1\p_2 G(1,2) \E
- \p_2 \bigg ( \sum _i  X^{(1)}_i \p_1 G(1,i) \E \bigg ) 
\no
\eea
The coefficients of the double derivative terms combine as follows,
\bea
-2 (k_1 \cdot  k_2 ) (  \epsilon _1 \cdot  \epsilon _2 )  
+ X_2 ^{(1)} + X_1 ^{(2)}
= 2 (f_1f_2) = - 2 f_1 ^{\mu \nu } f_2 ^{\mu \nu}
\eea
so that, up to total derivative terms,  the full contribution takes the form,
\bea
s \W_{(R^2)}  = 
2 (f_1 f_2) \p_1 \p_2 G_s (1,2) \, \E
- 2 \sum _{i,j} ^{} k_i ^\mu f_1 ^{\mu \nu} f_2 ^{\nu \rho} k_j ^\rho
\p_1 G (1,i) \p_2 G (2,j) \, \E
\eea

\subsubsection{$(t-u)\, \W_{(R^2)} (1,2)$ in terms of $f_i$}

To begin, we derive the following product re-arrangements,
\bea
(t-u) \, \epsilon ^\mu _1 
& = &
+ 2 f_1 ^{\mu \nu} k_{34} ^\nu + 2 (\epsilon _1 \cdot k_{34}) k_1 ^\mu
\no \\
(t-u) \, \epsilon ^\mu _2 
& = &
- 2 f_2 ^{\mu \nu} k_{34} ^\nu - 2 (\epsilon _2 \cdot k_{34}) k_2 ^\mu
\eea
using the notation 
\bea
\label{kij}
k_{ij} ^\mu = k_i ^\mu - k_j ^\mu.
\eea
These formulas are the starting point of our rearrangements,
namely replacing  in each term half  by rearranging $\epsilon _1$ 
and the other half by rearranging $\epsilon _2$ according to the above
formulas, we get
\bea
(t-u) \W_{(R^2)} 
& = &
\bigg ( 
\epsilon _2 ^\mu f_1 ^{\mu \nu} k_{34} ^\nu
- \epsilon _1 ^\mu f_2 ^{\mu \nu} k_{34} ^\nu
+ (\epsilon _1 \cdot k_{34}) (\epsilon _2 \cdot k_1 ) 
 \\ && \hskip 1.55in
- (\epsilon _2 \cdot k_{34}) (\epsilon _1 \cdot k_2 ) \bigg ) \p_1 \p_2 G(1,2) \, \E
\no \\ &&
- \sum _{i,j} \bigg \{ 
k_i ^\mu f_1 ^{\mu \nu} k_{34} ^\nu + (\epsilon _1 \cdot k_{34}) (k_1 \cdot k_i) 
\bigg \} (\epsilon _2 \cdot k_j) \p_1 G(1,i) \p_2 G(2,j) \, \E
\no \\ &&
+ \sum _{i,j} \bigg \{ 
k_j ^\mu f_2 ^{\mu \nu} k_{34} ^\nu + (\epsilon _2 \cdot k_{34}) (k_2 \cdot k_j) 
\bigg \} (\epsilon _1 \cdot k_i) \p_1 G(1,i) \p_2 G(2,j) \, \E \quad
\no
\eea
Using the derivative formulas,
\bea
- \sum _{i,j}   (k_1 \cdot k_i) 
 (\epsilon _2 \cdot k_j) \p_1 G(1,i) \p_2 G(2,j) \, \E
& = & 
\p_1 \bigg ( \sum _j  (\epsilon _2 \cdot k_j) \p_2 G(2,j) \, \E \bigg )
\no \\ &&
- (\epsilon _2 \cdot k_1) \p_1 \p_2 G(1,2) \, \E
\no \\
+ \sum _{i,j}   (k_2 \cdot k_j) 
 (\epsilon _1 \cdot k_i) \p_1 G(1,i) \p_2 G(2,j) \, \E
& = & 
- \p_2 \bigg ( \sum _i  (\epsilon _1 \cdot k_i) \p_1 G(1,i) \, \E \bigg )
\no \\ &&
+ (\epsilon _1 \cdot k_2) \p_1 \p_2 G(1,2) \, \E
\eea
and neglecting the total derivative terms, we have 
\bea
(t-u) \W_{(R^2)} 
& = &
\bigg ( 
\epsilon _2 ^\mu f_1 ^{\mu \nu} k_{34} ^\nu
- \epsilon _1 ^\mu f_2 ^{\mu \nu} k_{34} ^\nu  \bigg ) \p_1 \p_2 G(1,2) \, \E
\no \\ &&
- \sum _{i,j}  
( k_i ^\mu f_1 ^{\mu \nu} k_{34} ^\nu)  (\epsilon _2 \cdot k_j) 
\p_1 G(1,i) \p_2 G(2,j) \, \E
\no \\ &&
+ \sum _{i,j} 
( k_j ^\mu f_2 ^{\mu \nu} k_{34} ^\nu) (\epsilon _1 \cdot k_i) 
\p_1 G(1,i) \p_2 G(2,j) \, \E \quad
\eea
In this intermediate formula  half of the $\epsilon$'s
were replaced by $f$'s but half were not.

\medskip

We now make a key observation which allows us to also recast
the remaining $\epsilon$'s in terms of $f$'s without introducing 
kinematic poles. The observation is that
\bea
k_i ^\mu f_1 ^{\mu \nu} k_{34} ^\nu & = & k_2 \cdot p_i
\no \\
k_j ^\mu f_2 ^{\mu \nu} k_{34} ^\nu & = & k_1 \cdot q_j
\eea
for all $i,j$ and where $p_i$ and $q_j$ are polynomial
in the momenta. Of course, the $p_i$ and $q_j$, as defined
by the above equations are not unique, since shifts of 
$p_i$ by $k_2$ and $\epsilon_2$ (and $q_j$ by $k_1$ and $\epsilon _1$)
leave the relation invariant. But, any choice will do. 
Since $k_{34}^\mu f_1^{\mu\nu}k_{34}^\nu=0$ by the oddness
of $f^{\mu\nu}_1$, we have
\bea
k_3^\mu f_1^{\mu\nu}k_{34}^\nu
=
k_4^\mu f_1^{\mu\nu}k_{34}^\nu
=
\half
(k_3+k_4)^\mu f_1^{\mu\nu}k_{34}^\nu
=
-
\half
(k_1+k_2)^\mu f_1^{\mu\nu}k_{34}^\nu
=
-
\half
k_2^\mu f_1^{\mu\nu}k_{34}^\nu,
\nonumber
\eea
and hence we can take
\bea
p_1 ^\mu =0 &&
\quad p_2^\mu = f_1 ^{\mu \nu } k_{34} ^\nu
\qquad p_3 ^\mu = p_4 ^\mu = - \half f_1 ^{\mu \nu} k_{34} ^\nu
\no \\
q_2 ^\mu = 0 &&
\quad q_1 ^\mu = f_2 ^{\mu \nu} k_{34} ^\nu
\qquad q_3 ^\mu = q_4 ^\mu = - \half f_2 ^{\mu \nu} k_{34} ^\nu.
\eea
We now use the following  rearrangement,
\bea
( k_i ^\mu f_1 ^{\mu \nu} k_{34} ^\nu)  (\epsilon _2 \cdot k_j) 
& = & 
(k_2 \cdot p_i) (\epsilon _2 \cdot k_j) 
=
k_j ^\mu f_2 ^{\mu \nu} p_i ^\nu + (\epsilon _2 \cdot p_i) (k_2 \cdot k_j)
\no \\
( k_j ^\mu f_2 ^{\mu \nu} k_{34} ^\nu) (\epsilon _1 \cdot k_i) 
& = &
(k_1 \cdot q_j) (\epsilon _1 \cdot k_j)
=
k_i ^\mu f_1 ^{\mu \nu} q_j ^\nu + (\epsilon _1 \cdot q_j) (k_1 \cdot k_i)
\eea
Finally, using the following total derivative formulas,
\bea
- \sum _{i,j} (\epsilon _2 \cdot p_i) (k_2 \cdot k_j) \p_1 G(1,i) \p_2 G (2,j) \, \E
& = & 
\p_2 \bigg ( \sum _i (\epsilon_2 \cdot p_i) \p_1 G(1,i) \, \E \bigg )
\no \\ &&
- (\epsilon_2 \cdot p_2) \p_1 \p_2G(1,2) \, \E 
\no \\
+ \sum _{i,j} (\epsilon _1 \cdot q_j) (k_1 \cdot k_i) \p_1 G(1,i) \p_2 G(2,j) \, \E
& = & 
- \p_1 \bigg ( \sum _j (\epsilon_1 \cdot q_j) \p_2 G(2,j) \, \E \bigg )
\no \\ &&
+ (\epsilon_1 \cdot q_1) \p_1 \p_2G(1,2) \, \E 
\eea
Combining first the coefficients of the double derivative terms, we have
\bea
\epsilon _2 ^\mu f_1 ^{\mu \nu} k_{34} ^\nu
- \epsilon _1 ^\mu f_2 ^{\mu \nu} k_{34} ^\nu
- \epsilon_2 \cdot p_2 + \epsilon_1 \cdot q_1=0
\eea
while the remaining terms combine as follows, up to total derivative terms,
\bea
(t-u) \W_{(R^2)} 
& = &
+ \sum _{i,j}  
( k_i ^\mu f_1 ^{\mu \nu} q_j ^\nu ) \p_1 G(1,i) \p_2 G(2,j) \, \E
\no \\ &&
- \sum _{i,j} 
( k_j ^\mu f_2 ^{\mu \nu} p_i ^\nu) \p_1 G(1,i) \p_2 G(2,j) \, \E
\eea
Replacing $p_i$ and $q_j$ by their actual values, we have 
\bea
(t-u) \W_{(R^2)} 
& = &
+\half  \sum _{i}  
( k_i ^\mu f_1 ^{\mu \nu} f_2 ^{\nu \rho} k_{34} ^\rho ) \p_1 G(1,i) 
\p_2 \bigg \{ 2G(2,1) -  G(2,3)  -  G(2,4) \bigg \} \, \E
\no \\ &&
- \half \sum _{j} 
( k_j ^\mu f_2 ^{\mu \nu} f_1 ^{\nu \sigma} k_{34} ^\rho)  \p_2 G(2,j) 
\p_1 \bigg \{  2G(1,2) -  G(1,3) -  G(1,4) \bigg \} \, \E
\no \\
\eea

\subsubsection{$s_{ij}\W_{(R^2)}(1,2)$ in terms of $f_i^{\mu\nu}$}

Since $s+t+u=0$, the expressions for $s\W_{(R^2)}$
and for $(t-u)\W_{(R^2)}$ imply the following
formulas for $s_{ij}\W_{(R^2)}$, using again the notation (\ref{kij}),
\bea
\label{stuWR2}
s \W_{(R^2)}  & = & 
2 (f_3 f_4) \p_3 \p_4 G (3,4) \, \E
- 2 \sum _{i,j} ^{} k_i ^\mu f_3 ^{\mu \nu} f_4 ^{\nu \rho} k_j ^\rho
\p_3 G (3,i) \p_4 G (4,j) \, \E
 \\
t \W_{(R^2)} 
& = &
- (f_3 f_4) \p_3 \p_4 G (3,4) \, \E
+ \sum _{i,j} ^{} k_i ^\mu f_3 ^{\mu \nu} f_4 ^{\nu \rho} k_j ^\rho
\p_3 G (3,i) \p_4 G (4,j) \, \E
\no \\ &&
+{ 1 \over 4}  \sum _{i}  
( k_i ^\mu f_3 ^{\mu \nu} f_4 ^{\nu \rho} k_{12} ^\rho ) \p_3 G(3,i) 
\p_4 \bigg \{ 2G(3,4) -  G(1,4)  -  G(2,4) \bigg \} \, \E
\no \\ &&
-{ 1 \over 4}\sum _{j} 
( k_j ^\mu f_4 ^{\mu \nu} f_3 ^{\nu \sigma} k_{12} ^\rho)  \p_4 G(4,j) 
\p_3 \bigg \{  2G(3,4) -  G(1,3) -  G(2,3) \bigg \} \, \E
\no \\
u \W_{(R^2)} 
& = &
- (f_3 f_4) \p_3 \p_4 G (3,4) \, \E
+ \sum _{i,j} ^{} k_i ^\mu f_3 ^{\mu \nu} f_4 ^{\nu \rho} k_j ^\rho
\p_3 G (3,i) \p_4 G (4,j) \, \E
\no \\ &&
-{ 1 \over 4}  \sum _{i}  
( k_i ^\mu f_3 ^{\mu \nu} f_4 ^{\nu \rho} k_{12} ^\rho ) \p_3 G(3,i) 
\p_4 \bigg \{ 2G(3,4) -  G(1,4)  -  G(2,4) \bigg \} \, \E
\no \\ &&
+{ 1 \over 4}\sum _{j} 
( k_j ^\mu f_4 ^{\mu \nu} f_3 ^{\nu \sigma} k_{12} ^\rho)  \p_4 G(4,j) 
\p_3 \bigg \{  2G(3,4) -  G(1,3) -  G(2,3) \bigg \} \, \E
\no
\eea

\subsection{The $R^4$ amplitude}

We need first to compute the Heterotic chiral 4 vector amplitude
for fixed internal loop momenta $p_I^\mu$,
\bea
\W _{(R^4)} ^{(p)} (1,2,3,4) 
= 
\left \< Q(p_I) \prod _{j=1} ^4 \epsilon _j ^\mu \p x_+ ^\mu (z_j) 
e^{i k_j \cdot x_+(z_j)} \right \>
\eea
Omitting the overall bosonic exponential factor
$\prod_{i<j}E(z_i,z_j)^{k_i\cdot k_j}$, we have
\bea
\W _{(R^4)} ^{(p)} (1,2,3,4) 
& \sim &
\Q_1 \Q_2 \Q_3 \Q_4 
 \\ 
&&
+ \epsilon _1 \cdot \epsilon _2 \, \p_1 \p_2 \ln E(1,2)  ~
\epsilon _3 \cdot \epsilon _4 \, \p_3 \p_4 \ln E(3,4)  
\no \\
&&
+ \epsilon _1 \cdot \epsilon _3 \, \p_1 \p_3 \ln E(1,3)  ~
\epsilon _2 \cdot \epsilon _4 \, \p_2 \p_4 \ln E(2,4)  
\no \\
&&
+ \epsilon _1 \cdot \epsilon _4 \, \p_4 \p_2 \ln E(1,4)  ~
\epsilon _2 \cdot \epsilon _3 \, \p_2 \p_3 \ln E(2,3)  
\no \\
&& 
+ \epsilon _1 \cdot \epsilon _2 \, \p_1 \p_2 \ln E(1,2) \, \Q_3 \Q_4 
+ \epsilon _3 \cdot \epsilon _4 \, \p_3 \p_4 \ln E(3,4) \, \Q_1 \Q_2
\no \\
&& 
+ \epsilon _1 \cdot \epsilon _3 \, \p_1 \p_3 \ln E(1,3) \, \Q_2 \Q_4 
+ \epsilon _2 \cdot \epsilon _4 \, \p_2 \p_4 \ln E(2,4) \, \Q_1 \Q_3 
\no \\
&& 
+ \epsilon _1 \cdot \epsilon _4 \, \p_1 \p_4 \ln E(1,2) \, \Q_2 \Q_3 
+ \epsilon _2 \cdot \epsilon _3 \, \p_2 \p_3 \ln E(2,3) \, \Q_1 \Q_4 
\no
\eea
where we introduce the following convenient notations,
\bea
\Q _i 
& = & 
2 \pi \epsilon _i ^\mu p_I ^\mu \omega _I(z_i)
- i \sum _j \epsilon _i ^\mu k^\mu _j \p_i \ln E(i,j)
\no \\
\P _i ^\mu
& = &
\sum _{j \not= i}  k^\mu _j \p_i G (i,j)
\eea
Assembling left and right chiralities, and shifting $p \to \tilde p$ using
(\ref{ptilde}) amounts to making the following replacement, 
\bea
\Q_j \to i \epsilon _j \cdot \P_j + 2 \pi \epsilon _j \cdot  \tilde p_I  \omega _I (z_j)
\eea
Combining these contributions with the bosonic chiral
amplitude factors, and integrating out $\tilde p$ produces
\bea
\int d^{20} p \, \W _{(R^4)} ^{(p)}  \,  
e^{- 2 \pi \{ p_I p_J \Im \Omega _{IJ} 
+2 \sum _i k_i \Im \int ^i \! \! \omega _I  \} }
\prod _{i<j} |E(i,j) | ^{2 k_i \cdot k_j}
=
({\rm det}\,{\rm Im}\,\Omega)^{-5}\,\W_{(R^4)}(1,2,3,4)
\nonumber\\
\eea
where $\W_{(R^4)}(1,2,3,4)$ is given explicitly by
\bea
\W_{(R^4)} (1,2,3,4)
& = &
+ \epsilon _1 \cdot \epsilon _2 \, \p_1 \p_2 G_s (1,2)  ~
\epsilon _3 \cdot \epsilon _4 \, \p_3 \p_4 G_s (3,4)  \, \E
\no \\
&&
+ \epsilon _1 \cdot \epsilon _3 \, \p_1 \p_3 G_s (1,3)  ~
\epsilon _2 \cdot \epsilon _4 \, \p_2 \p_4 G_s (2,4)  \, \E
\no \\
&&
+ \epsilon _1 \cdot \epsilon _4 \, \p_1 \p_4 G_s (1,4)  ~
\epsilon _2 \cdot \epsilon _3 \, \p_2 \p_3 G_s (2,3)  \, \E
\no \\
&& 
- \epsilon _1 \cdot \epsilon _2 \, \p_1 \p_2 G_s (1,2) \, 
(\epsilon_3 \cdot  \P_3) (\epsilon_4 \cdot \P_4) \, \E
\no \\
&&
- \epsilon _3 \cdot \epsilon _4 \, \p_3 \p_4 G_s (3,4) \, 
(\epsilon_1 \cdot  \P_1) (\epsilon_2 \cdot \P_2) \, \E
\no \\
&& 
- \epsilon _1 \cdot \epsilon _3 \, \p_1 \p_3 G_s (1,3) \, 
(\epsilon_2 \cdot  \P_2) (\epsilon_4 \cdot \P_4)  \, \E 
\no \\
&&
- \epsilon _2 \cdot \epsilon _4 \, \p_2 \p_4 G_s (2,4) \, 
(\epsilon_1 \cdot  \P_1) (\epsilon_3 \cdot \P_3) \, \E 
\no \\
&& 
- \epsilon _1 \cdot \epsilon _4 \, \p_1 \p_4 G_s (1,4) \, 
(\epsilon_2 \cdot  \P_2) (\epsilon_3 \cdot \P_3) \, \E 
\no \\
&&
- \epsilon _2 \cdot \epsilon _3 \, \p_2 \p_3 G_s (2,3) \, 
(\epsilon_1 \cdot  \P_1) (\epsilon_4 \cdot \P_4) \, \E 
\no \\
&&
+ (\epsilon_1 \cdot  \P_1) (\epsilon_2 \cdot \P_2)
(\epsilon_3 \cdot  \P_3) (\epsilon_4 \cdot \P_4) \, \E
\eea
and the factor ${\cal E}$ was defined in (\ref{calE}).
This formula for $\W_{(R^4)}(1,2,3,4)$ is identical to the one obtained by computing 
the non-chiral correlator of the scalar field $x(z,\bar z)$, as follows,
\bea
\W _{(R^4)}  (1,2,3,4) 
= 
\left \<  \prod _{j=1} ^4 \epsilon ^\mu _j \p x ^\mu  (z_j) 
e^{i k_j \cdot x (z_j)} \right \>
\eea
using the propagator $G(z_i,z_j)$, defined in (\ref{scalarprop}). 

\subsection{The $R^4$ amplitude in terms of $f_i$}

The right-moving superstring chiral amplitude factor $\bar \Y_S$ 
consists of  terms linear in $s$, $t$, and $u$ multiplying holomorphic 
differentials in the insertion points $z_i$. We shall now show how
the left-moving Heterotic chiral amplitude, when multiplied by
$\bar \Y_S$ may be expressed in terms of the gauge invariant 
objects $f_i^{\mu \nu}$. It suffices to work out the case of $s$,
the cases of $t$ and $u$ being obtained by permuting the legs.

\medskip

Thus, the starting point is 
\bea
s\, \W _{(R^4)}  (1,2,3,4) 
= 
- 2 k_1 \cdot k_2 \, \left \<  \prod _{j=1} ^4 \epsilon ^\mu _j \p x ^\mu  (z_j) 
e^{i k_j \cdot x (z_j)} \right \>
\eea
The object is to converting all $\epsilon _i$ factors into $f_i$ factors,
keeping coefficients which are polynomial in the momenta $k_i$.
Two legs may be converted very easily, using the following remark
at the operator level,
\bea
k_i ^\nu \epsilon _i ^\mu \p_i x^\mu (z_i) e^{i k_i \cdot x(z_i)}
=
f_i ^{\mu \nu} \p_i x^\mu (z_i) e^{i k_i \cdot x(z_i)}
+ \p_i \bigg (-i  \epsilon _i ^\nu \,  e^{i k_i \cdot x(z_i)} \bigg )
\eea
Discarding the total derivative for the usual reasons, we see that
we may convert $\epsilon_1, \, \epsilon _2$ into $f_1, \, f_2$
using this simple formula. We obtain in this way
\bea
s \, \W_{(R^4)} (1,2,3,4)
=
\sum _{a=1} ^9 \L _a
\eea
where the $\L_a$ are defined as follows,
\bea
\L_1 & = &
+ 2 (f _1 f_2) \, \p_1 \p_2 G (1,2)  ~
\W _{(R^2)} (3,4)
\no \\
\L_2 & = &
+2  \epsilon _3 \cdot \epsilon _4 \, \p_3 \p_4 G (3,4) \, 
(f_1 ^{\mu \rho}   \P_1^\mu ) (f_2 ^{\nu \rho}  \P_2 ^\nu ) \, \E
\no \\
\L_3 & = &
-2 (f_1 ^{\mu \rho} \epsilon _3 ^\mu) \, \p_1 \p_3 G (1,3)  ~
(f_2 ^{\nu \rho} \epsilon _4^\nu) \, \p_2 \p_4 G (2,4)  \, \E
\no \\
\L_4 & = &
-2 (f_1 ^{\mu \rho} \epsilon _4 ^\mu) \, \p_1 \p_4 G (1,4)  ~
(f_2 ^{\nu \rho} \epsilon _3 ^\nu) \, \p_2 \p_3 G (2,3)  \, \E
\no \\
\L_5 & = &
+ 2 (f_1 ^{\mu \rho} \epsilon _3 ^\mu) \, \p_1 \p_3 G (1,3) \, 
(f_2 ^{\nu \rho}  \P_2 ^\nu) (\epsilon_4 \cdot \P_4)  \, \E 
\no \\
\L_6 & = &
+2 (f_2 ^{\mu \rho}  \epsilon _4 ^\mu) \, \p_2 \p_4 G (2,4) \, 
(f_1 ^{\nu \rho}  \P_1 ^\nu) (\epsilon_3 \cdot \P_3) \, \E 
\no \\
\L_7 & = &
+2  (f_1 ^{\mu \rho} \epsilon _4 ^\mu) \, \p_1 \p_4 G (1,4) \, 
(f_2 ^{\nu \rho}  \P_2 ^\nu) (\epsilon_3 \cdot \P_3) \, \E 
\no \\
\L_8 & = &
+2  (f_2 ^{\mu \rho} \epsilon _3 ^\mu) \, \p_2 \p_3 G (2,3) \, 
(f_1 ^{\nu \rho}  \P_1 ^\nu) (\epsilon_4 \cdot \P_4) \, \E 
\no  \\
\L_9 & = &
-2 (f_1 ^{\mu \rho}  \P_1 ^\mu) (f_2 ^{\nu \rho}  \P_2 ^\nu)
(\epsilon_3 \cdot  \P_3) (\epsilon_4 \cdot \P_4) \, \E
\eea

\subsubsection{Recasting $\L_9$}

Begin by concentrating on the last term and consider the factor
involving $f_1f_2$. It may be recast in the following way,
\bea
-2 (f_1 ^{\mu \rho}  \P_1 ^\mu) (f_2 ^{\nu \rho}  \P_2 ^\nu)
=
2 \sum _{i,j}  C_{ij} \p_1 G(1,i) \p_2 G(2,j)
\eea
The coefficients are defined as follows,
\bea
C_{ij} = k_i ^\mu f_1 ^{\mu \nu} f_2 ^{\nu \rho} k_j ^\rho
\eea
The key observation is that, for any $i,j$, there exists a matrix $C_{ij}^{\mu \nu}$
which is linear in $f_1$ and linear in $f_2$ and otherwise $k$-independent,
so that
\bea
\label{lemma1}
C_{ij} = k_3 ^\mu k_4 ^\nu C_{ij} ^{\mu \nu}
\eea
This relation manifestly holds for $C_{34}$ and $C_{43}$.
To prove that it also holds for the remaining $i,j$, we make 
use of the following auxiliary equations,
\bea
\label{aux1}
f_1 ^{\mu \nu} f_2 ^{\nu \rho} k_1 ^\rho & = & \half (f_1f_2) k_1^\mu
\no \\
f_2 ^{\mu \nu} f_1 ^{\nu \rho} k_2 ^\rho &=& \half (f_1f_2) k_2^\mu
\eea
and show that
\bea
C_{21} =
 \half (f_1 f_2) k_3 \cdot k_4
\eea
and $C_{31} = C_{24}$, $C_{41} = C_{23}$ and $C_{33} - C_{44}$.
It is now clear that (\ref{lemma1}) also holds for $C_{21}$.
With the help of these relations, the momentum conservation 
equations, namely $ \sum _i C_{ij} = \sum _i C_{ji} =0$,
may be solved uniquely in terms of $C_{21}$, $C_{34}$ and $C_{43}$. We find
\bea
C_{41}&=&\half (-C_{21}+C_{34}-C_{43}),
\no\\
C_{31}&=&\half(-C_{21}-C_{34}+C_{43}),
\no\\
C_{33}&=&\half(+C_{21}-C_{34}-C_{43}),
\eea 
and hence,
\bea
C_{31} ^{\mu \nu} = C_{24} ^{\mu \nu} =
- {1 \over 4} \eta ^{\mu \nu} (f_1 f_2) -  L^{\mu \nu} _-
& \qquad & 
C_{21} ^{\mu \nu} =  \half \eta ^{\mu \nu} (f_1 f_2)
\no \\
C_{41} ^{\mu \nu} = C_{23} ^{\mu \nu} = 
- {1 \over 4} \eta ^{\mu \nu} (f_1 f_2) +  L^{\mu \nu} _-
& \qquad &
C_{34} ^{\mu \nu} = f_1 ^{\mu \rho} f_2 ^{\rho \nu}
\no \\
C_{33} ^{\mu \nu} = C_{44} ^{\mu \nu} =
+ {1 \over 4} \eta ^{\mu \nu} (f_1 f_2) - L^{\mu \nu} _+
& \qquad & 
C_{43} ^{\mu \nu} = f_2 ^{\mu \rho} f_1 ^{\rho \nu}
\eea
where we have introduced the following combinations,
\bea
L_\pm ^{\mu \nu} 
= \half \bigg ( f_1 ^{\mu \rho} f_2 ^{\rho \nu} \pm  f_2 ^{\mu \rho} f_1 ^{\rho \nu}.
\bigg )
\eea
Using (\ref{lemma1}), the term $\L_9$ may be recast in the following way,
\bea
\L_9 = 2 \R ^{\mu \nu} 
(k_3 ^\mu \epsilon ^\rho _3 \P _3 ^\rho)
(k_4 ^\nu  \epsilon ^\sigma _4 \P _4 ^\sigma ) \, \E
\eea
where we have defined,
\bea
\R ^{\mu \nu} = \sum _{i,j} C_{ij} ^{\mu \nu} \p_1 G (1,i) \p_2 G (2,j).
\eea
Expressing 
$k_3 ^\mu \epsilon _3 ^\rho= f_3 ^{\rho \mu} + k_3 ^\rho \epsilon _3 ^\mu$,
and recognizing $k_3 \cdot \P_3 \, \E = - \p_3 \E$, we find
\bea
\L_9 
=
- 2 \R ^{\mu \nu} f_3 ^{\mu \rho} \P_3 ^\rho 
(k_4 ^\nu  \epsilon ^\sigma _4 \P _4 ^\sigma ) \, \E
- 2 \R^{\mu \nu} \epsilon _3 ^\mu (k_4 ^\nu  \epsilon ^\sigma _4 \P _4 ^\sigma ) \, \p_3 \E
\eea
and, by making a total derivative in $\p_3$ explicit,
\bea
\label{Lnine}
\L_9
& = &
- 2 \R ^{\mu \nu} f_3 ^{\mu \rho} \P_3 ^\rho 
(k_4 ^\nu  \epsilon ^\sigma _4 \P _4 ^\sigma ) \, \E
+ 2 (\p_3 \R^{\mu \nu}) \epsilon _3 ^\mu (k_4 ^\nu  \epsilon ^\sigma _4 \P _4 ^\sigma ) \, \E \no \\ &&
+ 2 \R^{\mu \nu} \epsilon _3 ^\mu k_4 ^\nu  \epsilon ^\sigma _4 
k_3 ^\sigma \p_3 \p_4 G_s(3,4)  \, \E
-2 \p_3 \big (   \R^{\mu \nu} \epsilon _3 ^\mu 
(k_4 ^\nu  \epsilon ^\sigma _4 \P _4 ^\sigma ) \,  \E \big )
\eea
The last term is a total derivative and may be omitted. 
Next, we start converting also $\epsilon_4$ into $f_4$. To do so,
we use
\bea
\label{pee4}
k_4 ^\nu \epsilon ^\rho \P ^\rho _4 \E
=
- f_4 ^{\nu \rho} \P _4 ^\rho \E  - \epsilon _4 ^\nu \E
\eea
Using this equation on the first term in (\ref{Lnine}) only, 
and omitting the total derivative term $\p_4 (2 \R^{\mu \nu}
f_3 ^{\mu \rho} \P^\rho _3 \epsilon _4 ^\nu \E)$, we 
obtain,
\bea
\L_9
& = &
 2 \R ^{\mu \nu} f_3 ^{\mu \rho} \P_3 ^\rho f_4 ^{\nu \sigma} \P_4 ^\nu \E
 - 2 \R^{\mu \nu} f_3 ^{\mu \rho} k_4 ^\rho \epsilon _4 ^\nu \p_3 \p_4 G(3,4)
\no \\ &&
+ 2 (\p_3 \R^{\mu \nu}) \epsilon _3 ^\mu 
    (k_4 ^\nu  \epsilon ^\sigma _4 \P _4 ^\sigma ) \, \E 
- 2 (\p_4 \R^{\mu \nu}) f_3 ^{\mu \rho} \P_3 ^\rho \epsilon _4 ^\nu \E
\no \\ &&
+ 2 \R^{\mu \nu} \epsilon _3 ^\mu k_4 ^\nu  \epsilon ^\sigma _4 
k_3 ^\sigma \p_3 \p_4 G(3,4)  \, \E
\eea
Next, we use the rearrangement 
$\epsilon _3 ^\mu k_3 ^\sigma = f_3 ^{\mu \sigma}
+ k_3 ^\mu \epsilon _3 ^\sigma$ in the last term, and  the identity
\bea
\R^{\mu \nu} k_3 ^\mu k_4 ^\nu
& = &
\sum _{i,j} k_3 ^\mu C_{ij}^{\mu \nu} k_4 ^\nu \p_1 G(1,i) \p_2 G(2,j)
\no \\
& = & 
f_1 ^{\mu \rho } \P_1^\mu \, f_2 ^{\rho \nu  } \P_2^\nu \, 
\eea
where we have used the defining equation for $C_{ij}$,
namely $k_3 ^\mu C_{ij}^{\mu \nu} k_4 ^\nu = k_i ^\mu f_1 ^{\mu \rho}
f_2 ^{\rho \nu} k_j ^\nu$. Thus, we are left with 
\bea
\L_9
& = &
 2 \R ^{\mu \nu} f_3 ^{\mu \rho} \P_3 ^\rho f_4 ^{\nu \sigma} \P_4 ^\nu \E
 + 2 \R^{\mu \nu} f_3 ^{\mu \rho} f_4 ^{\rho \nu} \p_3 \p_4 G(3,4) \E
 \no \\ &&
 + f_1 ^{\mu \rho } \P_1^\mu \, f_2 ^{\rho \nu  } \P_2^\nu 
    \epsilon _3 \cdot \epsilon_4  \p_3 \p_4 G(3,4)
\no \\ &&
+ 2 (\p_3 \R^{\mu \nu}) \epsilon _3 ^\mu 
    (k_4 ^\nu  \epsilon ^\sigma _4 \P _4 ^\sigma ) \, \E 
- 2 (\p_4 \R^{\mu \nu}) f_3 ^{\mu \rho} \P_3 ^\rho \epsilon _4 ^\nu \E
\eea
Finally, we make the terms in $\p _3 \R$ and $\p_4 \R$ symmetric,
by using again (\ref{pee4}) on the term in $\p_3 \R$,
\bea
2 (\p_3 \R^{\mu \nu}) \epsilon _3 ^\mu 
    (k_4 ^\nu  \epsilon ^\sigma _4 \P _4 ^\sigma ) \, \E
& = &
- 2 (\p_3 \R^{\mu \nu}) \epsilon _3 ^\mu f_4 ^{\nu \sigma} \P_4 ^\sigma \E
\no \\ &&
+ 2 \epsilon _3 ^\mu \epsilon _4 ^\nu 
\bigg ( 
C_{34} ^{\mu \nu} \p_1 \p_3 G(1,3) \p_2 \p_4 G(2,4) 
\no \\ && \qquad +
C_{43} ^{\mu \nu} \p_1 \p_4 G(1,4) \p_2 \p_3 G(2,3) \bigg ) \E
\eea
up to a total derivative term 
$\p_4 (-2 \R^{\mu \nu} \epsilon _3 ^\mu \epsilon _4 ^\nu)$.
By inspection, the expression for $\L_9$ may be recast as follows,
\bea
\label{ellnine}
\L _9 
= 
-\L_2 + \L_3 + \L_4 + \L_3 ' + \L_4 ' + \L_A + \L_B
\eea
where 
\bea
\L _3 ' & = &
- 2 \p_3 R^{\mu \nu} f_4 ^{\nu \sigma} \P_4 ^\sigma \epsilon _3 ^\mu \, \E
\no \\ 
\L_4 ' & = &
- 2 \p_4 R^{\mu \nu} f_3 ^{\mu \sigma} \P_3 ^\sigma \epsilon _4 ^\nu \, \E
\no \\ 
\L_A & = & 
+ 2 \R^{\mu \nu} f_3 ^{\mu \rho} f_4 ^{\nu \sigma } \P _3 ^\rho \P_4 ^\sigma \, \E
\no \\
\L_B & = &
+ 2 \R^{\mu \nu} f_3 ^{\mu \rho} f_4 ^{\rho \nu} \p_3 \p_4 G(3,4) \, \E
\eea
Note that the terms $\L_A$ and $\L_B$ are already completely expressed in 
terms of $f_i$'s alone. 

\subsubsection{Simplifications}

We shall now simplify $\L_3'$ and $\L_4'$,
as follows. The derivatives of $\R$ are given by
\bea
\p_3 \R ^{\mu \nu} 
& = & 
\p_1 \p_3 G (1,3) \sum _j C_{3j} ^{\mu \nu}  \p_2 G (2,j)
+
\p_2 \p_3 G (2,3) \sum _j C_{j3} ^{\mu \nu}  \p_1 G (1,j)
\no \\
\p_4 \R ^{\mu \nu} 
& = & 
\p_1 \p_4 G (1,4) \sum _j C_{4j} ^{\mu \nu}  \p_2 G (2,j)
+
\p_2 \p_4 G (2,4) \sum _j C_{j4} ^{\mu \nu}  \p_1 G (1,j) \qquad
\eea
The following terms can be grouped into total derivatives.
Let $[\p_3\R^{\mu\nu}]_{j=4}'$ and $[\p_3\R^{\mu\nu}]_{j=4}''$
be the contributions to ${\cal L}_3'$ from the two terms
in the preceding equation with $j=4$. Let $[{\cal L}_5]_{j=4}$
be the contribution from ${\cal L}_5$ with $j=4$, 
if we write its factor $f_2^{\nu\rho}{\cal P}_2^\nu$
as $f_2^{\nu\rho}\sum_{j\not=2}k_j^\nu\p_2G(2,j)$.
Then we have,
\bea
[\p_3\R^{\mu\nu}]_{j=4}'
+
{\cal L}_3
+
[{\cal L}_5]_{j=4}
=
2\p_4\bigg\{\epsilon_3^\mu f_1^{\mu\rho}f_2^{\rho\nu}\epsilon_4^\nu\,
\p_1\p_3G(1,3)\,
\p_2 G(2,4)\,{\cal E}\bigg\}.
\eea
Similarly, with analogous notations, we have
\bea
[\p_3\R^{\mu\nu}]_{j=4}''
+
{\cal L}_4
+
[{\cal L}_8]_{j=4}
&=&
2\p_4\bigg\{\epsilon_3^\mu f_2^{\mu\rho}f_1^{\rho\nu}\epsilon_4^\nu\,
\p_2\p_3G(2,3)\,
\p_1 G(1,4)\,{\cal E}\bigg\}
\nonumber
\eea
\bea
[\p_4\R^{\mu\nu}]_{j=3}'
+
{\cal L}_4
+
[{\cal L}_7]_{j=3}
&=&
2\p_3\bigg\{\epsilon_3^\mu f_2^{\mu\rho}f_1^{\rho\nu}\epsilon_4^\nu\,
\p_1\p_4G(1,4)\,
\p_2 G(2,3)\,{\cal E}\bigg\}
\nonumber
\eea
\bea
[\p_3\R^{\mu\nu}]_{j=4}''
+
{\cal L}_3
+
[{\cal L}_6]_{j=3}
&=&
2\p_3\bigg\{\epsilon_3^\mu f_1^{\mu\rho}f_2^{\rho\nu}\epsilon_4^\nu\,
\p_2\p_4G(2,4)\,
\p_1 G(1,3)\,{\cal E}\bigg\}.
\nonumber
\eea

Omitting the total derivative terms,
we are then left only with the contributions from $j=1,3$ in $\L_3'$, $\L_5$, $\L_8$, and the contributions from $j=2,4$ in $\L_4'$, $\L_6$, and $\L_7$.
All the terms in $\L_2$, $\L_3$, $\L_4$, $\L_5$,
$\L_6$, $\L_7$, and $\L_8$ have been either
absorbed in total derivatives or completely cancelled. What remains is 
\bea 
s \, \W_{(R^4)} (1,2,3,4)
=
\L_1 + \L_A + \L_B + \L_3 '' + \L_4 ''
\eea
where we have set
\bea
\L_3 '' 
& = &
+ 2 \p_1 \p_3 G(1,3) \sum _{j=1,3} \p_2 G(2,j) 
\epsilon _3 ^\mu \P_4 ^\sigma \bigg (
f_1 ^{\mu \rho} f_2 ^{\nu \rho} k_j ^\nu \epsilon _4 ^\sigma 
    - C_{3j} ^{\mu \nu} f_4 ^{\nu \sigma} \bigg )
\no \\ &&
+ 2 \p_2 \p_3 G(2,3) \sum _{j=2,3} \p_1 G(1,j) 
\epsilon _3 ^\mu \P_4 ^\sigma \bigg (
f_2 ^{\mu \rho} f_1 ^{\nu \rho} k_j ^\nu \epsilon _4 ^\sigma 
    - C_{j3} ^{\mu \nu} f_4 ^{\nu \sigma} \bigg )
\eea
and 
\bea
\L_4 '' 
& = &
+ 2 \p_1 \p_4 G(1,4) \sum _{j=1,4} \p_2 G(2,j) 
\epsilon _4 ^\mu \P_3 ^\sigma \bigg (
f_1 ^{\mu \rho} f_2 ^{\nu \rho} k_j ^\nu \epsilon _3 ^\sigma 
    - C_{4j} ^{\nu \mu} f_3 ^{\nu \sigma} \bigg )
\no \\ &&
+ 2 \p_2 \p_4 G(2,4) \sum _{j=2,4} \p_1 G(1,j) 
\epsilon _4 ^\mu \P_3 ^\sigma \bigg (
f_2 ^{\mu \rho} f_1 ^{\nu \rho} k_j ^\nu \epsilon _3 ^\sigma 
    - C_{j4} ^{\nu \mu} f_3 ^{\nu \sigma} \bigg )
\eea

\subsubsection{Simplifying the kinematical factors}

Next, we show that the kinematical coefficients 
can be re-written as,
\bea
\epsilon _3 ^\mu \P_4 ^\sigma \bigg (
f_1 ^{\mu \rho} f_2 ^{\nu \rho} k_j ^\nu \epsilon _4 ^\sigma 
    - C_{3j} ^{\mu \nu} f_4 ^{\nu \sigma} \bigg )
=
\pm f_3 ^{\mu \nu} L_- ^{\mu \nu} \P_4 ^\sigma \epsilon ^\sigma
\eea
where $j=1$ corresponds to $+$ and $j=3$ to $-$;
\bea
\epsilon _3 ^\mu \P_4 ^\sigma \bigg (
f_2 ^{\mu \rho} f_1 ^{\nu \rho} k_j ^\nu \epsilon _4 ^\sigma 
    - C_{j3} ^{\mu \nu} f_4 ^{\nu \sigma} \bigg )
=
\pm f_3 ^{\mu \nu} L_- ^{\mu \nu} \P_4 ^\sigma \epsilon ^\sigma
\eea
where $j=2$ corresponds to $+$ and $j=3$ to $-$;
\bea
\epsilon _4 ^\mu \P_3 ^\sigma \bigg (
f_1 ^{\mu \rho} f_2 ^{\nu \rho} k_j ^\nu \epsilon _3 ^\sigma 
    - C_{4j} ^{\nu \mu} f_3 ^{\nu \sigma} \bigg )
=
\pm f_4 ^{\mu \nu} L_- ^{\mu \nu} \P_3 ^\sigma \epsilon _3 ^\sigma
\eea
where $j=1$ corresponds to $+$ and $j=4$ to $-$;
\bea
\epsilon _4 ^\mu \P_3 ^\sigma \bigg (
f_2 ^{\mu \rho} f_1 ^{\nu \rho} k_j ^\nu \epsilon _3 ^\sigma 
    - C_{j4} ^{\nu \mu} f_3 ^{\nu \sigma} \bigg )
=
\pm f_4 ^{\mu \nu} L_- ^{\mu \nu} \P_3 ^\sigma \epsilon _3 ^\sigma
\eea
where $j=2$ corresponds to $+$ and $j=4$ to $-$. We present the 
derivation only for the first case with $j=1$; the others are analogous.
From the definition of $C_{31}^{\mu \nu}$, we have 
\bea
\label{equation}
\epsilon _3 ^\mu \P_4 ^\sigma \bigg (
f_1 ^{\mu \rho} f_2 ^{\nu \rho} k_1 ^\nu \epsilon _4 ^\sigma 
    - C_{31} ^{\mu \nu} f_4 ^{\nu \sigma} \bigg )
& = &
\epsilon _3 ^\mu \P_4 ^\sigma \bigg (
- \half (f_1f_2) k_1 ^\mu \epsilon _4 ^\sigma 
+ {1 \over 4} (f_1f_2) f_4 ^{\mu \sigma} + L_- ^{\mu \nu } f_4 ^{\nu \sigma} \bigg ).
\nonumber
\\
\eea
In the second term on the right hand side,
we replace $f_4^{\mu\sigma}$ by its definition
$\epsilon_4^\mu k_4\sigma-\epsilon_4^\sigma k_4^\mu$
and observe that $k_4\P_4\E=-\p_4\E$. This last term is a total
derivative and can be dropped. Thus the first two terms on the right hand reduce to
\bea
\epsilon _3 ^\mu \P_4 ^\sigma
(-{1\over 2}k_1^\mu-{1\over 4}k_4^\mu)\epsilon_4^\sigma
\,
(f_1f_2)
&=&
\epsilon _3 ^\mu \P_4 ^\sigma
({1\over 4}(k_2+k_3)^\mu-{1\over 4}k_1^\mu)\epsilon_4^\sigma
\,(f_1f_2)
\nonumber\\
&=&
-{1\over 4}\epsilon _3 ^\mu \P_4 ^\sigma
(k_1-k_2)^\mu\epsilon_4^\sigma
\,
(f_1f_2).
\eea
On the other hand, again ignoring total derivatives and thus
replacing $f_4^{\nu\sigma}$ effectively by
$-\epsilon^\sigma k_4^\nu$, the third term on the right hand
side of (\ref{equation}) can be expressed as
\bea
-\half \epsilon_3^\mu\P_4^\sigma \epsilon_4^\sigma k_4^\nu L_-^{\mu\nu}
&=&
\half \epsilon_3^\mu\P_4^\sigma \epsilon_4^\sigma (k_1+k_2+k_3)^\nu
(f_1^{\mu\rho}f_2^{\rho\nu}
-
f_2^{\mu\rho}f_1^{\rho\nu})
\nonumber\\
&=&
{1\over 4}(f_1f_2)\epsilon_3^\mu \P_4^\sigma\epsilon_4^\sigma (k_1-k_2)^\mu
+
\half\P_4^\sigma\epsilon_4^\sigma
f_3^{\mu\nu} L_-^{\mu\nu},
\eea
where we made use of the definition of $L_-^{\mu\nu}$,
and of the identities (\ref{aux1}).
This establishes the desired formula.
Since we clearly have
\bea
\half f_3 ^{\mu \nu} L_- ^{\mu \nu} \P_4 ^\sigma \epsilon _4 ^\sigma 
= 
+ \half  f_1^{\mu \rho} f_2 ^{\rho \nu} f_3 ^{\mu \nu} 
    \P_4 ^\sigma \epsilon _4 ^\sigma 
= 
- \half (f_1f_2f_3) \P_4 ^\sigma \epsilon _4 ^\sigma,
\eea
the above results may now be summarized in a succinct manner,
\bea
\L_3 '' & = & 
(f_1f_2f_3)  \P_4 ^\sigma \epsilon _4 ^\sigma F(1,2;3)
\no \\
\L_4 '' & = & 
(f_1f_2f_4) \P_3 ^\sigma \epsilon _3 ^\sigma F(1,2;4)
\eea
with the function $F$ defined by
\bea
F(1,2;i) & = & 
+  \{ \p_1 G(1,2) - \p_1 G(1,i) \} \, \p_2 \p_i G(2,i) \, \E
\no \\ &&
-  \{ \p_2 G(1,2) - \p_2 G(2,i) \} \, \p_1 \p_i G(1,i) \, \E
\eea
for $i=3,4$. 

\subsubsection{Final expressions and summary}

A summary of the contributions so far is given by
\bea
\label{sWR4}
s \, \W_{(R^4)} (1,2,3,4)
=
\L_1 + \L_A + \L_B + \L_3 '' + \L_4''
\eea
where we have 
\bea
\L_1 & = &
+ 2 (f _1 f_2) \, \p_1 \p_2 G (1,2)  ~ \W _{(R^2)} (3,4)
\no \\ 
\L_A & = & 
+ 2 \R^{\mu \nu} f_3 ^{\mu \rho} f_4 ^{\nu \sigma } \P _3 ^\rho \P_4 ^\sigma \, \E
\no \\
\L_B & = &
+ 2 \R^{\mu \nu} f_3 ^{\mu \rho} f_4 ^{\rho \nu} \p_3 \p_4 G(3,4) \, \E
\no \\
\L_3 '' & = & 
(f_1f_2f_3)  \P_4 ^\sigma \epsilon _4 ^\sigma F(1,2;3) 
\no \\
\L_4 '' & = & 
(f_1f_2f_4) \P_3 ^\sigma \epsilon _3 ^\sigma F(1,2;4) 
\eea
We have not quite achieved our goal yet, as $\L_1$,
$\L_3''$ and $\L_4''$ still exhibit explicit $\epsilon$-dependence.
Manifestly, each term by itself is gauge invariant, so one 
should not expect cancellations and recombinations of these 
terms with one another. Instead, they may be recast in terms of 
$f$'s entirely at the cost of multiplying either by $s$, or by $t$ or by $u$.
This was already established for $\W_{(R^2)}(3,4)$ in the section
on $R^2F^2$, and the presence of an extra factor $\p_1 \p_2 G(1,2)$
does not interfere with those arguments.

\medskip

For $\L_3 ''$ and $\L_4''$, the argument is as follows,
\bea
\P_3 ^\sigma \epsilon _3 ^\sigma 
& = & 
{1 \over (k_3 \cdot \ell_3)} \bigg ( f_3 ^{\sigma \rho } \ell_3 ^\rho 
    + (\epsilon _3 \cdot \ell_3 ) k_3 ^\sigma \bigg ) \P_3 ^\sigma
\no \\
\P_4 ^\sigma \epsilon _4 ^\sigma 
& = & 
{1 \over (k_4 \cdot \ell _4)} \bigg ( f_4 ^{\sigma \rho } \ell _4 ^\rho 
    + (\epsilon _4 \cdot \ell_4) k_4 ^\sigma \bigg )\P_4 ^\sigma
\eea
where $\ell _3$ and $\ell_4$ are arbitrary generic momenta. Clearly,
the second terms in the parentheses yield total derivatives
because $F(1,2;3)$ is independent of $z_4$ and $F(1,2;4)$ is
independent of $z_3$. Thus, the resulting terms may be neglected.

\medskip

Therefore, we have the following alternative expressions,
\bea
\label{stuL}
s \L_3 '' & = & -2 (f_1f_2f_3) f_4 ^{\sigma \rho} k_3 ^\rho \P_4 ^\sigma \, F(1,2;3) 
\no \\
t \L_3 '' & = & -2 (f_1f_2f_3) f_4 ^{\sigma \rho} k_1 ^\rho \P_4 ^\sigma \, F(1,2;3) 
\no \\
u \L_3 '' & = & -2 (f_1f_2f_3) f_4 ^{\sigma \rho} k_2 ^\rho \P_4 ^\sigma \, F(1,2;3) 
\no \\
s \L_4 '' & = & -2 (f_1f_2f_4) f_3 ^{\sigma \rho} k_4 ^\rho \P_3 ^\sigma \, F(1,2;4) 
\no \\
t \L_4 '' & = & -2 (f_1f_2f_4) f_3 ^{\sigma \rho} k_2 ^\rho \P_3 ^\sigma \, F(1,2;4) 
\no \\
u \L_4 '' & = & -2 (f_1f_2f_4) f_3 ^{\sigma \rho} k_1 ^\rho \P_3 ^\sigma \, F(1,2;4)  
\eea

\newpage

\section{Non-Renormalization of $F^4$, $R^2 F^2$ and $R^4$ Terms}
\setcounter{equation}{0}

We can discuss now the two-loop corrections to the low-energy
effective actions of the Type II and Heterotic superstrings. 
These are the corrections that remain when the  two-loop amplitude
is expressed in terms of $f_i$ only and the subsequent limit $k_i \to 0$ is taken. 

\medskip

The following fact is very important for some
of the cancellations found below: the (conformal) Green's function $G(z,w)$
defined in (\ref{scalarprop}) is the logarithm of a $(-1/2,-1/2)$-form in each 
variable $z$ or $w$. It is not the same as the (Riemannian) Green's function, 
which is defined as the inverse of the Laplacian orthogonal to constants, 
and requires a choice of metric. However, the two Green's functions differ 
only by expressions of the additive form $A(z)+A(w)$, which cancel out
in all expressions of interest to us, either by differentiation with respect to 
both $z$ and $w$, or by momentum conservation.
Thus we can effectively regard $G(z,w)$ as a single-valued, scalar function 
when we choose to.

\subsection{Type II $R^4$ Terms and Heterotic $F^4$ Terms}

The expression (\ref{TypeII}) for the Type II amplitude for the scattering 
of 4 gravitons depends only on the field strengths $f_i^{\mu\nu}$
and clearly tends to 0 as $k_i\to 0$. Thus there is no
two-loop renormalization for the $R^4$ term in the low-energy 
effective action for the Type II superstring.
The same remains valid for both Heterotic strings,
since the right sector contribution is that of the Type II
superstring which tends to 0, and the left sector is manifestly
independent of $\epsilon_i$.

\subsection{$R^2F^2$ Corrections in the Heterotic Strings}

Next, we consider the two-loop corrections to the $R^2F^2$ 
terms in the low-energy effective action for the Heterotic strings.
The key observation is that the contribution of the superstring 
from the right sector always includes an $\epsilon_i$-independent 
term factor proportional to either $s$, $t$, or $u$. 
The scattering amplitude of the two gravitons on the left sector results in 
the term $\W_{(R^2)}$, and all three terms $s\W_{(R^2)}$, $t\W_{(R^2)}$,
$u\W_{(R^2)}$ have been shown to admit expressions in terms of the $f_i$ 
alone (\ref{stuWR2}). In these expressions, all the terms involve extra 
momentum factors obviously tend to $0$ as $k_i\to 0$, and do not 
contribute to the low-energy effective action. In view of the anti-holomorphicity 
in each insertion point of the amplitude $\Y_S$ from the superstring side, 
the only terms which remain from (\ref{stuWR2}) 
are proportional to terms of the form
\bea
\int
\ \overline{\omega_I(3)}\,
\p_3\p_4\, G(3,4)
\,
{\rm exp}(-\sum_{i<j}k_i\cdot k_j G(i,j)).
\eea
However, at $k_i \cdot k_j \to 0$, this reduces to
\bea
\int \overline{\omega_I (3)}\, \p_3 \p_4 G_s(3,4) = 0,
\eea
because $\p_3\p_4 G(3,4)$ can be identified with
the derivative of a scalar single-valued function.

\subsection{$R^4$ Corrections in the Heterotic Strings}

As in the preceding case, the superstring in the right chiral sector 
always contributes from the factor $\bar \Y_S$, a factor of $s$, $t$, or $u$ 
multiplying an anti-holomorphic 1-form is each vertex insertion point. 
We need only consider $s\W_{(R^4)}$, since $t\W_{(R^4)}$ and 
$u\W_{(R^4)}$ are similar after  exchanging the legs. 
Now $s\W_{(R^4)}$ can be re-written as  in (\ref{sWR4}), where all 
the $\epsilon_i$ have been eliminated in favor of the $f_i^{\mu\nu}$ 
(\ref{stuL}).  We examine the contributions of all the terms
$\L_1$, $\L_A$, $\L_B$, $\L_3''$, $\L_4''$.

\medskip

The contribution $\L_A$ is expressed solely in terms of $f_i$,
but already has two extra momentum factors arising from 
$\P_3 ^\rho \P_4 ^\sigma$, and will therefore not 
contribute to the $R^4$ terms in the low energy effective action. 

\medskip

The term $\L_B$ is also in terms of $f_i$, but does not have these extra momentum factors. Recalling the formula for $\R$,
\bea
\R ^{\mu \nu} = \sum _{i,j} C_{ij} ^{\mu \nu} \p_1 G_s (1,i) \p_2 G_s (2,j)
\eea
and setting $\E=1$ in the limit $k_i \to 0$, we see that of the 
sum over $i,j$ in $\R$, any term that does not depend on both 
$z_3$ and $z_4$ will not contribute as then either $z_3$ or $z_4$
is a total derivative and yields 0 when integrated against anti-holomorphic
differentials on the superstring side. The two remaining terms in $\R$
that do depend on both $z_3$ and $z_4$ are for $(i,j)=(3,4)$ and
$(i,j) = (4,3)$. But these are total derivatives in both $z_1$ and $z_2$
and again lead to vanishing contributions upon integration against 
anti-holomorphic Abelian differentials on the superstring side. Thus, 
$\L_B$ integrates to 0 to this order and does not yield a $R^4$
contribution either.

\medskip

Next, we consider the term $\L_1$.  This term contains the factor 
$\W_{(R^2)}(3,4)$, which does not admit by itself a regular expression in terms
of the $f_i^{\mu\nu}$ alone. However, we have seen that
$s\W_{(R^2)}(3,4)$, $t\W_{(R^2)}(3,4)$, and
$u\W_{(R^2)}(3,4)$ do (see (\ref{WR2})).
Now if we expand ${\cal E}$ for $s,t,u$ near $0$, the contribution 
from the constant term $1$  vanishes upon integration against an 
anti-holomorphic form, just as we saw in the previous 
subsection\footnote{The integrals involved are only conditionally
convergent. This requires in principle a more detailed argument,
but the outcome is the same.}.
Now the other terms brought down from $\E$
always contain a factor of $s$, $t$, or $u$.
We can then make use of the $\epsilon_i$-free expression
for $s\W_{(R^2)}$, $t\W_{(R^2)}(3,4)$, and $u\W_{(R^2)}(3,4)$
in (\ref{stuWR2}). For $s\W_{(R^2)}(3,4)$, this for example yields contributions
proportional to $(f_1 f_2) (f_3 f_4)$ with coefficient,
\bea
(G_s(1,2) + G_s(3,4)) \p_1 \p_2 G_s(1,2) \p_3 \p_4 G_s(3,4)
\eea
whose integral against a form anti-holomorphic in $z_1,z_2,z_3,z_4$ vanishes.
The argument for terms brought down from $\E$ linear in $t$ or $u$
is analogous. Hence the contribution from terms in $\L_1$
also vanishes.

\medskip

The only terms left for consideration are $\L_3 ''$ and $\L_4''$.
Their behavior is similar to that of $\L_1$. Again, we only know that 
$s\L_3''$, $t\L_3''$, $u\L_3''$, $s\L_4''$, $t\L_4''$, $u\L_4''$ can be 
expressed directly in terms of the $f_i^{\mu\nu}$. Thus we expand again the 
exponential factor $\E$. Upon integration against the anti-holomorphic 
form coming from the superstring in the right sector, the constant
brought down from $\E$ contributes $0$, since we can integrate by 
parts in $z_4$ in $\L_3''$ and in $z_3$ in $\L_4''$. The other contributions
result in a factor of $s$ $t$, or $u$, so now the $\epsilon_i$-free formulas
(\ref{stuL}) become available. But each of them has an extra momentum 
factor and tends to 0 as $k_i\to 0$. Thus the terms $\L_3''$ and $\L_4''$ also
do not contribute, and the $R^4$ term in the low-energy effective action
for the Heterotic string receives no two-loop corrections.

\newpage

\begin{appendix}

\section{Riemann Surface Formulas for genus 2}
\setcounter{equation}{0}

In this section, we review basic formulas for the $\tet$-functions, prime
form, Green functions and Fay trisecant relations needed in this paper.
Standard references are \cite{fay} and \cite{dp88}.
The basic objects on a Riemann surface $\Sigma$, from which all 
others may be reconstructed, are the holomorphic
Abelian differentials, the Jacobi $\tet$-function, and the prime form. We
choose a canonical homology basis $A_I$, $B_I$, $I=1,\cdots ,2$, with
canonical intersection matrix $\# (A_I,B_J) = \delta _{IJ}$. Modular
transformations are defined to leave the intersection form invariant and
form the group $Sp(4,{\bf Z})$.
The {\sl holomorphic Abelian differentials} $\omega _I$ are holomorphic
1-forms which may be normalized on $A_I$ cycles, and whose integrals
on $B_I$ cycles produce the period matrix,
\bea
\label{abdiff}
\oint _{A_I} \omega _J = \delta _{IJ}
\hskip 1in 
\oint _{B_I} \omega _J = \Omega _{IJ}
\eea
The Jacobian is then defined as $J(\Sigma) \equiv {\bf C}^2
/ \{ {\bf Z}^2 + \Omega {\bf Z}^2\}$.
Given a base point $z_0$, the {\sl Abel map} sends $d$ points $z_i$, with
multiplicities $q_i \in {\bf Z}$, $i=1,\cdots,d$ and divisor $D=q_1
z_1 + \cdots q_d z_d$ of degree $q_1 + \cdots + q_d$ into ${\bf C}^2$ by
\bea
\label{abelmap}
q_1 z_1 + \cdots + q_d z_d 
\equiv 
\sum _{i=1} ^d q_i \int _{z_0} ^{z_i} (\omega _1, \omega _2)
\eea
The Abel map onto ${\bf C}^2$ is multiple valued, but it is single
valued onto $J(\Sigma)$. 

\subsection{Jacobi $\tet$-functions}

The {\sl Jacobi $\tet$-functions} are defined on 
$\zeta = (\zeta _1, \zeta _2)^t \in {\bf C}^2$ by 
\bea
\tet [\kappa] (\zeta, \Omega) 
\equiv  
\sum _{n - \kappa ' \in {\bf Z}^2  } 
\exp \biggl (i \pi n  ^t \Omega n + 2\pi i n ^t  (\zeta + \kappa '') \biggl ) \, .
\eea
Here, $\kappa  = \left ( \kappa ' | \, \kappa '' \right )$ will corresponds to a 
spin structure, and thus be valued in $\kappa', \kappa '' \in ({\bf Z}/2{\bf Z}) ^2$. 
The parity of the $\tet$-functions depends on $\kappa$ and is even 
or odd depending on whether  $4\kappa ' \cdot \kappa ''$ is even or odd, i.e.
$\kappa$ is referred to as an {\sl even or odd spin structure}.  
The standard $\tet$-function is defined by $\tet (\zet, \Omega)= \tet[0]
(\zet, \Omega)$, and is related to $\tet [\kappa]$ by
\bea
\label{charrel}
\tet [\kappa ] (\zet , \Omega)
=
\tet (\zet + \kappa '' + \Omega \kappa ', \Omega)
\ \exp \{\pi i \kappa ' \Omega \kappa ' +  2 \pi i \kappa '(\zet + \kappa '') \}
\eea
We have the following periodicity relations for
$\tet[\kappa ](\zet,\Omega)$, in which $M,N \in {\bf Z}^2$
\bea
\label{tetper}
\tet [\kappa ] (\zet + M + \Omega N, \Omega )
&=&
\tet [\kappa ](\zet , \Omega) \ \exp \{ -i \pi N \Omega N
- 2 \pi i N (\zet +\kappa '') + 2 \pi i \kappa ' M \}
\nonumber \\
\tet [\kappa ' +N, \kappa '' +M ] (\zet , \Omega)
&=&
\tet [\kappa ] (\zet, \Omega) \ \exp \{2 \pi i \kappa ' M\}
\eea
The periodicity formula simplifies when $2\kappa$ is a period; we note it here for
later use,
\bea
\label{periodshift}
\tet [\delta] (2 \kappa)
=
 \< \delta | \kappa \> ~ e^{- 4 \pi i  \kappa ' \Omega \kappa '} ~ \tet [\delta]
(0)
\eea
where $\kappa$ is a half-period, which may be even or odd. The signature
symbol is given by
\bea
\label{signature}
\< \kappa | \lambda \>
\equiv \exp \{ 4 \pi i ( \kappa ' \lambda '' - \lambda ' \kappa '' \}
\eea
For $\kappa, \lambda$ both half-integer characteristics,
$\< \kappa |\lambda \> = \pm 1$.

\medskip

Under a modular transformation $U \in Sp(4,{\bf Z})$, 
$\kappa = (\kappa ' | \, \kappa ")$ transforms as,
\bea
\left ( \matrix{ \tilde \kappa ' \cr \tilde \kappa ''} \right )
=
\left ( \matrix{D & -C \cr -B & A} \right )
\left ( \matrix{ \kappa ' \cr \kappa ''} \right )
+ \half {\rm diag} \left ( \matrix{CD^t \cr AB^t} \right ) 
\hskip .7 in
U= \left ( \matrix{A & B \cr C & D} \right )
\eea
The period matrix transforms as 
\bea
 \tilde \Omega = (A\Omega + B) (C \Omega + D)^{-1}
\eea 
while the $\tet$-function transforms as, with $\epsilon ^8=1$,
\bea
\tet [\tilde \kappa ] (\{ (C\Omega +D)^{-1} \}^t  \zeta , \tilde
\Omega) =
\epsilon (\kappa, U) \det (C\Omega + D)^{\half} \tet [\kappa ](\zeta ,
\Omega)
\eea

\subsection{The Riemann relations}

The Riemann relations may be
expressed as  quadrilinear {\sl sum over all spin structures},
\bea
\label{Riemann}
\sum _\lambda \<\kappa | \lambda \>
\prod _{a=1}^4  \tet [\lambda ](\zet _a ) 
= 4\,
\prod _{a=1}^4  \tet [\kappa ](\zet _a ') 
\eea
where the signature symbol $\< \kappa | \lambda \>$ was introduced
in (\ref{signature}). There is one Riemann relation for each spin
structure $\kappa$. The vectors $\zet$ and $\zet '$, are related by a 
matrix $M$, which satisfies $M ^2 = I$ and $2 M$ has only integer entries,
\bea
\label{LambdaM}
\left ( \matrix{
\zet _1 ' \cr  \zet _2 ' \cr \zet _3 ' \cr \zet _4 ' \cr} \right )
= M
\left ( \matrix{
\zet _1  \cr  \zet _2  \cr \zet _3  \cr \zet _4  \cr} \right )
\qquad \qquad
M =
\half \left (\matrix{
 1 &  1 &  1 &  1 \cr
 1 &  1 & -1 & -1 \cr
 1 & -1 &  1 & -1 \cr
 1 & -1 & -1 &  1 \cr} \right )
\eea
In the special case where at least one of the $\zeta _a$, $a=1,2,3,4$ 
vanishes, then only even spin structures  $\lambda $ will contribute to 
the sum and we have one Riemann identity for each odd spin structure 
$\kappa$. 
When only the {\sl sum over even spin structures} $\delta$ is needed, 
the following modification may be used,
\bea
\label{Riemanneven}
\sum _{\delta ~ {\rm even}} \<\delta | \kappa \>
\prod _{a=1}^4 \tet [\delta ](\zet _a ) 
& = &
\half \sum _\lambda \<\kappa | \lambda \>
\bigg ( \tet [\lambda ](\zet _1 ) + \tet [\lambda] (- \zet_1) \bigg )
\prod _{a=2}^4 \tet [\delta ](\zet _a )
\no \\ &=& 
 2\, \prod _{a=1}^4  \tet [\kappa ] (\zet _a ^+) 
+ 2\,\prod _{a=1}^4  \tet [\kappa ] (\zet _a ^-)
\eea
with the following relations between the vectors $\zet$ and $\zet ^\pm$,
expressed in terms of a matrix $M$, defined in (\ref{LambdaM}),
\bea
\label{Lambdaeven}
\left ( \matrix{
\zet _1 ^\pm \cr  \zet _2 ^\pm  \cr \zet _3 ^\pm  \cr \zet _4 ^\pm \cr} \right )
= M
\left ( \matrix{
\pm \zet _1  \cr  \zet _2  \cr \zet _3  \cr \zet _4  \cr} \right )
\eea
Clearly, the sign flip may be effected on any odd number of the $\zet_a$,
since both sides are manifestly invariant under permutations of the $\zet_a$.
In the special case where at least one of the $\zet_a$ vanishes, this formula
automatically reduces to (\ref{Riemann}).

\subsection{The Riemann vanishing Theorem}

The {\sl Riemann vector} $\Delta \in {\bf C}^h$, which depends on the base
point $z_0$ of the Abel map, enters the {\sl Riemann vanishing Theorem},
which states that $\tet (\zeta , \Omega )=0$ if and only if there
exist $h-1$ points $p_1, \cdots , p_{h-1}$ on $\Sigma$, so that $\zeta =
\Delta - p_1 \cdots -  p_{h-1}$. The explicit form of $\Delta$ may be
found in \cite{dp88}, formula (6.37) and will not be needed here.

\subsection{The prime form}

The {\sl prime form} is constructed as follows \cite{fay}. For any odd
spin structure
$\nu$, all the 2 zeros of the holomorphic 1-form $\sum _I \p_I
\tet [\nu ](0,\Omega) \omega _I(z)$ are double and the form admits a
unique (up to an overall sign) square root $h_\nu (z)$ which is a
holomorphic 1/2 form. The prime form is a $-1/2$ form in both variables
$z$ and $w$, defined by
\bea
\label{prime}
E(z,w) \equiv {\tet [\nu ] (z-w, \Omega) \over h_\nu (z) h_\nu (w)}
\eea
where the argument $z-w$ of the $\tet$-functions stands for the Abel map
of (\ref{abelmap}) with $z_1=z$, $z_2=w$. The form
$E(z,w)$ defined this way is actually independent of
$\nu$. It is holomorphic in $z$ and $w$ and has a unique simple zero at
$z=w$. It is single valued when $z$ is moved around $A_I$ cycles, but has
non-trivial monodromy when $z\to z'$ is moved around $B_I$ cycles,
\bea
\label{primemonod}
E(z',w) = - \exp \biggl ( -i \pi \Omega _{II} + 2 \pi i \int ^z _w \!
\omega _I \biggr ) E(z,w) \, .
\eea
The combination $\p_z \p_w \ln E(z,w)$ is a single valued meromorphic
differential (Abelian of the second kind) with a single double pole at
$z=w$. Its integrals around homology cycles are given by
\bea
\label{prf}
\oint _{A_I} \! dz \p_z \p_w \ln E(z,w) & = & 0
\nonumber \\
\oint _{B_I} \! dz \p_z \p_w \ln E(z,w) & = & 2 \pi i \omega _I(w)
\eea

\subsection{Green functions}

The Szeg\" o kernel $S_\delta (z,w)$ for even spin structure $\delta$ is a 
$(\half,0)$ form is each $z$ and $w$, with a single simple pole at $z=w$
and is given by 
\bea
S_\delta (z,w) = 
{\vartheta [\delta ]\big (z-w  \big ) \over 
 \vartheta [\delta ]\big ( 0 \big ) \ E(z,w)} \, .
\eea
The Green's function $G(z;z_1,z_2;p_1,p_2)$ is the Abelian differential 
of the third kind in $z$, with  simple poles at $z_1$ and $z_2$, and zeros at
$p_1$ and $p_2$. Thus, it satisfies,
\bea
\p _{\bar z} G(z;z_1,z_2;p_1,p_2)  = + 2 \pi \delta (z,z_1) - 2\pi
\delta (z,z_2)
\eea
and is explicitly given by the following expression
\bea
\label{green1}
G(z;z_1,z_2;p_1,p_2)
=
{\tet(z-z_1-z_2+p_1+p_2-\Delta)E(z,p_1)E(z,p_2)E(z_1,z_2)\sigma(z)
\over
\tet(-z_2+p_1+p_2-\Delta)E(z,z_1)E(z_1,p_1)E(z_1,p_2)E(z,z_2)\sigma(z_1)}
\quad 
\nonumber\\
\eea
where the nowhere-vanishing 1-form $\sigma (z)$ is defined by the 
ratio,
\bea
\label{sigma}
{\sigma(z)\over\sigma(w)}
={\tet(r_1+r_2-z-\Delta)E(w,r_1)E(w,r_2)
\over
\tet(r_1+r_2 -w-\Delta)E(z,r_1)E(z,r_2)} 
\eea
where $r_1,r_2$, are arbitary points on the surface. Note
that $\sigma(z)$ is single valued around $A_I$ cycles but multivalued
around $B_I$ cycles in the following way
\bea
\label{sigmamonod}
\sigma (z') &=& \sigma (z) \exp \bigl \{ -i\pi  \Omega _{II} + 2 \pi i
(z-\Delta) \bigr \}
\eea
A very useful alternative formula for $G$, in which the ratio of $\sigma$'s 
has been expressed in terms of $\tet$'s and prime forms is as follows,
\bea
\label{green2}
G(z;z_1,z_2;p_1,p_2)
=
{\tet(z-z_1-z_2+p_1+p_2-\Delta) \tet(p_1 + p_2 - \Delta -z) E(z_1,z_2)
\over
\tet(p_1+p_2 -z_1 -\Delta) \tet(p_1+p_2-z_2-\Delta)  E(z,z_1)E(z,z_2)} \quad
\eea

\subsection{The Fay trisecant identity and its variants}

Some of the identities needed involve a summation over products
with more than four Szeg\"o kernels.
At first sight, it would seem that the Riemann identities can not inform
us on such sums, since each term is a product of more than four
$\tet$-functions. The key is the additional use of the Fay trisecant identity,
\bea
\label{Fay}
&&
\tet [\delta ] (z_1 + z_2 - w_1 - w_2) \tet [\delta ](0) E(z_1,z_2) E(w_1,w_2)
\no \\ && \hskip 1in
=
+ \tet [\delta ] (z_1 - w_2) \tet [\delta ](z_2 - w_1) E(z_1,w_1) E(z_2,w_2)
\no \\ && \hskip 1.16in
- \tet [\delta ] (z_1 - w_1) \tet [\delta ](z_2 - w_2) E(z_1,w_2) E(z_2,w_1)
\eea
This identity  is equivalent to
\bea
\label{Fay0}
&&
S_\delta (z_1, w_2) S_\delta (z_2 , w_1)
-
S_\delta (z_1, w_1) S_\delta (z_2 , w_2)
\no \\
&& \hskip .6in
=
{\tet [\delta ] (z_1 + z_2 - w_1 - w_2) E(z_1, z_2) E(w_1, w_2)
\over
\tet [\delta ](0) E(z_1, w_1) E(z_2, w_1)E(z_1, w_2) E(z_2, w_2)}
\eea
We shall use this identity often for ``adjacent" Szeg\"o kernels, which
have an argument in common. Simply setting two points equal
in (\ref{Fay}) leads to a trivial result. Therefore,  we take the derivative
in $w_2$ of (\ref{Fay}), set $w_2 = z_2$, use the fact that
$\p _I \tet [\delta] (0)=0$ for even $\delta$, divide by
$\tet [\delta](0)^2 E(z_1, z_2) E(z_2, w_1) E(z_1,w_1)$,
and set $w_1 = z_3$. The result is,
\bea
\label{Fay1}
S_\delta (z_1, z_2) S_\delta (z_2, z_3)
=
- \omega _I (z_2) {\p _I \tet [\delta ] (z_1  - z_3) \over \tet [\delta ](0)
E(z_1,z_3)}
+ S_\delta (z_1  , z_3)   \p _{z_2} \ln { E(z_3,z_2) \over E(z_1, z_2)}
\eea
An immediate variant of (\ref{Fay1}) may be obtained by letting $z_3 \to z_1$,
and we obtain,
\bea
\label{Fay2}
S_\delta (z,w)^2 = \p_z \p _w \ln E(z,w) + \omega _I (z) \omega _J (w)
{ \p _I \p _J \tet [\delta ](0) \over \tet [\delta ](0)}
\eea
Notice that the first term on the right hand side is independent of $\delta$.
Finally,
a formula with ``adjacent" Szeg\"o kernels may also be derived from
(\ref{Fay0}), again by first taking a derivative in $w_2$ and then setting
$w_2=w_1$. The result is
\bea
\label{Fay3}
S_\delta (z_1, w) \p _w S_\delta (w,z_2) -
S_\delta (z_2, w) \p _w S_\delta (w,z_1) =
- { \tet [\delta] (z_1 + z_2 - 2 w) E(z_1, z_2) \over
\tet [\delta ](0) E(z_1, w)^2 E(z_2, w)^2}
\eea
The three Fay idenitities, (\ref{Fay1}), (\ref{Fay2}), (\ref{Fay3}) all have the
property that the product of two Szeg\"o kernels is reduced to a  single
$\tet [\delta]$ combination.

\newpage
\section{Simple geometric properties of the unitary gauge}
\setcounter{equation}{0}

In this appendix, we gather some simple useful facts about Riemann surfaces $\Sigma$ of genus 2.

\medskip

It is a consequence of the Riemann-Roch theorem that holomorphic
forms on $\Sigma$ must have 2 zeroes, counting multiplicities.
A first useful fact is the following: if $\omega(z)$ and $\tilde\omega(z)$ are two
holomorphic forms which have one
common zero (and neither form vanishes identically), then
they are proportional. In particular, they have the same zeroes.
Indeed, if their second zeroes are distinct, then
the ratio $f(z)=\omega(z)/\tilde\omega(z)$ is a meromorphic function with a
single pole. Then $f(z)-c$ has a single zero
for each complex number $c$, which implies that $f:\Sigma\to S^2$
is a biholomorphism from $\Sigma$ to the sphere. This contradicts
the fact that $\Sigma$ has genus 2.

\medskip
Assume now that $q_1$, $q_2$ are the zeroes of 
a holomorphic form $\varpi(z)$.
We deduce immediately the following identity
\bea
\Delta(q_1,q_2)=0.
\eea
This follows from the fact that $\Delta(z,q_2)$ is a holomorphic form in $z$,
which vanishes at $z=q_2$. Thus it must vanish also at $z=q_1$, since $q_1,q_2$
are the zeroes of $\varpi(z)$.

\medskip
These properties are even easier to read off from the hyperelliptic representation of $\Sigma$. If we view $\Sigma$
as the Riemann surface of the function $s^2=\prod_{i=1}^6(x-p_j)$,
then the holomorphic forms on $\Sigma$ are given by
\bea
\omega={ax+b\over s}dx,
\eea
from which it is clear that the zeroes of $\omega$ always
occur as the two points $q_1,q_2$ lying above the same point
$-b/a$ in the complex sphere.
\newpage

\section{Proof of summation identities involving $\Z[\delta]$}
\setcounter{equation}{0}

In this and the next two appendices, we give the proof of all
the identities for $I_1$ to $I_{16}$ summarized in section \S 3.  
They hold in unitary gauge, where $q_1, q_2$  satisfy the relation,
\bea
\label{unitaryq}
q_1 + q_2 - 2 \Delta = 2 \kappa
\eea
Here, $ 2 \kappa$ is a full period ($\kappa$ may be an even or odd
half-period or itself a full period). The $\delta$-dependence of 
$\Z[\delta]$ then simplified,  and reduces to 
\bea
\label{Zdelta}
\Z[\delta ] = \Z_0 \, E(q_1,q_2) \, e^{4 \pi i \kappa ' \Omega \kappa '}
 \< \kappa | \delta \> \tet [\delta] (0)^4
\eea

\subsection{Identities using the Riemann relations}

We begin by establishing the vanishing of the sums $I_1$ to
$I_7$. For these, we need only the Riemann relations.
Retaining only the $\delta$-dependent parts, the sums
$I_1$ to $I_7$ may all be recast in the following form
\bea
\label{ZRiemann}
I = \sum _\delta \Z [\delta] \,
\prod _{a=1}^4 {  \tet [\delta ] (\zet _a) \over  \tet [\delta] (0)}
\eea
where
\bea
I_1
& \qquad & \zet _1 = q_1 - q_2, ~
\zet _2 = \zet _3 = \zet _4 =0
\no \\
I_2
& \qquad & \zet _1 = q_1 - q_2, ~
\zet _2 = z_1-z_2, ~\zet _3 = z_2 - z_1, ~\zet _4 =0
\no \\
I_3
& \qquad & \zet _1 = q_1 - q_2, ~
\zet _2 = z_1-z_2, ~ \zet _3 = z_2 - z_3, ~ \zet _4 =z_3 - z_1
\no \\
I_4
& \qquad & \zet _1 = q_1 - z_1, ~
\zet _2 = z_1 -q_2, ~ \zet _3 =    \zet _4 = 0
\no \\
I_5
& \qquad & \zet _1 = q_1 - z_1, ~
\zet _2 = z_1-q_2, ~ \zet _3 = z_2 - z_3, ~ \zet _4 =z_3 - z_2
\no \\
I_6
& \qquad & \zet _1 = q_1 - z_1, ~
\zet _2 = z_1-z_2, ~ \zet _3 = z_2 - q_2, ~ \zet _4 = 0
\no \\
I_7
& \qquad & \zet _1 = q_1 - z_1, ~
\zet _2 = z_1-z_2, ~ \zet _3 = z_2 - z_3, ~ \zet _4 =z_3 - q_2
\eea
Making use of $\Z[\delta ] \sim \< \kappa | \delta \> \tet [\delta] (0)^4$
in unitary gauge, $I$ is seen to be proportional to
\bea
\label{Riemann2}
\sum _\delta \< \kappa | \delta \> 
\prod _{a=1}^4 
\tet [\delta ] (\zet _a) 
=
 2  \prod _{a=1} ^4 \tet [\kappa ] (\zet _a ^+) 
+
2 \prod _{a=1} ^4 \tet [\kappa ] (\zet _a ^-) 
\eea
The relation above is precisely the Riemann relation  (\ref{Riemanneven})
with $\zet ^\pm$ in terms of $\zet$ given by (\ref{Lambdaeven}).
To calculate the $I_i =0$ for $i =1,\cdots, 7$, we only use the following $\zet
^\pm$,
\bea
I_1
& \qquad & \zet _1^\pm = \zet _2 ^\pm = \zet _3 ^\pm  = \zet _4 ^\pm
= \pm (q_1 - \Delta - \kappa)
\\
I_2
& \qquad &  \zet _1^\pm  = \zet _2 ^\pm  = \pm (q_1 - \Delta - \kappa)
\no \\
I_3
& \qquad &  \zet _1^\pm  = \pm (q_1 - \Delta - \kappa)
\no \\
I_4
& \qquad &  \zet _1^\pm  = \zet _2 ^\pm  = \pm (q_1 - \Delta - \kappa)
\no \\
I_5
& \qquad &  \zet _1^+ = \zet _2 ^+ = q_1 - \Delta - \kappa, ~
\zet _1 ^- = \zet _2 ^- = z_1 - \Delta - \kappa
\no \\
I_6
& \qquad & \zet _1^\pm  = \pm (q_1 - \Delta - \kappa), ~
\zet _2 ^\pm  = \pm (\Delta - z_2 + \kappa), ~ \zet _4 ^\pm  =
\pm (\Delta - z_1 + \kappa)
\no \\
I_7
& \qquad &
\zet _1^+ = q_1 - \Delta - \kappa, ~ \zet _2^+ = \Delta - z_2 + \kappa, ~
\zet _1^- = z_1 - \Delta - \kappa, ~ \zet _4^- = z_3 - \Delta - \kappa
\no
\eea
For each case, at least one $\zet ^+$ and one $\zet ^-$ are of the form
$p-\Delta - \kappa$ for $p \in \{ q_1, z_1, z_2, z_3 \}$.
Using the Riemann vanishing theorem, we have $\tet [\kappa ] ( p - \Delta - \kappa) 
\sim \tet (p - \Delta) =0$ and this proves that $I_i =0$ for all $i = 1, \cdots ,7$.

\subsection{Identities using the derivative Riemann relations}

Making use of the Fay identity (\ref{Fay1}) in $I_8,I_9,I_{10}$ on the 
following  ``adjacent" pairs of Szego kernels, 
\bea
I_8 ~ & \qquad & S_\delta (z_1, z_2) S_\delta (z_2, z_3)
\no \\
I_9 ~ & \qquad & S_\delta (q_1, z_1) S_\delta (z_1, z_2)
\no \\
I_{10} & \qquad & S_\delta (z_2, z_3) S_\delta (z_3, z_4)
\eea
the  contributions from the second term in (\ref{Fay1})
vanish in view of $I_7 =0$, $I_5=0$, and $I_5=0$ respectively.
The $\delta$-dependence in the remaining sums is then,
\bea
I_8 \sim \sum _\delta \< \kappa |\delta \> ~
\p _I \tet [\delta ] (z_1  - z_3) ~ \tet [\delta] (q_1- z_1)   ~
\tet [\delta]  (z_3 - z_4) \tet [\delta]  (z_4 - q_2)
\no \\
I_9 \sim \sum _\delta \< \kappa |\delta \> ~
\p _I \tet [\delta ] (q_1  - z_2) ~\tet [\delta] (z_2- q_2)  ~
\tet [\delta]  (z_3 - z_4) ~ \tet [\delta]  (z_4 - z_3)
\no \\
I_{10} \sim \sum _\delta \< \kappa |\delta \> ~
\p _I \tet [\delta ] (z_2  - z_4) ~ \tet [\delta] (z_4- z_2)  ~
\tet [\delta]  (q_1 - z_1) ~ \tet [\delta]  (z_1 - q_2)
\eea
To evaluate these sums, we use a first derivative of the Riemann
identities (\ref{Riemann2}) with respect to $\zet $ of the formula,
where we have
\bea
I_8 & \qquad &
\zet _1 = 2\zet + z_1 - z_3, ~
\zet _2 = q_1 - z_1, ~
\zet _3 = z_3 - z_4, ~
\zet _4 = z_4 - q_2
\no \\
I_9 & \qquad &
\zet _1 = 2\zet + q_1 - z_2, ~
\zet _2 = z_2 - q_2, ~
\zet _3 = z_3 - z_4, ~
\zet _4 = z_4 - z_3
\no \\
I_{10} & \qquad &
\zet _1 = 2\zet + z_2 - z_4, ~
\zet _2 = z_4 - z_2, ~
\zet _3 = q_1 - z_1, ~
\zet _4 = z_1 - q_2
\eea
To obtain $\zet ^\pm$, we change the sign of $\zet _4$, so that
\bea
I_8 & \qquad &
\zet _1 ^+ =  \zet + q_1 - \Delta - \kappa , ~
\zet _2 ^+ =  \zet - z_3 + \Delta + \kappa
\no \\ &&
\zet _1 ^- =  \zet - z_4 + \Delta + \kappa, ~
\zet _3 ^- =  \zet + z_1 - \Delta - \kappa
\no \\
I_9 & \qquad &
\zet _1 ^+  =   \zet + q_1 - \Delta - \kappa, ~
\zet _2 ^+  =   \zet + q_1 - \Delta - \kappa
\no \\ &&
\zet _3 ^-   =   \zet + \Delta + \kappa - z_2, ~
\zet _4 ^-   =   \zet + \Delta + \kappa - z_2
\no \\
I_{10} & \qquad &
\zet _1 ^+ =   \zet + q_1 - \Delta - \kappa, ~
\zet _2 ^+ =   \zet -  q_1 + \Delta + \kappa
\no \\ &&
\zet _1 ^- =   \zet -  z_1 + \Delta + \kappa, ~
\zet _2 ^- =   \zet + z_1 - \Delta - \kappa
\eea
In each case, both $\zet ^+$ and $\zet ^-$ produce a zero in
$\theta [\kappa] (\zet _k ^\pm)$; when a single  derivative is applied,
at least one zero will remain. Thus, we have $I_8=I_9=I_{10}=0$.

\subsection{Identities with 5  Szeg\"o kernels}

Next, we evaluate the objects $I_{11}$ and $I_{12}$ with 5 Szego kernels, 
needed for the 4-point function. Both are single-valued 1-forms in each $z_i$
which are holomorphic in each $z_i$, since singularities at coincident
$z_i$'s cancel using $I_2=0$ for $I_{11}$, and $I_3=0$ for $I_{12}$.

\subsubsection{ Calculation of $I_{11}$}

Using (\ref{Fay2}) and $I_2=0$, we readily derive that
\bea
I_{11} =
\omega _I(z_1) \omega _J (z_2) \omega _K (z_3) \omega _L (z_4)
\sum _\delta \Z [\delta] S_\delta (q_1, q_2)
{ \p_I \p_J \tet [\delta ] (0) \p_K \p_L \tet [\delta ] (0)
\over  \tet [\delta ](0)^2 }
\eea
which makes the holomorphicity, as well as some of the symmetries, 
of $I_{11}$ manifest. To calculate it, we use  (\ref{Zdelta}). 
The above sum then becomes,
\bea
\label{sum1}
I_{11} = \Z_0  ~ e ^{4 \pi i \kappa ' \Omega \kappa '}
\sum _\delta \< \kappa |\delta \> ~
\p_I \p_J \tet [\delta ] (0) ~ \p_K \p_L \tet [\delta ] (0) ~
\tet [\delta] (q_1 - q_2) ~ \tet [\delta ](0)
\eea
To perform the sum, we use the Riemann identity (\ref{Riemann}) and 
the relation 
\bea
\label{kappashift}
\tet [\kappa] (\zet - \kappa)
& = &
\tet (\zet) ~ \exp\{- i \pi \kappa ' \Omega \kappa ' + 2 \pi i \kappa ' \zet \}
\no \\
\tet [\kappa] (\zet + \kappa)
& = &
\tet (\zet) ~ \exp \{- i \pi \kappa ' \Omega \kappa ' - 2 \pi i \kappa ' \zet +
4 \pi i \kappa ' \kappa '' \}
\eea
One thus obtains,
\bea
&&
\sum _\delta \< \kappa |\delta \> ~
\tet [\delta] (\zet _1) ~ \tet [\delta] (\zet _2) ~
\tet [\delta] (q_1 - q_2) ~ \tet [\delta ](0)
\no \\ && \hskip .3in
=
4  e^{ - 4 \pi i \kappa '( \Omega \kappa ' -2 q_1 +2 \Delta )}
\prod _{\alpha, \beta  = \pm} \tet \left ( \half (\zet _1 + \alpha \zet _2) + \beta (q_1 -
\Delta) \right )
\eea
In the above product, each of the 4 factors will vanish when $\zet _1 = \zet_2
=0$.
Thus, the 4 derivatives needed in (\ref{sum1}) will have to be applied
one on each factor to obtain a non-vanishing contribution.  Of the six different
terms that arise from taking these 4 derivatives, two terms arise from applying
the $\zet_2$ derivatives to factors with the same $\alpha$ and thus produce a
$+$ sign, while four terms arise from applying the $\zet_2$ derivatives
to factors with opposite $\alpha$ and thus produce a $-$ sign. Hence,
\bea
&&
e ^{4 \pi i \kappa ' \Omega \kappa '} \sum _\delta \< \kappa |\delta \> ~
\p_I \p_J \tet [\delta ] (0) ~ \p_K \p_L \tet [\delta ] (0) ~
\tet [\delta] (q_1 - q_2) ~ \tet [\delta ](0)
\\ && \hskip .5in
=
- 2 ~ e^{8 \pi i \kappa ' (q_1-\Delta)} ~
\p _I \tet (q_1 - \Delta) ~ \p _J \tet (q_1 - \Delta) ~
\p _K \tet (q_1 - \Delta) ~ \p _L \tet (q_1 - \Delta)
\no
\eea
Combining all factors for $I_{11}$, we get the result in (\ref{1-12}),
using $\varpi(z)$, the holomorphic differential in $z$, which
vanishes at $q_{1,2}$, defined in (\ref{varpi}).

\subsubsection{Calculation of $I_{12}$}

To evaluate $I_{12}$, we relate it to $I_{11}$. 
To do so, we make use of one of its symmetries,
$I_{12}  (z_1, z_2, z_3, z_4) = I_{12}  (z_3, z_2, z_1, z_4)$.
Since $I_{12}$ is symmetric and holomorphic in $z_1$ and $z_3$, 
no information is lost by setting $z_3=z_1$. This is because
there is an isomorphism between biholomorphic one-forms and
holomorphic two-forms, which we may schematically denote by
$ \omega _{\{ I} (z) \omega _{J \} } (w) ~ \leftrightarrow ~
\omega _I (z) \omega _J (z)$.
Setting $z_3 = z_1$, and using (\ref{Fay2}), as well as the vanishing of
$I_1$ and $I_2$, we have
\bea
I_{12} (z_1, z_2, z_1, z_4)
& = &
\sum _\delta \Z [\delta ] S_\delta (q_1, q_2) \omega _I (z_1) \omega _J (z_2)
\omega _K (z_1) \omega _L (z_4)
\no \\ && \hskip 1in \times
\tet [\delta ](0)^{-2} \p _I \p_J \tet [\delta ](0) \p _K \p_L \tet [\delta ](0)
\no
\eea
The original $I_{12} (z_1, z_2, z_3, z_4)$ is recovered by letting
$\omega _I (z_1) \omega _K (z_1) ~ \to ~ \omega _{\{ I} (z_1) \omega _{K \} }
(z_3)$,  and thus
\bea
I_{12} (z_1, z_2, z_3, z_4)
= \half I_{11} (z_1, z_2; z_3, z_4) + \half I_{11} (z_3, z_2; z_1, z_4)
=
I_{11}(z_1, z_2; z_3, z_4)
\eea
The last equality holds because by explicit calculation, we have found that
$I_{11}$ is totally symmetric in all its arguments $z_i$.

\subsection{Identities involving the fermionic stress tensor}

In this subsection, we evaluate the summation identities for 
$I_{13}$, $I_{14}$, $I_{15}$ and $I_{16}$, all of which 
invove the insertion of the fermionic stress tensor.

\subsubsection{Calculation of $I_{13}$}

Exhibiting the $\delta$-dependence of $\Z[\delta]$ and using (\ref{varphi}),
the sum is reduced to
\bea
I_{13} & = &
 { \Z_0 ~ e^{4 \pi i \kappa ' \Omega \kappa '}
\over E(z_1,w)^2 E(z_2,w)^2 }
\sum _\delta \< \kappa |\delta \> \tet [\delta ] (q_1-q_2) \tet [\delta]
(z_1-z_2)
\no \\ && \hskip 2.5in \times
\tet [\delta ](z_1 + z_2 - 2 w) \tet [\delta ](0)
\eea
It suffices to use the Riemann relations (\ref{Riemann}) and  
(\ref{kappashift}) to obtain expressions in terms of
$\tet$-functions without characteristics,
\bea
I_{13} =  4 \Z_0 e^{8 \pi i \kappa ' (q_1- \Delta)} 
\prod _{\alpha = \pm} \prod _{\beta=1,2}
{ \tet (\alpha (q_1 - \Delta) - z_\beta  + w) 
\over  E(z_\beta , w) }
\eea
Formula (\ref{I13}) is recovered by using the following useful identity,
\bea
\label{tetbil}
\prod _{\alpha = \pm} \tet (\alpha (q_1 - \Delta) + z - w)   
=
- \varpi (z) \varpi (w) E(z,w)^2 \, e^{ - 4 \pi i \kappa ' (q_1 - \Delta)}
\eea

\subsubsection{Calculation of $I_{14}$}

Exhibiting the $\delta$-dependence of $\Z[\delta]$ and using (\ref{varphi}),
the sum is reduced to
\bea
I_{14} & = &
- { \Z_0 ~ e^{4 \pi i \kappa ' \Omega \kappa '} E(z_1, z_2)
\over E(z_1,w)^2 E(z_2,w)^2   E(z_2,z_3)  E(z_3, z_1) }
\\ \no &&
 \\ && \qquad \times
\sum _\delta \< \kappa |\delta \> \tet [\delta ] (q_1-q_2)
\tet [\delta] (z_2-z_3)  \tet [\delta ](z_3-z_1)
\tet [\delta ](z_1 + z_2 - 2 w)
\no
\eea
It suffices to use the Riemann relations to obtain,
\bea
I_{14} =
- { 2 \Z_0 ~ e^{4 \pi i \kappa ' \Omega \kappa '} E(z_1, z_2)
\over E(z_1,w)^2 E(z_2,w)^2  E(z_2,z_3) E(z_3, z_1) }
\left ( \prod _{i=1} ^4 \tet [\kappa ] (\zet _i ^+)
+  \prod _{i=1} ^4 \tet [\kappa ] (\zet _i ^-) \right )
\eea
where
\bea
\zet_1 ^\pm & = & \pm (q_1 - \Delta - \kappa) + z_2 - w
\no \\
\zet_2 ^\pm & = & \pm (q_1 - \Delta - \kappa) - z_3 + w
\no \\
\zet_3 ^\pm & = & \pm (q_1 - \Delta - \kappa) - z_1 - z_2 + z_3  + w
\no \\
\zet_4 ^\pm & = & \pm (q_1 - \Delta - \kappa) + z_1 - w
\eea
The interchange between $\zet ^+ _i$ and $\zet _i ^-$ is equivalent 
to  $q_1 \leftrightarrow q_2$, in view of the unitary gauge relation
(\ref{unitaryq}). Using the relations (\ref{kappashift}), we have
\bea
\label{I_14}
I_{14} & = &
- 2 \Z_0 e^{8 \pi i \kappa ' (q_1 - \Delta)}
{ \tet (q_1 - \Delta + z_1 -w) \tet (q_1 - \Delta + z_2 -w)  E(z_1, z_2)
\over
E(z_1,w)^2 E(z_2,w)^2  E(z_2,z_3) E(z_3, z_1) }
\\ && \hskip .5in \times
\tet (q_1 - \Delta - z_3 + w) \tet (q_1 - \Delta - z_1 - z_2 + z_3  + w)
+ (q_1 \leftrightarrow q_2)
\qquad
\no
\eea
As a function of $z_3$, $I_{14}$ has  simple poles at $z_1$ and $z_2$
with opposite non-vanishing residues, given by $\pm I_{13}$.
Thus, it is natural to seek an alternative formula in terms of a Green
function $G(z;z_1,z_2;p_1,p_2)$ of (\ref{green1}) and (\ref{green2}),
 with poles at $z_1$ and $z_2$, and zeros at $p_1$ and $p_2$.
Inspection of (\ref{I_14}) indicates that $I_{14}$ vanishes at $q_1$ and $w$
as a  function of $z_3$; therefore, we choose $p_1 = q_1$ and $p_2 = w$.
Expression $\tet (z_3-z_1-z_2 + q_1 + w - \Delta )$ in (\ref{I_14}) in terms
of $G$, and using (\ref{tetbil}), and the fact that bilinears in $\varpi$ are invariant
under $q_1 \leftrightarrow q_2$, we recover (\ref{I14}).

\subsection{A relation between $I_{15}$ and $I_{16}$}

We establish a simple relation between $I_{15}$ and $I_{16}$.
Recall their expressions,
\bea
I_{15} (w; z_1,z_2,z_3,z_4)
& = &
\sum _\delta \Z [\delta] S_\delta (q_1, q_2) \varphi [\delta] (w;z_1,z_2)
S_\delta (z_2, z_3) S_\delta (z_3, z_4) S_\delta (z_4, z_1)
\no \\
I_{16} (w; z_1,z_2; z_3, z_4)
& = &
\sum _\delta \Z [\delta] S_\delta (q_1, q_2) \varphi [\delta] (w;z_1,z_2)
S_\delta (z_2, z_1) S_\delta (z_3, z_4)^2
\eea
We now use the full Fay identity (\ref{Fay0}), applied to $I_{15}$
and $I_{16}$ with 
\bea
&&
S_\delta (z_2, z_1) S_\delta (z_3 , z_4)
=
S_\delta (z_2, z_4) S_\delta (z_3 , z_1)
\no \\
&& \hskip 1in
+
{\tet [\delta ] (z_2 + z_3 - z_1 - z_4) E(z_2, z_3) E(z_4, z_1)
\over
\tet [\delta ](0) E(z_2, z_4) E(z_3, z_4)E(z_2, z_1) E(z_3, z_1)}
\eea
Inserting this expression for the combination
$S_\delta (z_2, z_1) S_\delta (z_3 , z_4)$ in $I_{16}$, it is manifest that
the first term on the rhs equals $- I_{15} (w;z_1,z_2,z_4,z_3)$, while the
second term may be recast in the following form, using the explicit
factorized expression for $\varphi [\delta]$,
\bea
&&
I_{16} (w;z_1,z_2,z_3,z_4) + I_{15} (w;z_1,z_2,z_4,z_3)
\\ && \hskip .1in =
- { \Z_0 ~ e ^{4 \pi i \kappa ' \Omega \kappa '} ~ E(z_2,z_3) E(z_1,z_4)
\over
E(z_2,z_4) E(z_3,z_4)^2 E(z_3,z_1)  E(z_1,w)^2 E(z_2,w)^2}
\no \\ && \hskip .4in \times
\sum _\delta \< \kappa |\delta \>
\tet [\delta ] (q_1-q_2)
\tet [\delta ] (z_1 + z_2 - 2w)
\tet [\delta ] (z_3-z_4)
\tet [\delta ] (z_2+z_3-z_1-z_4)
\no
\eea
The $\delta$-sum is carried out using the Riemann identities (\ref{Riemanneven}),
with the help of the following $\zet^\pm$,
\bea
\zet ^\pm _1 & = & \pm (q_1 - \Delta - \kappa) -z_2 -z_3+z_4+w
\no \\
\zet ^\pm _2 & = & \pm (q_1 - \Delta - \kappa) -z_1+ z_3- z_4 +w
\no \\
\zet ^\pm _3 & = & \pm (q_1 - \Delta - \kappa) + z_2 - w
\no \\
\zet ^\pm _4 & = & \pm (q_1 - \Delta - \kappa) + z_1 - w
\eea
Clearly, the contributions from $\zet^+$ and $\zet^-$ may be
obtained from one another by simple interchange of $q_1$ and $q_2$.
Using the form for the Green function $G$ given in (\ref{green2}) 
and the relation (\ref{tetbil}), we obtain (after ample simplifications of 
prime forms and $\sigma$ functions),
\bea
\label{I15I16}
&&
I_{16} (w;z_1,z_2,z_3,z_4) + I_{15} (w;z_1,z_2,z_4,z_3)
\\ && \hskip .5in
=
- 2 \Z_0 \varpi (z_1) \varpi (z_2 ) \varpi (w)^2
G(z_3;z_4,z_1;q_1,w) G(z_4;z_3,z_2;q_1,w)
\no \\ && \hskip .67in
- 2 \Z_0 \varpi (z_1) \varpi (z_2 ) \varpi (w)^2
G(z_3;z_4,z_1;q_2,w) G(z_4;z_3,z_2;q_2,w)
\no
\eea
This equation will yield the expression for $I_{16}$ of (\ref{I16}), 
as soon as $I_{15}$ will be known.
The latter will be evaluated next.

\subsection{Calculating the antisymmetric part  $I_{15}^A$}

Given the symmetry of $I_{16} $ under  $z_1 \leftrightarrow z_2$
(as well as under the interchange of $z_3$ and $z_4$), we can
eliminate $I_{16}$ in relation (\ref{I15I16}) above and directly 
obtain the antisymmetric part $I_{15}^A$, producing the 
second equation in (\ref{I15SA}).

\subsection{Calculation of the symmetric part  $I_{15}^S$}

Recall the definition of $I_{15}^S $, expressed as a sum over $\delta$,
\bea
2 I_{15} ^S &=&
\sum _\delta \Z [\delta] S_\delta (q_1, q_2) \varphi [\delta] (w;z_1,z_2)
S_\delta (z_3, z_4)
\bigg \{
S_\delta (z_2, z_3) S_\delta (z_4, z_1)
\no \\ && \hskip 2.8in
 - S_\delta (z_1, z_3) S_\delta (z_4, z_2)
\biggr \}
\no
\eea
The combination in braces may be recast using the Fay identity (\ref{Fay0})
and equals
\bea
{ \tet [\delta ] (z_1+z_2-z_3-z_4) E(z_1,z_2) E(z_3,z_4)
\over \tet [\delta ](0) E(z_1,z_3) E(z_1,z_4) E(z_2,z_3) E(z_2,z_4)}
\eea
Using also the explicit expression for $\varphi [\delta]$, we have
\bea
\label{I15int}
2 I_{15} ^S &= &
- { \Z _0 e^{4 \pi i \kappa ' \Omega \kappa '} E(z_1,z_2)^2
\over
 E(z_1,z_3) E(z_1,z_4) E(z_2,z_3) E(z_2,z_4)
E(z_1,w)^2 E(z_2,w)^2} \times S
\\
S & \equiv &
\sum _\delta \< \kappa |\delta \>
\tet [\delta ](q_1-q_2)
\tet [\delta ](z_1+z_2-2w)
\tet [\delta ](z_3-z_4)
\tet [\delta ](z_1+z_2-z_3-z_4)
\no
\eea
The last sum may be evaluated using the Riemann identities, with
\bea
\zet ^\pm _1 & = & \pm (q_1 - \Delta - \kappa) + z_1 + z_2 - z_4 -w
\no \\
\zet ^\pm _2 & = & \pm (q_1 - \Delta - \kappa) + z_4 -w
\no \\
\zet ^\pm _3 & = & \pm (q_1 - \Delta - \kappa) - z_1 - z_2 + z_3 + w
\no \\
\zet ^\pm _4 & = & \pm (q_1 - \Delta - \kappa) - z_3 +w
\eea
so that
\bea
S =
2 \prod _{i=1} ^4 \tet [\kappa ] (\zet ^+ _i)
+
2 \prod _{i=1} ^4 \tet [\kappa ] (\zet ^- _i)
=
2 \prod _{i=1} ^4 \tet [\kappa ] (\zet ^+ _i)   + (q _1 \leftrightarrow q_2)
\eea
Working out the first product, we have
\bea
\label{product1}
\prod _{i=1} ^4 \tet [\kappa ] (\zet ^+ _i) & = &
 e^{- 4 \pi i \kappa ' \Omega \kappa ' + 8 \pi i \kappa ' (q_1 - \Delta)}
 \\ && \times
\tet (q_1-\Delta + z_1 + z_2 - z_4 -w) \tet (q_1-\Delta + z_4 -w)
\no \\ && \times
\tet (q_1-\Delta - z_1 - z_2 + z_3 +w)
\tet (q_1-\Delta - z_3 + w)
\no
\eea
The two $\tet$-function factors whose arguments are a sum of 6 terms
may each be recast in terms of the Green function $G$ using (\ref{green2}).
There are two different Green functions $G$ that can enter: the one with 
zeros at $w$ and $q_1$, and the one with zeros at $w$ and $q_2$. 
To recast the above results in this form, we perform the following operations
on the first two $\tet$-functions in (\ref{product1}),
\bea
\tet (q_1-\Delta + z_1 + z_2 - z_4 -w)
& = &
\tet (q_2 - \Delta - z_1 - z_2 + z_4 +w)
\\ &&  \times
\exp \{ 4 \pi i \kappa ' (q_2 - \Delta -z_1 - z_2 + z_4 + w - \Omega \kappa ') \}
\no
\eea
and 
\bea
\tet (q_1-\Delta + z_4 -w) =
\tet (q_2 - \Delta - z_4 + w) e^{ 4 \pi i \kappa ' (q_2 - \Delta - z_4 + w -
\Omega \kappa ')}
\no
\eea
Furthermore, we make use of the following rearrangement formula (\ref{tetbil})
to recast the product in the following final form in terms of the Green functions $G$,
\bea
\label{product3}
\prod _{i=1} ^4 \tet [\kappa ] (\zet ^+ _i) & = &
- e^{ - 4 \pi i \kappa '  \Omega \kappa'  }
 \prod _{i=1}^2 { E(z_i,w)^2 E(z_3,z_i) E(z_4,z_i) \over E(z_1,z_2)}
 \\ && \times
 \varpi (z_1) \varpi (z_2) \varpi (w)^2
 G (z_3;z_1,z_2;q_1,w)   G (z_4;z_1,z_2;q_2,w)
\no
\eea
Using this result for the product in the symmetrized expression for $I_{15}$,
and combining with the prefactor to $S$ in (\ref{I15int}), we recover 
the expression for $I_{15}^S$ in (\ref{I15SA}).

\newpage

\section{Proof of summation identities involving $\Xi_6[\delta](\Omega)$}
\setcounter{equation}{0}

In this section, we prove the $\delta$-summation identities for 
$I_{17}$, $I_{18}$, $I_{19}$, $I_{20}$ and $I_{21}$, which all
involve $\Xi_6 [\delta]$, introduced in \cite{IV} and studied there.
We recall here the expression for $\Xi_6 [\delta]$ in terms of $\tet$-constants.
Let $\delta = \nu _1 + \nu _2 + \nu_3$ the decomposition of $\delta$
as a sum of three distinct odd spin structures $\nu_1,\nu_2,\nu_3$; then
we have
\bea
\Xi_6 [\delta] 
=
\sum _{1 \leq i < j \leq 3}
\< \nu_i |\nu_j\> \prod _{k=4,5,6} \tet [\nu_i + \nu _j + \nu_k](0)^4
\eea
where the sum is over the remaining 3 mutually distinct odd spin structures.

\subsection{The vanishing of $I_{17}$, $I_{18}$, $I_{19}$}

The vanishing of $I_{17}$ is one of the fundamental identities established 
in \cite{IV}. It may also be expressed as follows,
\bea
0= \sum _{l\not= m} \< \nu _l | \nu _m\> \prod _{n\not=l,m}
\tet [\nu_l + \nu _m +\nu _n](0)^4
\eea
Differentiating with respect to $\Omega _{IJ}$, an operation that we
denote by $\p _{IJ}$, we find
\bea
0= \sum _{l\not= m} \< \nu _l | \nu _m\> \sum _{n\not=l,m} \p_{IJ}
\tet [\nu _l + \nu _m + \nu _n ](0)^4
\prod _{p\not=l,m,n}  \tet [\nu_l + \nu _m +\nu _p](0)^4
\eea
Symmetrizing this equation in $l,m,n$,  we recognize the result as the sum over
all triples $(l,m,n)$ of distinct $l,m,n$, which we may identity with the
even spin structures $\delta = \nu _l+\nu_m+\nu_n$, of $\p_{IJ} \tet [\delta]$ 
multiplied by the sum of three terms which precisely add up to be
$\Xi _6 [\delta]$. Using now the heat equation for the $\tet$-functions, we have
$4 \pi i\p_{IJ} \tet [\delta ](0)= \p _I \p_J \tet [\delta](0)$, 
\bea
\sum _\delta \Xi _6 [\delta ] \ \tet [\delta ](0)^3 \p_I \p_J \tet [\delta ](0)  =0
\eea
Using the Fay identity (\ref{Fay2}), this readily proves also that $I_{18}=0$.
To prove that also $I_{19}=0$,  we notice that it is holomorphic in each $z_i$, $i=1,2,3$, since the poles in the Szeg\"o kernels have residues proportional to 
$I_{18}$, which vanishes. Thus, $I_{19}$ must be of the form,
\bea
I_{19} (z_1, z_2,z_3) =
\sum _{IJK} C_{IJK} \omega _I (z_1) \omega _J (z_2) \omega _K (z_3)
\eea
where $C_{IJK}$ are independent of the $z_i$. From its definition, $R_3$ is odd
under the interchange of any two $z$'s, so $C_{IJK}$ must
accordingly be a completely antisymmetric object under the interchange of
any two labels. Since $I,J,K=1,2$ only, such object cannot exist, and hence
$I_{19} =0$.

\subsection{Calculation of $I_{20}$}

Recall that $I_{20}$ is defined by
\bea
I_{20} (x,y;u,v)) =
 \sum _\delta \Xi _6 [\delta ] \, \tet [\delta ](0)^4 S_\delta (x,y)^2 S_\delta (u,v)^2
\eea
It is convenient to evaluate $I_{20}$ using the hyperelliptic realization of 
Riemann surfaces of genus $2$. As usual, the hyperelliptic formulation
leaves an overall sign to be determined, which we fix by studying the 
asymptotics in the $\tet$-function formulation.
Let the surface $\Sigma$ be the Riemann surface of the function
\bea
s^2=\prod_{j=1}^6(x-p_j).
\eea
A spin structure $\delta$ corresponds to the grouping of the 6 branch points
$\{p_j\}$ into two sets $A\cup B$, with
$A=\{a_1,a_2,a_3\}$ and $B=\{b_1,b_2,b_3\}$.
In this realization, the holomorphic differentials $\omega_I(z)$, 
the $\tet$-constants, and the Szeg\"o kernel are given by
\bea
2 \pi i \omega _I (z)
& = &
\sum _{J=1,2} \sigma _{IJ} {x^{J-1} dx \over s(x)}
\no \\
\tet [\delta ](0)^8
& = &
{1 \over (\det \sigma)^4} \prod _{i <j} (a_i-a_j)^2 (b_i - b_j)^2
\no \\
S_\delta (x,y)
& = &
\half { s_A(x) s_B(y) + s_A(y) s_B (x) \over x-y}
\left ( {dx \over s(x)} \right )^\half  \left ( {dy \over s(y)} \right )^\half,
\eea
where the matrix $\sigma_{IJ}$ of change of bases of holomorphic differentials
is defined by the equation given above, and we use the notation,
\bea
s_A(x)^2 & = &  (x-a_1) (x-a_2) (x-a_3) = x^3 - A_1 x^2 + A_2 x - A_3
\no \\
s_B(x)^2 & = &  (x-b_1) (x-b_2) (x-b_3) \, = x^3 - B_1 x^2 + B_2. x - B_3
\eea
First, we extract from the square of the Szeg\"o kernel the piece that
is holomorphic and $\delta$-dependent, neglecting $\delta$-independent
pieces. We find,
\bea
\label{Sdeltasquare}
S_\delta (x,y)^2
=
- {1 \over 4}
\bigg ( xy A_1B_1 - (x+y) (A_1 B_2 + B_1 A_2) + A_2 B_2 \bigg )
{dx \over s(x)}   {dy \over s(y)}
\eea
up to $\delta$-independent terms.
Next, we make the expression for $\Xi_6[\delta]\tet [\delta](0)^4$ in terms of
the hyperelliptic form explicit. To do so, we first express it
in terms of odd spin structures only, using $\delta = \nu_1 + \nu_2 + \nu_3$,
\bea
\Xi_6 [\delta] \tet [\delta ](0)^4
= \sum _{i \leq i < j \leq 3}
\< \nu _i |\nu _j \> \prod _{k \not= i,j} \tet [\nu _i + \nu_j + \nu _k](0)^4
\eea
Hence, $I_{20}$ may be recast in the following form,
\bea
I_{20} (x,y;u,v)
& = & \half \sum _{i<j} t[i,j]
\sum _{l \not= i,j} S_\delta (x,y)^2 S_\delta (u,v)^2
\no \\
t[i,j] & \equiv &
\< \nu _i |\nu _j \> \prod _{k \not= i,j} \tet [\nu _i + \nu_j + \nu _k](0)^4
\eea
where $\delta = \nu _i + \nu _j + \nu _l$. To compute this, we express $t[i,j]$
first in terms of hyperelliptic form with the help of the Thomae formulas,
\bea
t[i,j] = s_0(\det \sigma )^{-8}
(p_i-p_j)^4 \prod _{k \not \in \{i,j\}} (p_k-p_i)(p_k-p_j)
\prod _{k,l \not \in \{i,j\}} (p_k-p_l)^2
\eea
This formula is precise, up to an overall $i,j$-independent sign $s_0$,
which will be determined later by considering asymptotics.
Expressing $S_\delta (x,y)^2$ and $S_\delta (u,v)^2$ in terms
of hyperelliptic data with the help of (\ref{Sdeltasquare}), 
and carrying out the sum over $i<j$, we find
\bea
I_{20} (x,y;u,v) =
s_0
{ \prod _{i<j} (p_i-p_j)^2 ~ dx \, dy \, du \, dv
\over
4 (\det \sigma )^8 ~ s(x) s(y) s(u) s(v) }
\bigg ( (x-u)(y-v)+(x-v)(y-u) \bigg ) \quad
\eea
Converting the antisymmetric biholomorphic 1-form $\Delta$ to the hyperelliptic
form as well,
\bea
\Delta (x,y) = {\det \sigma  \over 4 \pi ^2} (x-y) { dx \, dy  \over s(x) s(y) }
\eea
we have
\bea
I_{20} (x,y;u,v) = 4 \pi ^4 s_0 (\det \sigma )^{-10} \prod _{i<j}(p_i-p_j)^2
\bigg ( \Delta (x,u) \Delta (y,v) +\Delta (x,v) \Delta (y,u) \bigg )
\eea
Using now again the Thomae formula, in the form
\bea
(\det \sigma )^{10} \Psi _{10} = \prod _{i<j} (p_i-p_j)^2
\eea
we obtain $I_{20}$ up to the sign $s_0$. In the subsequent subsection,
we shall show that $s_0=-1$; which leads to our final result,
\bea
\label{I20int}
I_{20} (x,y;u,v) = -4 \pi ^4  \Psi _{10}
\bigg ( \Delta (x,u) \Delta (y,v) +\Delta (x,v) \Delta (y,u)
\bigg )
\eea

\subsubsection{The sign $s_0$ from asymptotics}

The antisymmetrized expression $I_{20}(x,y;u,v) - I_{20} (x,u;y,v)$
may be written in two ways; first, via (\ref{I20int}); second via 
its expression in terms of $\tet$-functions. Omitting a common factor 
of $\Delta (x,v) \Delta (y,u)$, the equality between these 
gives the following equation,
\bea
\label{signeq}
12 \pi ^4 s_0 \Psi _{10} 
=
 \sum _{\delta } \Xi _6 [\delta ] \tet [\delta ](0)^2
\det \p_I \p_J \tet [\delta ](0) \quad
\eea
which will determine the sign $s_0$.
Note that $\det \p _I \p_J \tet [\delta ](0)$ does not transform as a
modular tensor, although its sum against $ \Xi _6 [\delta ] \tet [\delta ](0)^2$
does in view of $I_{17}=I_{18}=0$.

\medskip

In computing the proportionality coefficient between rhs and lhs,
we work in the separating degeneration limit. Throughout, we shall use the
notations and result of \cite{IV}, specifically, section 5.1.
The behavior of the modular form $\Psi_{10}$ is given by
\bea
\Psi _{10} = - 2^{14}  \pi^2 \tau^2 \eta (\tau_1)^{12} \eta (\tau_2)^{12}
+\O (\tau ^4)
\eea
It is of order $\tau^2$, so we shall have to expand the rhs
to order $\tau^2$ as well. The determinant is given by
\bea
\det \p _I \p _J \tet [\delta ] (0,\Omega ) =
4 \pi ^2 (\p _\tau \tet [\delta ])^2 -  16 \pi ^2 \p_{\tau _1}\tet [\delta ]
\p_{\tau _2} \tet [\delta ] 
\eea
Using the expansions for the $\tet$-constants derived in \cite{IV}, its
expansion to order $\tau^2$ is,
\bea
\label{detexpan}
\tet \left [ \matrix{  \mu_1 \cr \mu_2  \cr} \right ] ^2 \det \p _I \p _J
\tet \left [ \matrix{  \mu_1 \cr \mu_2  \cr} \right ]
&=&
- \pi^2 \p \tet _1 [\mu_1] (\tau_1)^4 \p \tet_1 [\mu _2] (\tau _2) ^4
\nonumber \\
&&
+ 12 \pi ^2 \tau ^2 (\p \tet _1 [\mu_1] (\tau_1)^2)^2 (\p \tet_1 [\mu _2]
(\tau _2) ^2)^2
\nonumber \\
&&
- 2 \pi ^2 \tau ^2 (\p \tet _1 [\mu_1] (\tau_1)^2)^2 \p ^2 \tet _1 [\mu
_2] (\tau _2)^4
\nonumber \\
&&
- 2 \pi ^2 \tau ^2 (\p \tet _1 [\mu_2] (\tau_2)^2)^2 \p ^2 \tet _1 [\mu
_1] (\tau _1)^4
\eea
At the leading order, we have $\tet [\delta  _0 ] \to 0$ and
\bea
 \Xi _6 \left [ \matrix{  \mu_1 \cr \mu_2  \cr} \right ]
= - 2^8 \< \mu _1 | \nu _0\> \< \mu _2 |\nu _0\> \eta (\tau _1)^{12} \eta
(\tau _2)^{12} +\O (\tau ^2)
\eea
Perform the sum over the first 9 even spin structures using the genus one
Riemann relation, we find that the leading order term in (\ref{detexpan}) 
indeed sums to 0.

\medskip

To determine the  order $\tau ^2$ terms asymptotics, we proceed as follows. 
In summing over the first 9 spin structures, the last two terms in 
(\ref{detexpan}) clearly sum to 0, and we shall henceforth drop them. 
Similarly, the $\O(\tau ^2)$ terms proportional to 
$\p \ln \eta (\tau _1)^{12}$ and $\p \ln \eta (\tau _2)^{12}$ cancel by the 
same identity. Collecting the remaining terms, and using the 
asymptotics of $\tet$-functions to order $\tau^2$, we have
\bea
&&
\sum  _{\mu _1, \mu _2} \Xi _6 \left [ \matrix{  \mu_1 \cr \mu_2  \cr}
\right ] \tet \left [ \matrix{  \mu_1 \cr \mu_2  \cr} \right ] ^2 \det
\p_I \p_J \tet \left [ \matrix{  \mu_1 \cr \mu_2  \cr} \right ]
 \\
&& \qquad =
2^8 \eta (\tau _1) ^{12} \eta (\tau _2) ^{12} \sum _{\mu _1, \mu_2}
\< \mu _1 |\nu_0\> \< \mu _2 |\nu _0\>
\biggl [
-1 + {3 \over 2} \tau ^2 \p \ln \tet _1 [\mu _1] (\tau _1)^4 \p \ln \tet
_1 [\mu _2] (\tau _2)^4 \biggr ]
\nonumber \\
&& \qquad \qquad \times \biggl [
- \pi ^2 \p \tet _1 [\mu _1] (\tau _1)^4 \p \tet _1 [\mu _2] (\tau _2)^4
+ 12 \pi ^2 \tau ^2 \biggl (\p \tet _1 [\mu _1] (\tau _1)^2 \p \tet _1
[\mu _2] (\tau _2)^2\biggr )^2 \biggr ]
\nonumber
\eea
Combining the two products and cancelling the leading behavior, we are left 
to perform the following genus 1 sum,
\bea
U(\tau) \equiv \sum _\mu \< \mu | \nu _0 \> \biggl ( \p _\tau \tet _1
[\mu ] (\tau) ^2 \biggr )^2
\eea
Using the modular transformation laws of genus 1,
it is readily established that $U(\tau)$ has  precisely the same 
transformation laws as $\eta (\tau )^{12}$.
Comparing asymptotics as $\tau \to i \infty$, we find
$U(\tau) = 4 \pi ^2 \eta (\tau )^{12}$.
Carrying out these sums, we find
\bea
\sum  _{\mu _1, \mu _2} \Xi _6 \left [ \matrix{  \mu_1 \cr \mu_2  \cr}
\right ] \tet \left [ \matrix{  \mu_1 \cr \mu_2  \cr} \right ] ^2 \det
\p_I \p_J \tet \left [ \matrix{  \mu_1 \cr \mu_2  \cr} \right ]
&=&
- 9 \cdot 2^{14} \pi ^6 \tau ^2 \eta (\tau _1)^{24} \eta (\tau _2)^{24}
+\O (\tau ^4)
\nonumber \\
&& \\
\Xi _6 [ \delta _{10}] \tet [\delta _{10}] ^2
\det \p_I \p_J \tet [\delta _{10}]
&=&
- 3 \cdot 2^{14} \pi ^6 \tau ^2 \eta (\tau _1)^{24} \eta (\tau _2)^{24}
+\O (\tau ^4)
\nonumber
\eea
Adding these together and comparing with the asymptotic
expansion of $\Psi _{10}$, we find the sign $s_0=-1$.

\subsubsection{Calculation  of $I_{21}$}

Finally, it remains to compute $I_{21}$, defined by,
\bea
I_{21} (z_1,z_2,z_3,z_4)
=
 \sum _\delta \Xi _6 [\delta ] \, \tet [\delta ](0)^4
S_\delta (z_1, z_2) S_\delta (z_2, z_3) S_\delta (z_3,z_4) S_\delta (z_4,z_1).
\eea
We shall relate $I_{21}$ to $I_{20}$, using the same reasoning
that was used to relate $I_{12}$ to $I_{11}$ in appendix \S C.3.2.
The quantity $I_{21} (z_1,z_2,z_3,z_4)$ is a holomorphic 1-form in 
each $z_i$,  since the poles of  the Szeg\"o kernels cancel in view of identity
$I_{19}=0$. Also, $I_{21}$ is symmetric under the interchange 
of  $z_1$ and $z_3$. In view of the one-to-one correspondence
between holomorphic 2-forms and symmetric biholomorphic 
1-forms, $I_{21} (z_1,z_2,z_3,z_4)$ may be completely
reconstructed from $I_{21} (z,z_2,z,z_4)$, which is given by
\bea
I_{21}(z,z_2,z,z_4) = I_{20} (z,z_2;z,z_4)
=  4 \pi ^4  \Psi _{10}
\Delta (z,z_2) \Delta (z,z_4)
\eea
Finally, using the map $ \omega _{\{ I} (z) \omega _{J \} } (w) ~ \leftrightarrow ~
\omega _I (z) \omega _J (z)$ to reconstruct $I_{21} (z_1,z_2,z_3,z_4)$
from $I_{21}(z,z_2,z,z_4)$, we have 
\bea
I_{21} (z_1,z_2,z_3,z_4)
& = &
 4 \pi ^4  \Psi _{10}
\ep ^{IJ} \ep ^{KL} \omega _{\{ I} (z_1) \omega _{K\}}(z_3)
\omega _J(z_2) \omega _L(z_4)
\no \\
& = &
 2\pi ^4  \Psi _{10} \bigg (
\Delta (z_1,z_2) \Delta (z_3,z_4) - \Delta (z_1,z_4) \Delta (z_2,z_3) \bigg ).
\eea
This completes the proof of the 21 $\delta$-summation identities.

\end{appendix}

\end{document}